\pdfoutput=1
\newcommand*{\ATLASLATEXPATH}{}
\documentclass[cernpreprint, atlasdraft=true, texlive=2016, UKenglish]{\ATLASLATEXPATH atlasdoc}

\usepackage{\ATLASLATEXPATH atlaspackage}
\usepackage{rotating}
\usepackage{multirow}
\usepackage{\ATLASLATEXPATH atlasbiblatex}

\usepackage[hepprocess, hepparticle]{\ATLASLATEXPATH atlasphysics}

\graphicspath{{./}{figures/}}

\RequirePackage{todonotes}
\RequirePackage[capitalize,noabbrev]{cleveref}
\usepackage{HIGG-2017-07-PAPER-defs}
\AtlasTitle{Cross-section measurements of the Higgs boson decaying into a pair of \Pgt{}-leptons in proton--proton collisions at $\sqrt{s}=\SI{13}{\TeV}$ with the ATLAS detector}

\author{The ATLAS Collaboration}

\PreprintIdNumber{CERN-EP-2018-232}

\AtlasJournalRef{\PRD 99 (2019) 072001}
\AtlasDOI{10.1103/PhysRevD.99.072001}

\HepDataRecord{https://www.hepdata.net/record/84820}

\AtlasJournal{\PRD}

\AtlasAbstract{%
  A measurement of production cross sections of the Higgs boson
  in proton--proton collisions is presented in the \Htautau decay channel.
  The analysis is performed using \SI{36.1}{\per\fb} of data
  recorded by the ATLAS experiment at the Large Hadron Collider
  at a center-of-mass energy of $\sqrt{s}=\SI{13}{\TeV}$.
  All combinations of leptonic ($\Pgt \to \Pl \Pgn \Pagn$ with $\Pl = \Pe, \Pgm$) 
  and hadronic ($\Pgt \to \text{hadrons}~ \Pgn$) \Pgt decays are considered.
  The \Htt signal over the expected background from other Standard Model processes
  is established with an observed (expected) significance of 4.4 (4.1) standard deviations.
  Combined with results obtained using data taken
  at 7 and \SI{8}{\TeV} center-of-mass energies, the observed (expected) significance amounts to 6.4 (5.4) standard deviations
  and constitutes an observation of \Htt decays.
  Using the data taken at $\sqrt{s}=\SI{13}{\TeV}$, the total cross section in the \Htautau decay channel 
  is measured to be 
  $\num[round-mode=figures,round-precision=3]{3.76558372941}\,^{+\num[round-mode=figures,round-precision=2]{0.601597224288}}_{-\num[round-mode=figures,round-precision=2]{0.591224858352}}\,\text{(stat.)}\,^{+\num[round-mode=figures,round-precision=2]{0.874736193936}}_{-\num[round-mode=figures,round-precision=2]{0.74335289208}}\,\text{(syst.)}\,\si{\pb}$,
  for a Higgs boson of mass \SI{125}{\GeV} assuming the relative contributions of its production modes as predicted by the Standard Model.
  Total cross sections in the \Htautau decay channel are determined separately for vector-boson-fusion production and gluon--gluon-fusion production to be 
  $\xsecVBF = \num[round-mode=figures,round-precision=2]{0.2800453818} \pm \num[round-mode=figures,round-precision=1]{0.09}\,\text{(stat.)}\,^{+\num[round-mode=figures,round-precision=2]{0.106}}_{-\num[round-mode=figures,round-precision=1]{0.093}}\,\text{(syst.)}\,\si{\pb}$
  and 
  $\xsecggF = \num[round-mode=figures,round-precision=2]{3.096016028} \pm \num[round-mode=figures,round-precision=2]{1.0}\,\text{(stat.)}\,^{+\num[round-mode=figures,round-precision=2]{1.63}}_{-\num[round-mode=figures,round-precision=2]{1.258}}\,\text{(syst.)}\,\si{\pb}$,
  respectively. 
  Similarly, results of a fit are reported in the framework of simplified template cross sections.
  All measurements are in agreement with Standard Model expectations.
}

\begin{document}

\maketitle

\section{Introduction}
\label{sec:intro}

The ATLAS and CMS collaborations discovered~\cite{HIGG-2012-27,CMS-HIG-12-028} a particle consistent with the
Standard Model (SM)~\cite{Glashow:1961tr,Weinberg:1967tq,Salam:1968rm} Higgs boson~\cite{Englert:1964et,Higgs:1964pj,Guralnik:1964eu,Higgs:1966ev,Kibble:1967sv} in 
2012. 
Several properties of this particle, such as its coupling strengths, spin and charge-parity (CP) quantum numbers, were 
studied with 7 and \SI{8}{\TeV} center-of-mass energy ($\sqrt s$) proton--proton 
collision data delivered by the Large Hadron Collider (LHC) in 2011 and 2012, 
respectively, referred to as \enquote{\RunOne{}}.
These results rely predominantly on studies of the bosonic decay modes~\cite{HIGG-2014-06,CMS-HIG-14-009,HIGG-2013-17,CMS-HIG-14-018} 
and have not shown any significant deviations from the SM expectations.

The coupling of the Higgs boson to the fermionic sector has
been established with the observation of the \Htt decay mode with a
signal significance of 5.5$\sigma$ from a combination of ATLAS and CMS results~\cite{HIGG-2013-32,CMS-HIG-13-033,HIGG-2015-07}
using LHC Run-1 data. 
A measurement performed by the CMS Collaboration with Run-2 data at $\sqrt s=\SI{13}{\TeV}$ reached a significance of 4.9$\sigma$
using \SI{35.9}{\per\fb} of integrated luminosity and 5.9$\sigma$ combined with data from \RunOne~\cite{CMS-HIG-16-043}. 
While the Higgs-boson coupling to 
other fermions such as top quarks~\cite{HIGG-2018-13,CMS_ttH_Observation} and bottom quarks~\cite{ATLAS_Hbb_obs,CMS_Hbb_obs} have been observed, 
only upper limits exist on its coupling to muons ~\cite{HIGG-2016-10,CMS_Hmm} 
and the \Htautau decay mode has been the only accessible leptonic decay 
mode. It was also used to 
constrain CP violation in the production via vector-boson fusion (\VBF)~\cite{HIGG-2015-06} 
and is unique in that it provides sensitivity to CP violation in the Higgs-boson coupling to leptons~\cite{HtautauCP_theory}.

This paper presents cross-section times branching-fraction measurements of Higgs bosons that decay into a pair of \Pgt{}-leptons in proton--proton (\pp) collisions at $\sqrt{s}=\SI{13}{\TeV}$ using data collected by the ATLAS experiment in 2015 and 2016, 
corresponding to an integrated luminosity of \SI{36.1}{\per\fb}.  
All combinations of leptonic ($\Pgt \to \Pl \Pgn \Pagn$ with $\Pl = \Pe,
\Pgm$) and hadronic ($\Pgt \to \text{hadrons} \ \Pgn$) \Pgt decays
are considered.\footnote{Throughout this paper
the inclusion of charge-conjugate decay modes is implied.
The symbol \Pl is used to denote electrons and muons, also referred to as \enquote{light leptons}.}
The corresponding three analysis channels are denoted by $\tlep\tlep$,
$\tlep\thad$ and $\thad\thad$ and are composed of different dominant
backgrounds.  While \Ztt is a dominant background in all channels, the
relative contributions from other backgrounds from top-quark and
other vector-boson decays, as well as from misidentified leptonic or hadronic \Pgt decays, vary considerably between the channels.
Two analysis categories
are defined that are predominantly sensitive to Higgs bosons produced via \VBF and
gluon--gluon fusion (\ggF). A maximum-likelihood fit is performed on data
using distributions of the reconstructed di-\Pgt mass in signal regions (SRs), simultaneously with event
yields from control regions (CRs) that are included to constrain
normalizations of major backgrounds estimated from simulation.
The dominant and irreducible \Ztt background is
estimated from simulation. This is different from the search for \Htt
decays in Run~1~\cite{HIGG-2013-32}, which used the embedding
technique~\cite{HIGG-2014-09}. A reliable modeling of this background is therefore
of crucial importance for this analysis. Validation regions (VRs) based on \Zll events are studied, but not included in the fit,
to verify as precisely as possible the modeling of the \Ztt background.

The paper is organized as follows. \Cref{sec:detector} describes the 
ATLAS detector.  This is followed in \cref{sec:samples}
by a description of the dataset and Monte Carlo (MC) simulated samples employed 
by this measurement. \Cref{sec:reco} details the reconstruction 
of particles and jets.
The event selection for each channel and event category as well as signal, control and validation regions are 
discussed in \cref{sec:selection}. Background estimation techniques 
and the systematic uncertainties of the analysis are described in 
\cref{sec:background,sec:systematics}, respectively.
The signal extraction procedure and the results of the Higgs cross-section measurements in the 
\Htt decay mode are presented in \cref{sec:results}.
\Cref{sec:conclusions} gives the conclusions.

\section{The ATLAS detector}
\label{sec:detector}

The ATLAS experiment~\cite{ATLASDetector} at the LHC is a multipurpose particle 
detector with a forward--backward symmetric cylindrical geometry and 
a near-$4\pi$ coverage in solid angle.\footnote{  
The ATLAS Collaboration uses a right-handed coordinate system with its origin at 
the nominal interaction point (IP) in the center of the detector
and the $z$-axis along the beam pipe. The $x$-axis points from the IP to the 
center of the LHC ring, and the $y$-axis points upwards. Cylindrical coordinates
$(r,\phi)$ are used in the transverse plane, $\phi$ being the azimuthal angle 
around the $z$-axis. The pseudorapidity is defined in terms of the polar angle
$\theta$ as $\eta=-\ln\tan(\theta/2)$. Angular distance 
is  measured in units of $\Delta R \equiv  \sqrt{(\Delta\eta)^2 + (\Delta\phi)^2} $.}
It consists of an inner tracking detector
surrounded by a thin superconducting solenoid,
electromagnetic and hadron calorimeters, and a muon spectrometer.
The inner tracking detector covers the pseudorapidity range $|\eta| < 2.5$. 
It consists of a silicon pixel detector, which
has an additional innermost layer (positioned at a radial distance of \SI{3.3}{\cm} from the beam line)
that was installed after Run~1~\cite{IBL,IBL2}, and a
silicon microstrip detector surrounding the pixel detector, both
covering $|\eta| < 2.5$,  followed by a transition radiation straw-tube tracker
covering $|\eta| < 2$.
The inner tracking detector is immersed in a \SI{2}{\tesla} axial magnetic field provided
by the solenoid.
Lead/liquid-argon (LAr) sampling calorimeters provide electromagnetic (EM) energy measurements
with high granularity.
A hadron (steel/scintillator-tile) calorimeter covers the central pseudorapidity range ($|\eta| < 1.7$).
The endcap and forward regions are instrumented with LAr calorimeters 
for both the EM and hadronic energy measurements up to $|\eta| = 4.9$. 
The muon spectrometer surrounds the calorimeters and is based on three large air-core 
toroidal superconducting magnets with eight coils each. 
The field integral of the toroids ranges between \num{2.0} and \SI{6.0}{\tesla\metre}
across most of the detector.
The muon spectrometer includes a system of precision tracking chambers and fast detectors for triggering.

Events are selected using a two-level trigger system.  The 
first-level trigger is implemented in hardware and uses a subset of the 
detector information to filter events that are then processed by a software-based 
high-level trigger.  This further reduces the average recorded collision rate to
approximately \SI{1}{\kilo\hertz}.

\section{Data and simulation samples}
\label{sec:samples}

The data used in this analysis were taken from \pp collisions at the LHC where
proton bunches are collided every \SI{25}{\ns} at $\sqrt{s} = \SI{13}{\TeV}$.
A combination of several triggers for single light leptons, two light leptons and two hadronically decaying \Pgt{}-leptons were used
to record the data for the analysis, depending on the analysis channel (see \cref{sec:presel}). 
After data quality requirements, the samples used for this measurement 
consist of \SI{3.2}{\ifb} of data recorded in 2015, with an average of 14 interactions per 
bunch crossing, and \SI{32.9}{\ifb} recorded in 2016, with an average of 25 interactions
per bunch crossing.

Samples of signal and background processes were simulated using various MC generators as summarized in \cref{tab:MCSamples}.
The signal contributions considered include the following four processes for Higgs-boson production at the LHC: \ggF, \VBF 
and associated production of a Higgs boson with a vector boson (\VH) or with a top--antitop quark pair (\ttH)
where all decay modes for the \Htt process are included.
Other Higgs production processes such as associated production with a bottom--antibottom quark pair and with a single top quark are found to be negligible.
Higgs decays into $\PW\PW$ are considered background and simulated similarly for these production processes.
The mass of the Higgs boson was assumed to be \SI{125}{\GeV}~\cite{HIGG-2014-14}.

\begin{table}
  \caption{Monte Carlo generators used to describe all signal and background processes together with the 
    corresponding PDF set and the model of parton showering, hadronization and underlying event (UEPS).
    In addition, the order of the total cross-section calculation is given. 
    The total cross section for \VBF production is calculated at approximate-NNLO QCD. More details are given in the text.}
  \label{tab:MCSamples}
  \centering
  \vspace{2mm}
  \small
  \begin{tabular}{lllll}
    \toprule
    Process & Monte~Carlo generator & PDF & UEPS & Cross-section order \\
    \midrule
    \ggF & \powhegbox v2 & PDF4LHC15 NNLO & \pythiaEightTwo & N$^3$LO QCD + NLO EW \\
    \VBF & \powhegbox v2 & PDF4LHC15 NLO  & \pythiaEightTwo & $\sim$NNLO QCD + NLO EW \\
    \VH  & \powhegbox v2 & PDF4LHC15 NLO  & \pythiaEightTwo & NNLO QCD + NLO EW \\
    \ttH & MG5\_aMC@NLO v2.2.2 & \textsc{NNPDF30LO} & \pythiaEightTwo & NLO QCD + NLO EW \\
    \midrule
    $\PW/\Zjets$ & \sherpa 2.2.1 & \textsc{NNPDF30NNLO} & \sherpa 2.2.1 & NNLO \\
    $VV$/$V\Pggx$ & \sherpa 2.2.1 & \textsc{NNPDF30NNLO} & \sherpa 2.2.1 & NLO \\
    \ttbar & \powhegbox v2 & CT10 & \pythiaSix & NNLO+NNLL \\
    $\PW\Pqt$ & \powhegbox v1 & CT10F4 & \pythiaSix & NLO \\
    \bottomrule  
  \end{tabular}
\end{table}

Higgs production by \ggF was simulated with the \powhegbox v2~\cite{Powheg_Nason,Powheg_Frixione,Powheg_Alioli,Powheg_Bagnaschi} NNLOPS program~\cite{Hamilton:2013fea} at next-to-leading-order (NLO) accuracy in quantum chromodynamics (QCD) using the MiNLO approach~\cite{Hamilton:2015nsa}, and reweighted to next-to-next-to-leading order (NNLO) in QCD in the Higgs rapidity.
The \VBF and \VH production processes were simulated at NLO accuracy in QCD using \powhegbox with the MiNLO approach.
The \ttH production process was simulated with \MCatNLO v2.2.2~\cite{McAtNLO} at NLO accuracy in QCD.
For these signal samples, the simulation was interfaced to the \pythiaEightTwo~\cite{Pythia8} model of parton showering, hadronization and underlying event (UEPS).
To estimate the impact of UEPS uncertainties, the \ggF, \VBF and \VH samples were also simulated with the \herwig~\cite{HerwigPP,Herwig7} UEPS model.
The \textsc{PDF4LHC15}~\cite{Butterworth:2015oua} parameterization of the parton
distribution functions (PDFs) was used for these production processes.
The
\textsc{AZNLO}~\cite{STDM-2012-23} set of tuned parameters was used, with the
\textsc{CTEQ6L1}~\cite{PDF-CTEQ6L1} PDF set, for the modeling of
non-perturbative effects.
For the \ttH production process the \textsc{NNPDF30LO}~\cite{PDF-NNPDF30} PDF parametrization was used in the matrix element and the \textsc{NNPDF23LO}~\cite{PDF-NNPDF} PDF parametrization for the UEPS model with the \textsc{A14}~\cite{ATL-PHYS-PUB-2014-021} set of tuned parameters for the modeling of non-perturbative effects.
\textsc{Photos++} version 3.52~\cite{PhotosPP} was used for QED emissions from electroweak (EW) vertices and charged leptons.

The overall normalization of the \ggF process is taken from a
next-to-next-to-next-to-leading-order (N$^3$LO)
QCD calculation with NLO EW corrections
included~\cite{Anastasiou:2015ema,Anastasiou:2016cez,Actis:2008ug,Anastasiou:2008tj}.
Production by \VBF is normalized to an approximate-NNLO QCD cross
section with NLO EW corrections included~\cite{Ciccolini:2007jr,Ciccolini:2007ec,Bolzoni:2010xr}.
The \VH samples are normalized to cross sections calculated at NNLO in QCD,
with NLO EW corrections included~\cite{Brein:2003wg,Altenkamp:2012sx,Denner:2011id}.
The \ttH process is normalized to a cross section calculated at NLO in QCD with
NLO EW corrections applied~\cite{ttH_NLO1,ttH_NLO2,ttH_NLO3, ttH_NLO4,ttH_NLO5,ttH_NLO6}. 

Background samples of EW production of $\PW/\PZ$ bosons from \VBF, $\PW/\PZ$-boson production with associated jets and diboson production processes were simulated
with the \sherpa 2.2.1~\cite{Sherpa} generator.
Matrix elements were calculated using the Comix~\cite{Gleisberg:2008fv} and
OpenLoops~\cite{Cascioli:2011va} matrix-element generators and merged
with the \sherpa UEPS model~\cite{Schumann:2007mg} using the
\textsc{ME+PS@NLO} prescription~\cite{Hoeche:2012yf}.
For \PW and \PZ production with associated jets the matrix elements were calculated for up to two partons at NLO and four partons at LO precision.
Their inclusive cross sections are normalized to NNLO calculations from \textsc{Fewz}~\cite{Melnikov:2006kv,Anastasiou:2003ds}.
In particular, the dominant \Ztt background is estimated using these simulations of \PZ-boson production.
For diboson production, the matrix elements were calculated for up to one additional parton at NLO and up to three additional partons at LO precision.
For all samples the \textsc{NNPDF30NNLO}~\cite{PDF-NNPDF30} PDF set was used together with
the \sherpa UEPS model.

The impact of UEPS uncertainties, and other modeling uncertainties such as LO/NLO precision comparison for leading jets, on the main background from \Ztt is studied in
an alternative sample which was simulated using \MCatNLO 2.2.2~\cite{McAtNLO} at leading order interfaced to the \pythiaEightOne UEPS model. The A14 set of tuned parameters~\cite{ATL-PHYS-PUB-2014-021} was used together with the \textsc{NNPDF23LO} PDF set~\cite{PDF-NNPDF}.

For the generation of \ttbar production, the \powhegbox
v2~\cite{Powheg_Nason,Powheg_Frixione,Powheg_Alioli,Alioli:2011as}
generator with the \textsc{CT10} PDF sets in the matrix element
calculations was used.
The predicted \ttbar cross section was calculated with the \textsc{Top++}2.0 program to NNLO in perturbative QCD, including soft-gluon resummation to next-to-next-to-leading-log order~\cite{Czakon:2011xx}.
Single top-quark production of $\PW\Pqt$ was simulated
using the \powhegbox v1~\cite{Alioli:2009je,Re:2010bp} generator.
This generator uses the four-flavor scheme for the NLO matrix-element calculations together 
with the fixed four-flavor PDF set \textsc{CT10F4}.
For all top-quark production processes, top-quark spin correlations
were preserved, using
MadSpin~\cite{Madspin} for the t-channel.
The parton shower, hadronization, and the
underlying event were simulated using \pythiaSix~\cite{Pythia6} with the \textsc{CTEQ6L1} PDF set and the
corresponding Perugia 2012 set of tuned parameters~\cite{Perugia2012}.
The top mass was assumed to be \SI{172.5}{\GeV}.
The EvtGen v.1.2.0 program~\cite{EvtGen} was used for the properties of \Pqb- and \Pqc-hadron decays.

For all samples, a full simulation of the ATLAS detector
response~\cite{ATLAS_Simulation} using the \textsc{Geant4}
program~\cite{GEANT4} was performed. 
The effect of multiple \pp interactions in the same and neighboring bunch
crossings (pileup) was included by overlaying minimum-bias events simulated
with \pythiaEightOne using the MSTW2008LO PDF~\cite{MSTW2008LO} 
and the A2~\cite{A2tune} set of tuned parameters
on each generated signal and background event.
The number of overlaid events was chosen such that the distribution of the average
number of interactions per \pp bunch crossing in the simulation matches that
observed in data.

\section{Object reconstruction}
\label{sec:reco}

Electron candidates are reconstructed from energy deposits in the
electromagnetic calorimeter associated with a charged-particle track
measured in the inner detector.  The electron candidates
are required to pass the \enquote{loose} likelihood-based identification
selection of Refs.~\cite{PERF-2013-05,ATLAS-CONF-2016-024},
to have transverse momentum
$\pt > \SI{15}{\GeV}$ and to be in the fiducial volume of the 
inner detector, $\abseta<2.47$. The transition region between the barrel 
and endcap calorimeters ($1.37<\abseta<1.52$) is excluded.
The trigger efficiency for single electrons selected in the analysis ranges between 90\% and 95\%~\cite{TRIG-2016-01}.
Electron candidates are ignored if they share their reconstructed track with a muon candidate defined below
or if their angular distance from a jet is within $0.2 < \Delta R < 0.4$.

Muon candidates are constructed by matching an inner detector track
with a track reconstructed in the muon spectrometer~\cite{PERF-2015-10}.
The muon candidates are required to have $\pt > \SI{10}{\GeV}$
and $\abseta<2.5$ and to pass the \enquote{loose} muon identification requirements of Ref.~\cite{PERF-2015-10}.
The trigger efficiency for single muons selected in the analysis is close to 80\% (70\%) in the barrel in the 2016 (2015) dataset and 90\% in the endcaps~\cite{TRIG-2016-01}.
Muon candidates are ignored if their angular distance from a jet is $\Delta R < 0.4$ with the following exceptions:
If $\Delta R < 0.2$ or the muon track is associated with the jet, and if the jet has either
less than three tracks or less than twice the transverse momentum of the
muon candidate, the jet is removed instead.
This recovers efficiency for muons that radiate a hard bremsstrahlung photon in the calorimeter.

In the \tll and \tlhad signal regions, events are selected only if the selected electron and muon candidates satisfy their respective \enquote{medium} identification criteria.
The reconstruction and identification efficiency for muons with the \enquote{medium} identification requirement has been measured in \Zmm events~\cite{PERF-2015-10}. It is well above 98\% over the full phase space, except for $|\eta| < 0.1$ where the reconstruction efficiency is about 70\%.
The combined identification and reconstruction efficiency for \enquote{medium} electrons ranges from 80\% to 90\% in the \pT range of \SIrange{10}{80}{\GeV} as measured in \Zee events~\cite{ATLAS-CONF-2016-024}.
In addition, the electrons and muons must satisfy the \enquote{gradient} isolation criterion, which requires that there are no additional high-\pT tracks in a cone around the track and no significant energy deposits in a cone around the calorimeter clusters of the object after correcting for pileup. The size of the respective cones depends on the \pT of the light lepton.
This isolation requirement rejects about 10\% of light leptons for low \pT and less than 1\% for $\pT > \SI{60}{\GeV}$~\cite{PERF-2015-10,ATLAS-CONF-2016-024}.

Jets are reconstructed from topological clusters in the calorimeter using 
the anti-$k_t$ algorithm~\cite{AntiKT,PERF-2014-07}, with a radius parameter 
value $R = 0.4$, and have $\pt > \SI{20}{\GeV}$ and  $\abseta < 4.9$. 
To reject jets from pileup a \enquote{Jet Vertex Tagger} (JVT)~\cite{PERF-2014-03} 
algorithm is used for jets with $\pt < \SI{50}{\GeV}$ and  $\abseta < 2.4$. It
employs a multivariate technique that relies on jet-tracking and calorimeter-cluster-shape variables 
to determine the likelihood that the jet originates from pileup.
Similarly, pileup jets in the forward region are suppressed with a \enquote{forward JVT}~\cite{PERF-2016-06}
algorithm, relying in this case only on calorimeter-cluster-shape variables, which is applied to all jets with $\pt < \SI{50}{\GeV}$ and $\abseta > 2.5$.
In the pseudorapidity range $\abseta<2.5$, $\Pqb$-jets are selected
using a multivariate algorithm~\cite{PERF-2012-04,ATL-PHYS-PUB-2016-012}.
A working point is chosen that corresponds to an efficiency of approximately 85\% for $\Pqb$-jets and rejection factors of 2.8 and 28 for $\Pqc$-jets and light-flavor jets, respectively, in simulated \ttbar events.
A jet is ignored if it is within $\Delta R = 0.2$ of an electron or hadronically decaying \Pgt candidate.

Leptonic \Pgt decays are reconstructed as electrons and muons. The reconstruction of the visible decay products of hadronic \Pgt decays (\tauhadvis)~\cite{ATLAS-CONF-2017-029} starts with a reconstructed jet
that has $\pt > \SI{10}{\GeV}$ and $|\eta| < 2.5$.
As in the case of electron reconstruction the transition region between the barrel and endcap calorimeters is excluded.
To discriminate \tauhadvis from jets initiated by light-quarks or gluons, an identification algorithm using multivariate techniques is applied to \tauhadvis candidates.
They have to pass the \enquote{loose} identification requirement of Ref.~\cite{ATLAS-CONF-2017-029}.
In addition, the \tauhadvis candidates are required to have $\pt > \SI{20}{\GeV}$, to have one or three associated tracks and an absolute electric charge of one.
Their energy is reconstructed by multivariate regression techniques using information about the associated tracks and calorimeter clusters, as well as the average number of collisions recorded.
The trigger efficiency per \tauhadvis selected in the analysis is 95\% and 85\% for 1-prong and 3-prong \Pgt{}-leptons, respectively~\cite{ATLAS-CONF-2017-061}.
The \tauhadvis candidates are ignored if they are within $\Delta R = 0.2$ of a muon or electron candidate
or if they have a high likelihood score of being an electron~\cite{ATLAS-CONF-2016-024}.
The requirement on the likelihood score corresponds to a \tauhadvis efficiency measured in \Ztt decays of 95\%~\cite{ATLAS-CONF-2017-029}.

In the \tlhad signal regions, events are selected only if the \tauhadvis candidate passes the \enquote{medium} identification requirement, corresponding to an efficiency of 55\% and 40\% for real 1-prong and 3-prong \tauhadvis, respectively~\cite{ATLAS-CONF-2017-029}. In addition,
if a 1-prong \tauhadvis candidate and an electron candidate are selected, a dedicated multivariate algorithm to reject electrons misidentified as \tauhadvis is applied to suppress \Zee events.
In the \thadhad signal regions, both selected \tauhadvis candidates have to fulfill the \enquote{tight} identification requirement, which corresponds to a selection efficiency of 45\% for real 1-prong \tauhadvis and 30\% for real 3-prong \tauhadvis~\cite{ATLAS-CONF-2017-029}.

The missing transverse momentum vector is calculated as the
negative vectorial sum of the \pt of the fully calibrated and
reconstructed physics objects~\cite{Aaboud:2018tkc}.
This procedure includes a soft
term, which is calculated from the inner detector tracks that originate
from the vertex associated with the hard-scattering process and that are not associated with any
of the reconstructed objects.
The missing transverse momentum (\met) is defined as the magnitude of this vector.

The Higgs-boson candidate is reconstructed from the visible decay products of the \Pgt{}-leptons
and from the \met, which is assumed to originate from the final-state neutrinos.
The di-\Pgt invariant mass (\mMMC) is determined using the missing-mass calculator (MMC)~\cite{MMC}.
The standard deviation of the reconstructed di-\Pgt mass is \SI{17.0}{\GeV}, \SI{15.3}{\GeV} and \SI{14.7}{\GeV} for signal events selected in the \tll, \tlhad and \thadhad channels, respectively.
The \pt of the Higgs-boson candidate (\pTH) is computed as the vector sum of the transverse momenta
of the visible decay products of the \Pgt{}-leptons and the missing transverse momentum vector.

\section{Event selection and categorization}
\label{sec:selection}

In addition to data quality criteria that ensure that the detector was functioning
properly, events are rejected if they contain reconstructed jets 
associated with 
energy deposits that can arise from hardware problems, beam-halo events
or cosmic-ray showers. Furthermore, events are required to have at least one reconstructed primary vertex with at least
two associated tracks with $\pT > \SI{0.5}{\GeV}$, which rejects non-collision events originating from cosmic rays or beam-halo
events.
The primary vertex is chosen as the \pp vertex candidate with the highest sum of the squared transverse momenta of all associated tracks.

The triggers and event selection for the three analysis channels are described in \cref{sec:presel}.
Selected events are categorized into exclusive signal regions, with enhanced signal-to-background 
ratios. In addition, control regions are defined where a specific background is dominant, and thus a CR  
facilitates the adjustment of the simulated prediction of a background contribution to match the observed data.
The signal and control regions are included in the fit described in \cref{sec:results}.
They are described in \cref{sec:categories} together with validation regions (VRs) used to validate the simulation of the dominant \Zjets background.

\subsection{Event selection}
\label{sec:presel}

Depending on the trigger, transverse momentum requirements are applied to selected electron, muon, and
\tauhadvis~candidates.
They are summarized in \cref{tab:triggers} and their per-object efficiencies are given in \cref{sec:reco}.
Due to the increasing luminosity and the different
pileup conditions, the \pT thresholds of the triggers were increased during
data-taking in 2016, which is taken into account in the \pT requirements of the event selection.
In the \tll channel, the triggers for multiple light leptons are used only if the
highest-\pT light lepton does not pass the corresponding single-light-lepton trigger
\pT requirement. This ensures that each trigger selects an exclusive set of events.

\begin{table}
  \caption{Summary of the triggers used to select events for the
    three analysis channels during 2015 and 2016
    data-taking and the corresponding \pt requirements applied in the
    analysis.
    For the electron+muon trigger the first number corresponds to the electron \pT requirement, the 
    second to the muon \pT requirement.
    For the \thadhad channel, at least one high-\pT jet in addition to the two \tauhadvis candidates is required for the 2016 dataset (see \cref{sec:presel}).}
  \label{tab:triggers}
  \centering
  \vspace{2mm}
    \begin{tabular}{llcc}
      \toprule
      Analysis & Trigger & \multicolumn{2}{c}{Analysis \pt requirement [\GeV]} \\
       channel &         & 2015 & 2016 \\
      \midrule
      \multirow{2}{*}{\tll \& \tlhad} & Single electron  & 25 & 27 \\
                & Single muon      & 21 & 27 \\
      \midrule
      \multirow{3}{*}{\tll} & Dielectron & 15 / 15 & 18 / 18 \\
                & Dimuon & 19 / 10 & 24 / 10 \\
                & Electron+muon  & 18 / 15 & 18 / 15 \\
      \midrule
      \thadhad & Di-\tauhadvis  & 40 / 30 & 40 / 30 \\
      \bottomrule
    \end{tabular}
\end{table}

All channels require the exact number of identified \enquote{loose}
leptons, i.e.\ electrons, muons and \tauhadvis, as defined in \cref{sec:reco}, corresponding to their respective final state.
Events with additional \enquote{loose} leptons are rejected.
The two leptons are required to be of opposite charge and they have to fulfill the \pT requirements of the respective trigger shown in \cref{tab:triggers}.
The selected \tauhadvis in the \tlhad channel is required to have $\pT > \SI{30}{\GeV}$.

The event selection for the three analysis channels is summarized in \cref{t:selection}.
Only events with $\met > \SI{20}{\GeV}$ are selected to reject events without neutrinos.
In the \tll channel with two same-flavor (SF) light leptons this requirement is further tightened to suppress the large \Zll background.
For the same reason, requirements are tightened on the invariant mass of two light leptons (\mll) and a requirement is introduced on the \met calculated only from the physics objects without the soft track term (\METHPTO).
Requirements on the angular distance between the visible decay products of the two selected \Pgt{}-lepton decays (\dRtt) and their
pseudorapidity difference ($|\detatt|$) are applied in all channels to reject non-resonant background events.
Requirements are applied to the fractions of the \Pgt{}-lepton momenta carried by each visible decay product
$x_i=p^\text{vis}_i/\left(p^\text{vis}_i+p^\text{miss}_i\right)$, where $p^\text{vis}_i$ and $p^\text{miss}_i$ are the visible and missing momenta of the $i$th \Pgt lepton, ordered in descending \pT, calculated in the collinear
approximation~\cite{CollinearMass}, to suppress events with \met that is incompatible with a di-\Pgt decay.
Low transverse mass (\mT), calculated from \MET and the momentum of the selected light lepton,
is required in the \tlhad channel to reject events with leptonic \PW decays.
A requirement on the di-\Pgt mass calculated in the collinear approximation (\mcoll) of $\mcoll > m_{\PZ}-\SI{25}{\GeV}$ is introduced in the \tll channel to suppress events from \Zll and to ensure orthogonality between this measurement and the measurement of $\HWW\to\Pl\Pgn\Pl\Pgn$~\cite{HIGG-2013-13}, which has a similar final state.

All channels require at least one jet ($j_1$) with $\pTjl > \SI{40}{\GeV}$ to select Higgs bosons produced by VBF and to suppress background from \Ztt events when selecting Higgs bosons produced through \ggF.
Since 2016 the di-\tauhadvis first-level trigger requires a jet with $\pT > \SI{25}{\GeV}$ calibrated at trigger level with $|\eta| < 3.2$ in addition to the two \tauhadvis candidates.
In the \thadhad channel the jet \pt requirement is thus raised to $\pTjl > \SI{70}{\GeV}$ to achieve uniform trigger selection efficiency as a function of \pTjl. The trigger efficiency for the additional jet ranges from 95\% to 100\% for these requirements.
In the \tll and \tlhad channels, the top-quark background is suppressed by requiring that no jet with $\pT>\SI{25}{\GeV}$ is tagged as a $\Pqb$-jet. 

\begin{table}
  \caption{Summary of the event selection requirements for the three analysis channels
    that are applied in addition to the respective lepton \pT requirements listed in \cref{tab:triggers}.
    \METHPTO is an alternative \met calculated only from the physics objects without the soft-track term.
    The transverse mass (\mT) is calculated from \MET and the momentum of the selected light lepton.
    The visible momentum fractions $x_1$ and $x_2$ of the respective \Pgt{}-lepton and the collinear di-\Pgt mass (\mcoll) are calculated in the collinear approximation~\cite{CollinearMass}.}
  \label{t:selection}
  \centering
  \vspace{2mm}
  \small
  \begin{tabular}{c|c c c}
    \toprule
    \multicolumn{2}{c}{\tll}                                 & \tlhad                               & \thadhad \\
    $\Pe\Pe$/$\Pgm\Pgm$        & $\Pe\Pgm$                    &                                      &          \\
    \midrule
    \multicolumn{2}{c}{$N^\text{loose}_{\Pe/\Pgm}=2$, $N^\text{loose}_{\tauhadvis}=0$} & $N^\text{loose}_{\Pe/\Pgm}=1$, $N^\text{loose}_{\tauhadvis}=1$ & $N^\text{loose}_{\Pe/\Pgm}=0$, $N^\text{loose}_{\tauhadvis}=2$ \\
    \multicolumn{2}{c}{$\Pe$/$\Pgm$ : Medium, gradient iso.} & $\Pe$/$\Pgm$ : Medium, gradient iso. &          \\
    \multicolumn{2}{c}{}                                     & \tauhadvis : Medium                  & \tauhadvis : Tight \\
    \multicolumn{2}{c}{Opposite charge}                      & Opposite charge                      & Opposite charge \\
    \multicolumn{2}{c}{$\mcoll > m_{\PZ}-\SI{25}{\GeV}$}     & $\mT < \SI{70}{\GeV}$                &          \\
    $30<\mll<\SI{75}{\GeV}$    & $30<\mll<\SI{100}{\GeV}$     &                                      &          \\
    $\met > \SI{55}{\GeV}$     & $\met > \SI{20}{\GeV}$       & $\met > \SI{20}{\GeV}$               & $\met > \SI{20}{\GeV}$ \\
    $\METHPTO > \SI{55}{\GeV}$ &                              &                                      &          \\
    \multicolumn{2}{c}{$\dRll < 2.0$}                        & $\dRlt < 2.5$                        & $0.8 < \dRtt < 2.5$ \\
    \multicolumn{2}{c}{$|\detall| < 1.5$}                    & $|\detalt| < 1.5$                    & $|\detatt| < 1.5$ \\
    \multicolumn{2}{c}{$0.1<x_1<1.0$}                        & $0.1<x_1<1.4$                        & $0.1<x_1<1.4$ \\
    \multicolumn{2}{c}{$0.1<x_2<1.0$}                        & $0.1<x_2<1.2$                        & $0.1<x_2<1.4$ \\
    \multicolumn{2}{c}{$\pTjl > \SI{40}{\GeV}$}              & $\pTjl > \SI{40}{\GeV}$              & $\pTjl > \SI{70}{\GeV}, |\eta_{j_1}| < 3.2$ \\
    \multicolumn{2}{c}{$N_\text{\Pqb-jets} = 0$}             & $N_\text{\Pqb-jets} = 0$             &          \\
    \bottomrule
  \end{tabular}
\end{table}

\subsection{Signal, control and validation regions}
\label{sec:categories}

To exploit signal-sensitive event topologies, a \enquote{\VBF} and a \enquote{boosted} analysis category are defined without any overlap in phase space.
The \VBF category targets events with a Higgs boson produced by \VBF and is characterized by the presence of a second high-\pT jet ($\pTjsl > \SI{30}{\GeV}$).
In addition, the two jets are required to be in opposite hemispheres of the detector with a large pseudorapidity separation of $|\detajj| > 3$ and their invariant mass (\mjj) is required to be larger than \SI{400}{\GeV}.
The selected leptons are required to have $\eta$-values that lie between those of the two jets (\enquote{central leptons}).
Although this category is dominated by \VBF production, it also includes significant contributions from \ggF production, amounting to up to 30\% of the total expected Higgs-boson signal.

The boosted category targets events with Higgs bosons produced through \ggF with additional recoiling jets, which is motivated by the harder \pT-spectrum of the \Htt signal compared to the dominant background from \Ztt.
It contains all events with $\pTH > \SI{100}{\GeV}$ that do not pass the \VBF selection.
In addition to events from \ggF, the boosted categories contain sizable contributions from \VBF and \VH production of 10--20\% of the expected signal.
Events that pass the event selection, detailed in \cref{t:selection}, but do not fall into the \VBF or boosted categories, are not used in the analysis.

\begin{figure}[htbp]
 \centering
 \includegraphics[width=\textwidth]{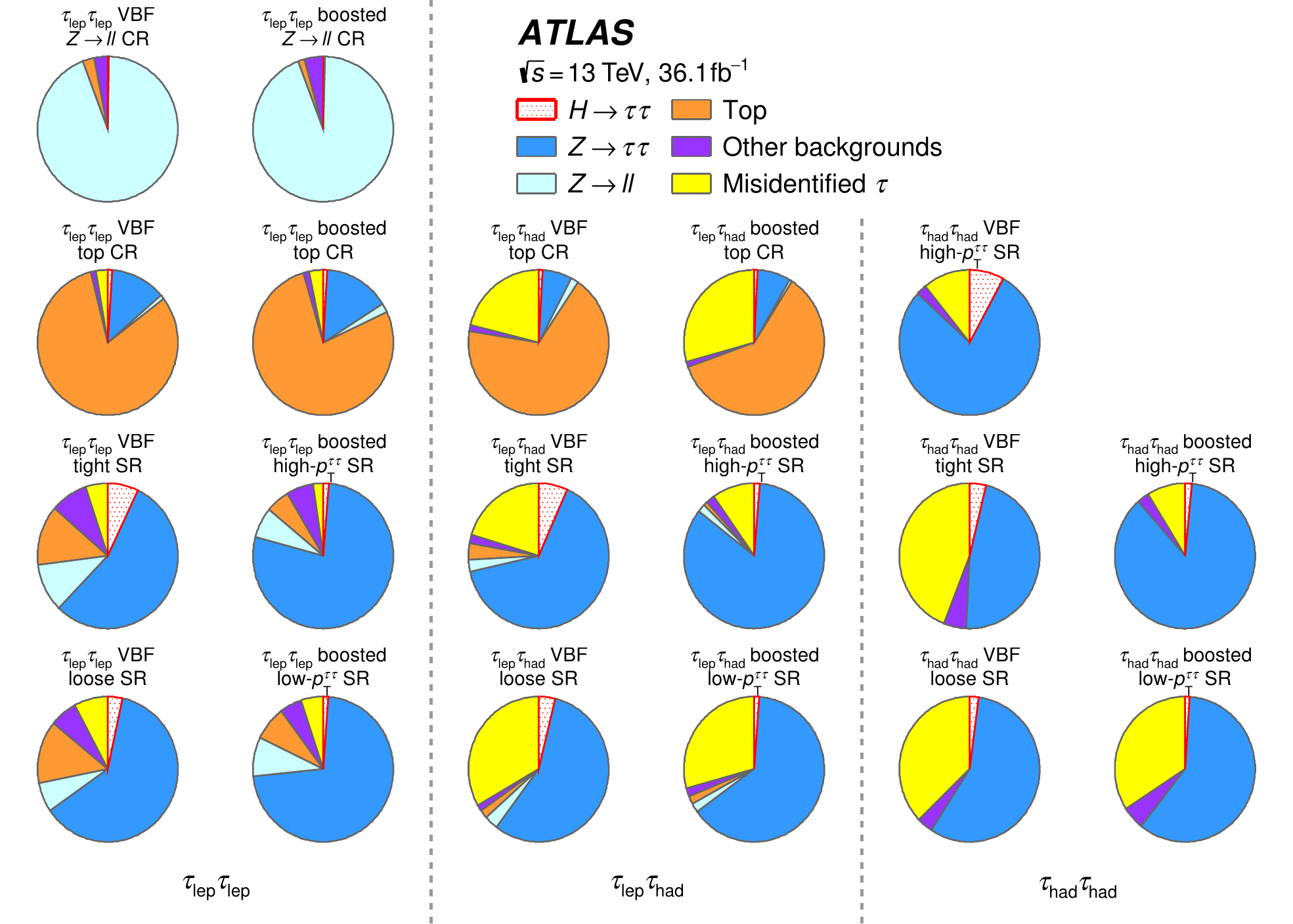}
 \caption{Expected signal and background composition in 6 control regions (CRs) and the 13 signal regions (SRs) used in the analysis.}
 \label{fig:bkgcomp}
\end{figure}

\begin{figure}[htbp]
   \centering
   \subfloat[]{\includegraphics[width=0.33\textwidth]{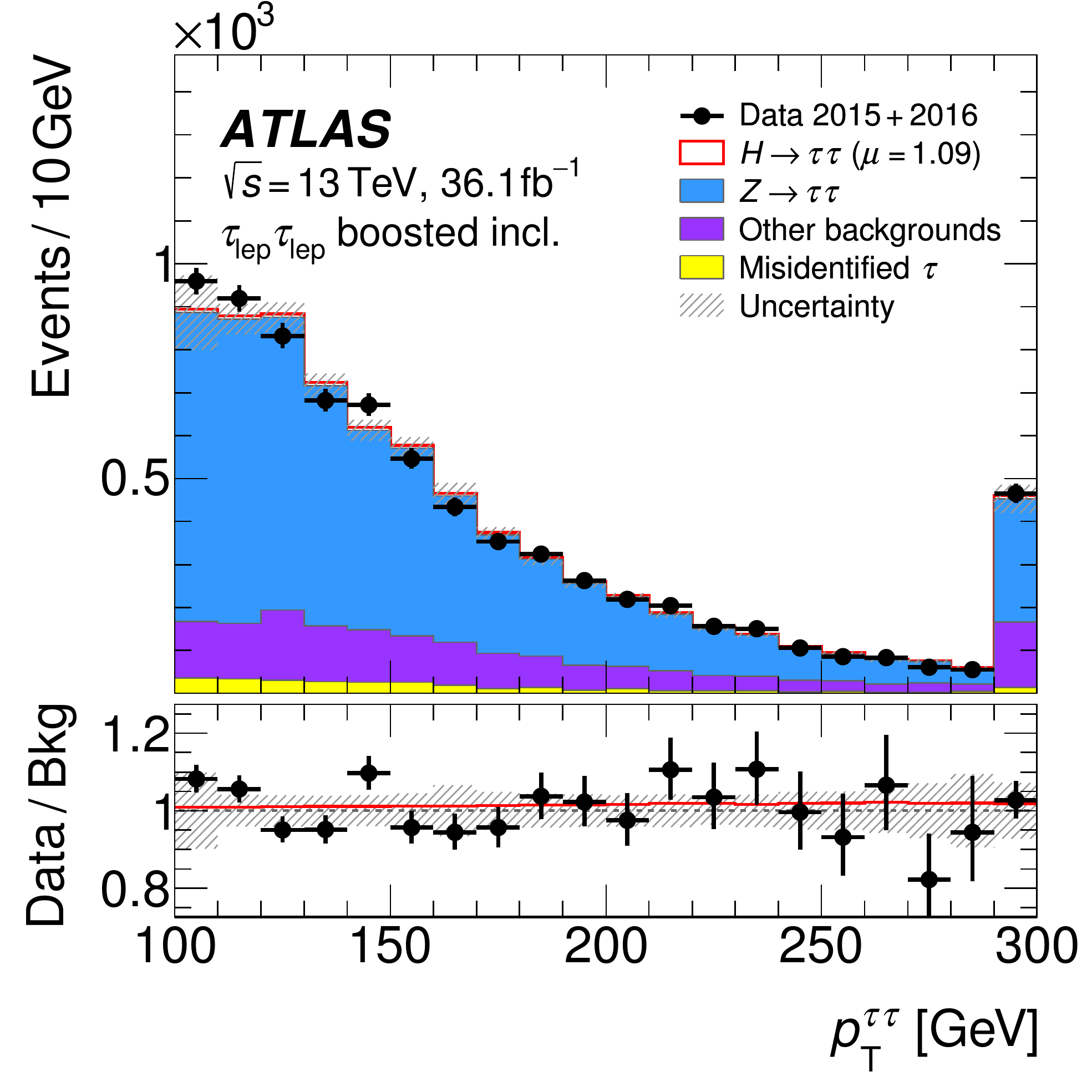} }
   \subfloat[]{\includegraphics[height=0.33\textwidth]{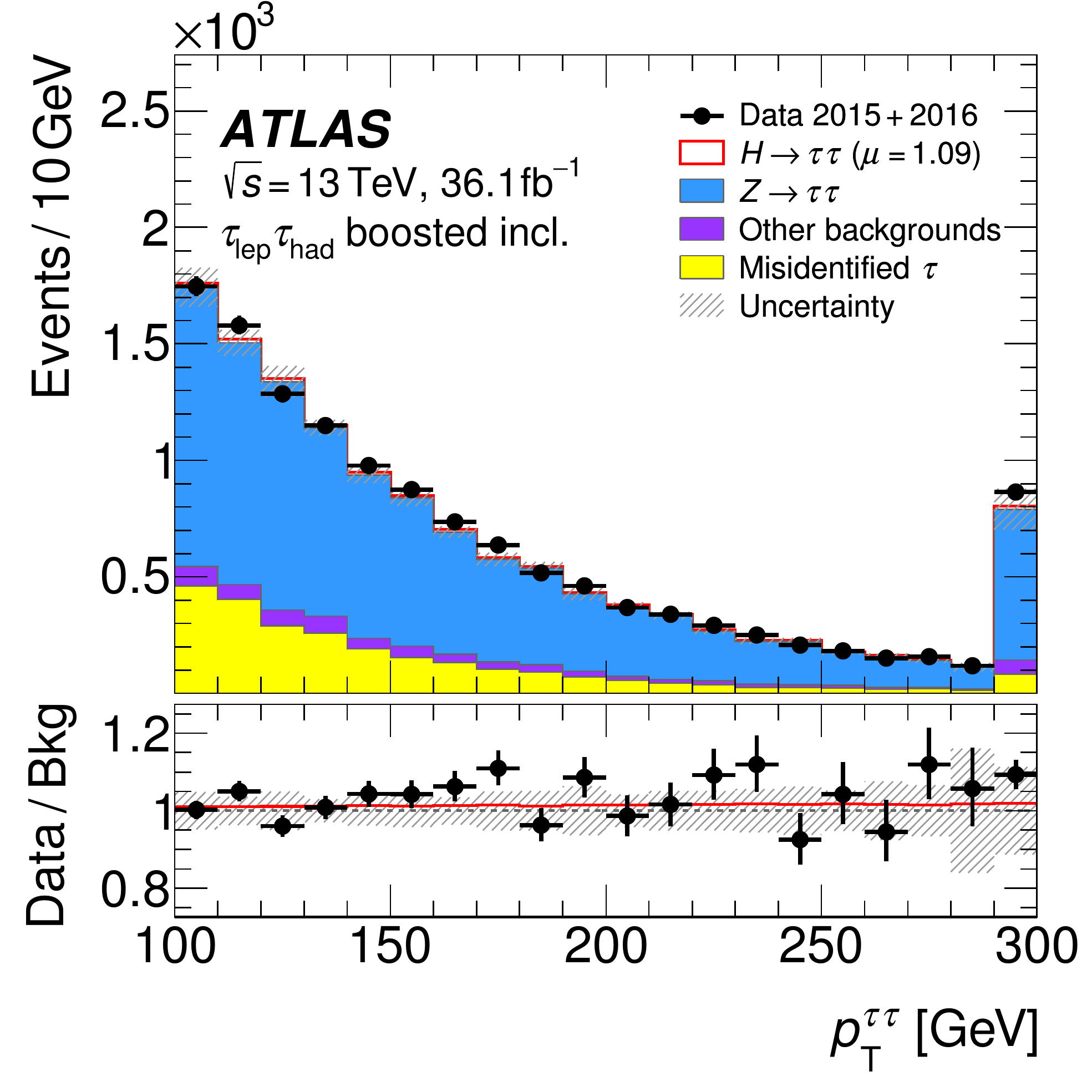} }
   \subfloat[]{\includegraphics[height=0.33\textwidth]{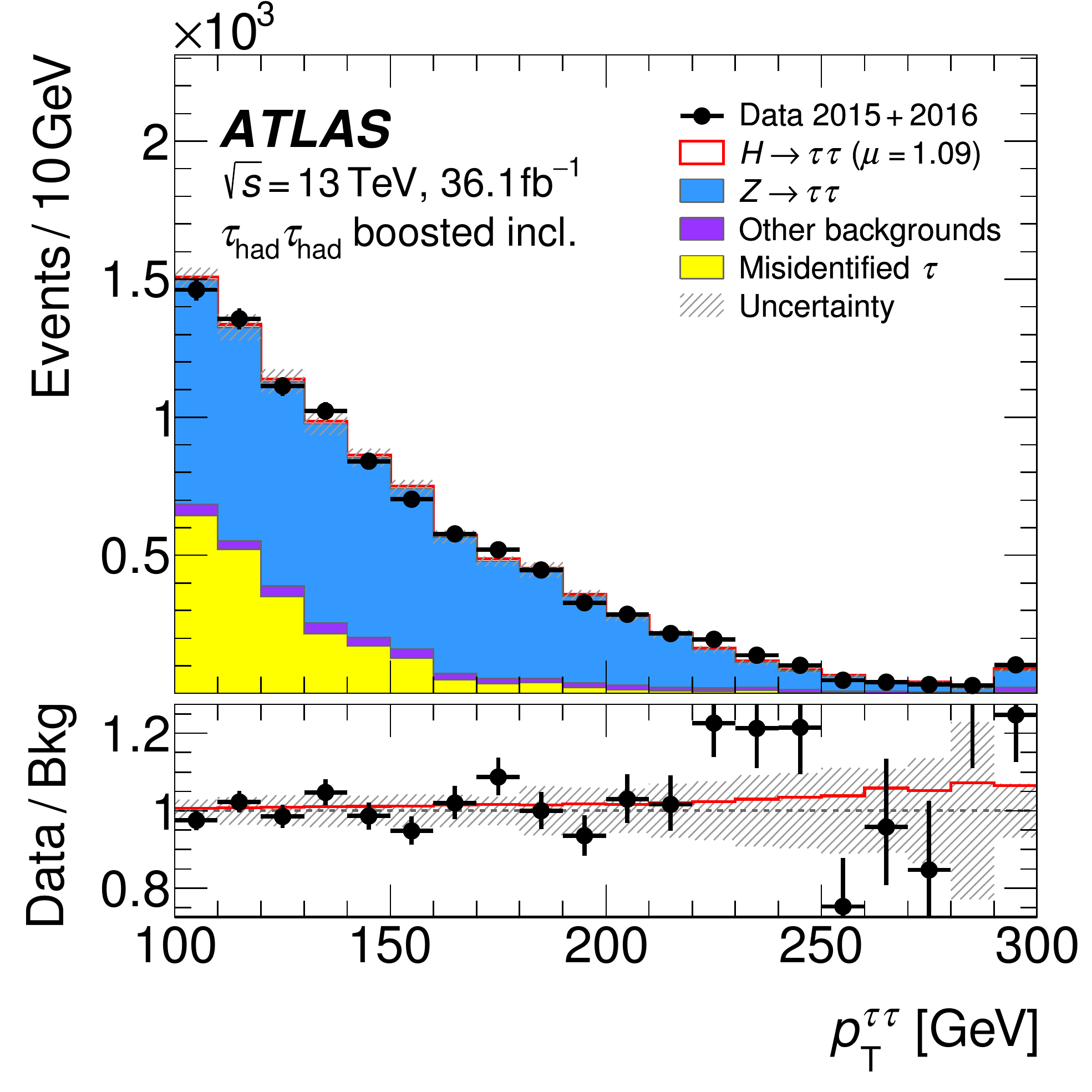} } \\
   \subfloat[]{\includegraphics[width=0.33\textwidth]{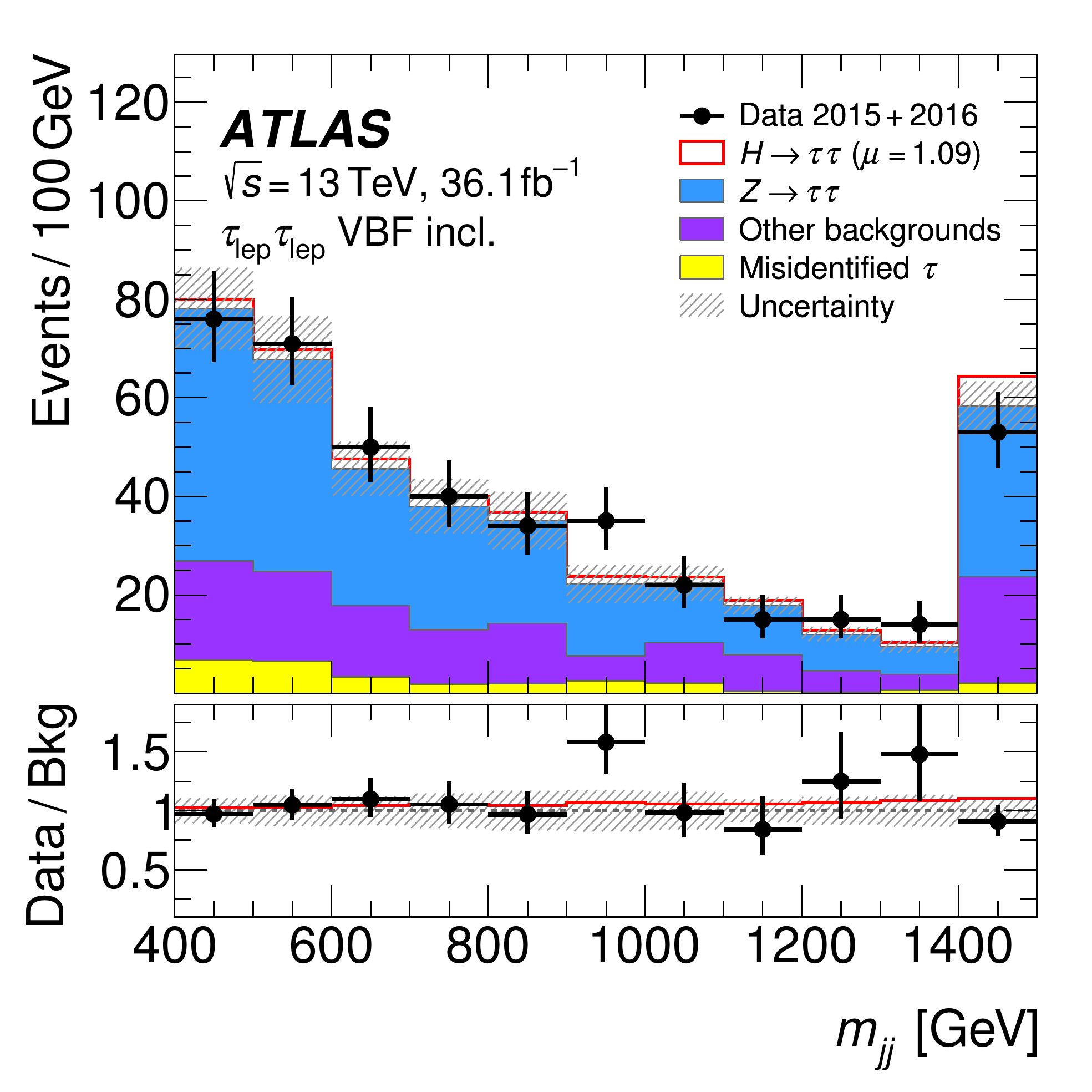} }
   \subfloat[]{\includegraphics[height=0.33\textwidth]{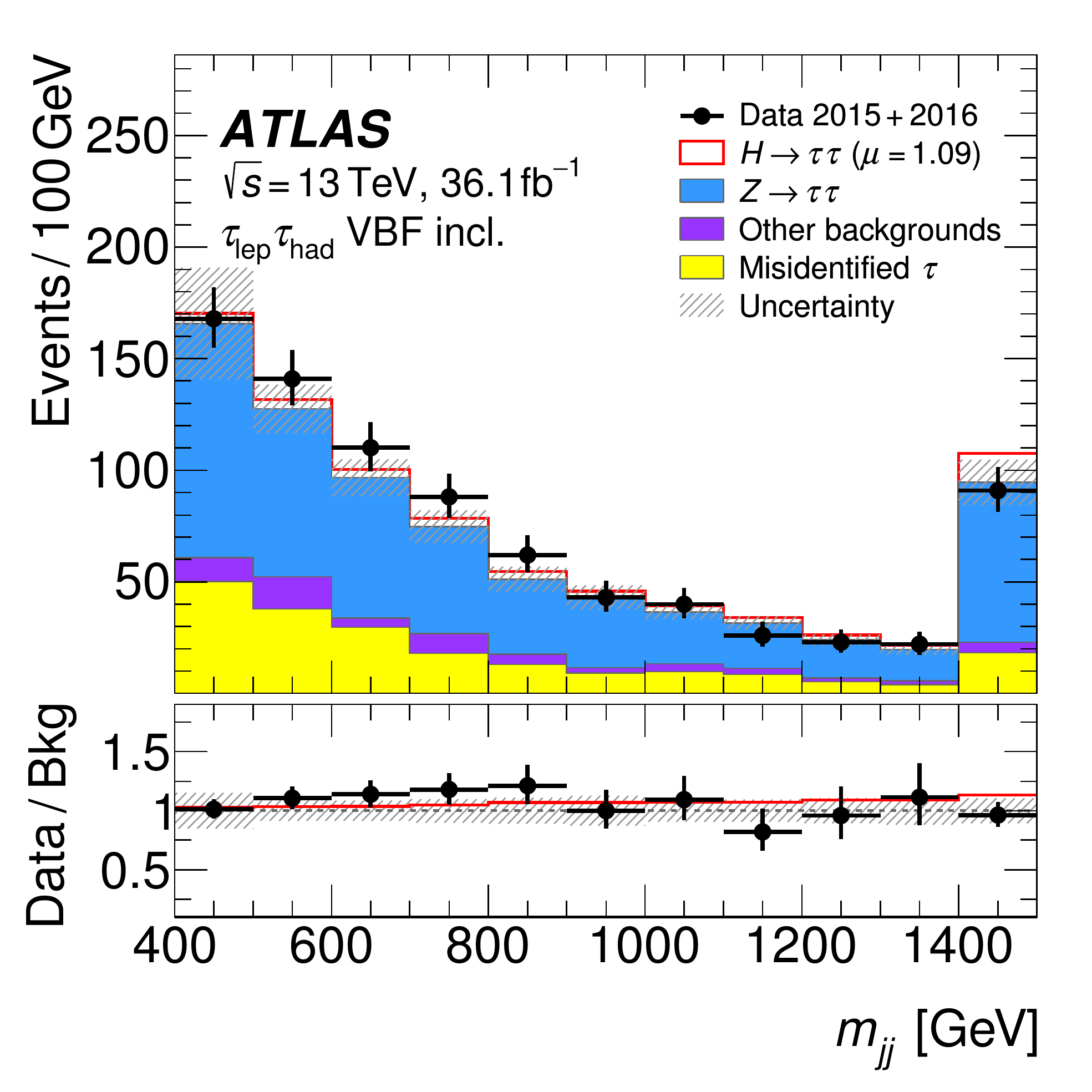} }
   \subfloat[]{\includegraphics[height=0.33\textwidth]{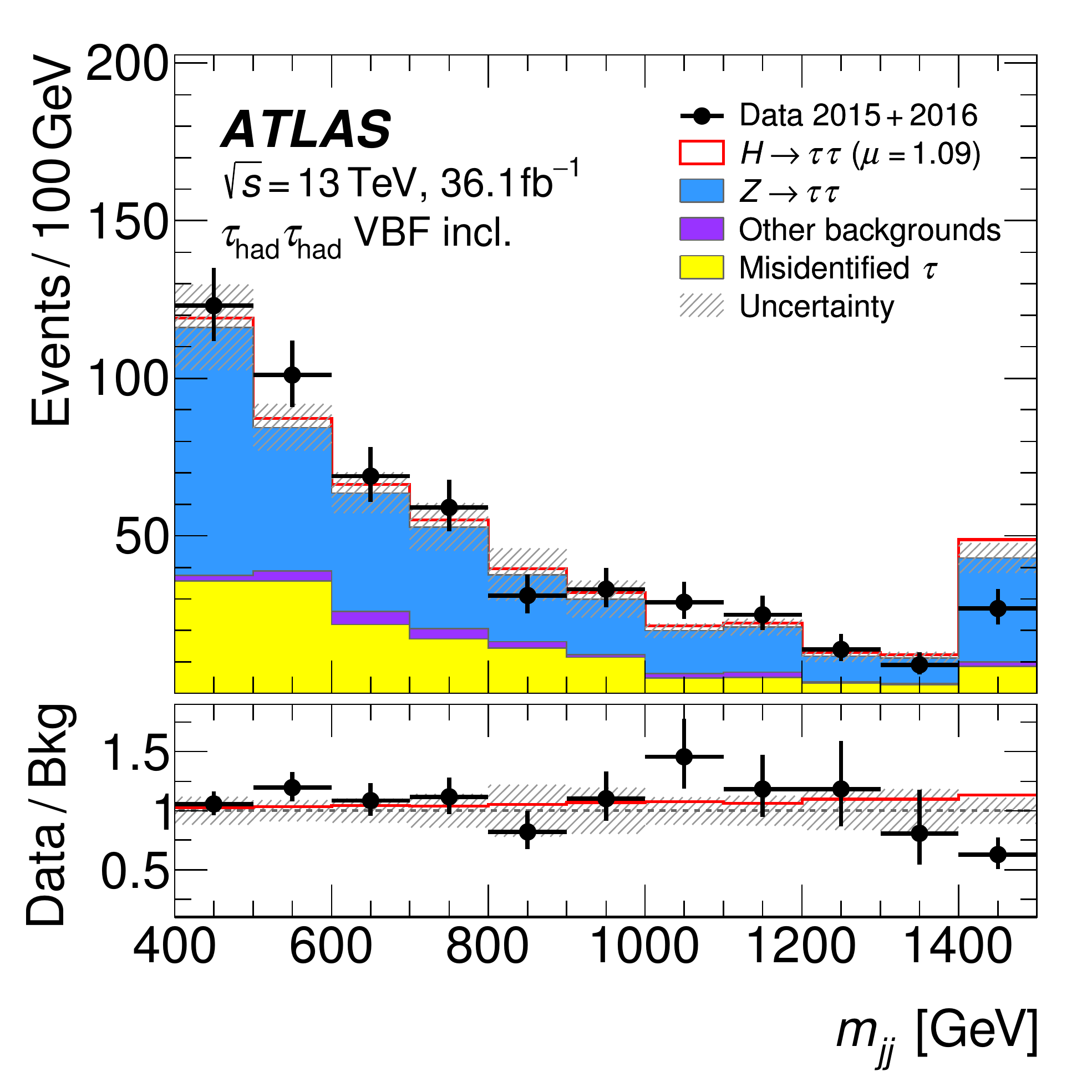} }
   \caption{Comparisons between data and predictions as computed by the fit of (top) the \pt of the Higgs-boson candidate (\pTH) in the boosted inclusive category and (bottom) the invariant mass of the two highest-\pT jets (\mjj) in the VBF inclusive category for (left) the \tll channel, (center) the \tlhad channel and (right) the \thadhad channel.
   The ratios of the data to the background model are shown in the lower panels.
   The observed Higgs-boson signal ($\mu = 1.09$) is shown with the solid red line.
   Entries with values that would exceed the $x$-axis range are shown in the last bin of each distribution.
   The size of the combined statistical, experimental and theoretical uncertainties in the background is indicated by the hatched bands.}
   \label{fig:SRmodelling}
 \end{figure}

Using \pTH, \dRtt and \mjj, the \VBF and boosted categories, referred to as \enquote{inclusive} categories, are split further into 13 exclusive signal regions with different signal-to-background ratios to improve the sensitivity.
\Cref{t:categorydef} summarizes the analysis categories and signal region definitions. \Cref{fig:bkgcomp} illustrates the expected signal and background composition in the signal and control regions of all analysis channels.
\Cref{fig:SRmodelling} compares for each analysis channel the observed distributions with predictions, as resulting from the fit described in \cref{sec:results}, for \pTH in the boosted inclusive categories, and for \mjj in the \VBF inclusive categories.
The observed data agree within the given uncertainties with the background expectation described in \cref{sec:background} for all distributions.

\begin{table}
 \begin{center}
 \caption{Definition of the \VBF and boosted analysis categories and of their respective signal regions (SRs).
 The selection criteria, which are applied in addition to those described in \cref{t:selection}, are listed for each channel. 
 The \VBF high-\pTH SR is only defined for the \thadhad channel, resulting in a total of seven \VBF SRs and six boosted SRs.
All SRs are exclusive and their yields add up to those of the corresponding \VBF and boosted inclusive regions.}
 \label{t:categorydef}
 \centering
 \vspace{2mm}
\small
 \begin{tabular}{cc|c|c|c|c}
  \toprule
  \multicolumn{2}{c|}{Signal Region} & Inclusive & \tll & \tlhad & \thadhad \\
  \midrule
  \parbox[t]{12pt}{\multirow{6}{*}{\rotatebox[origin=c]{90}{\VBF\quad}}} & \multirow{2}{*}{High-\pTH} &
  \multirow{6}{*}{\begin{tabular}[c]{c}\rule{0pt}{\normalbaselineskip+9pt}$\pTjsl>\SI{30}{\GeV}$\\
  $|\detajj|>3$ \\ $\mjj>\SI{400}{\GeV}$ \\ $\eta_{j_1}\cdot\eta_{j_2}<0$ \\ Central leptons \end{tabular}}
  & \multicolumn{2}{c|}{\multirow{2}{*}{---}} & $\pTH>\SI{140}{\GeV}$ \\
  &           &       & \multicolumn{2}{c|}{} & $\dRtt<1.5$ \\[2pt]
  \cline{2-2} \cline{4-6}
  \rule{0pt}{\normalbaselineskip} & \multirow{2}{*}{Tight} &      & $\mjj>\SI{800}{\GeV}$ & $\mjj>\SI{500}{\GeV}$ & Not \VBF high-\pTH \\[2pt]
  &           &       &                       & $\pTH>\SI{100}{\GeV}$ & $\mjj>(1550 - 250\cdot |\detajj|)\,\GeV$ \\[2pt]
  \cline{2-2} \cline{4-6}
  \rule{0pt}{\normalbaselineskip} & \multirow{2}{*}{Loose} &       & \multicolumn{2}{c|}{\multirow{2}{*}{Not \VBF tight}} & Not \VBF high-\pTH \\
  &           &       & \multicolumn{2}{c|}{} & and not \VBF tight \\
  \midrule
  \parbox[t]{12pt}{\multirow{3}{*}{\rotatebox[origin=c]{90}{Boosted\enspace}}} & \multirow{2}{*}{High-\pTH} & \multirow{3}{*}{\begin{tabular}[c]{c}\rule{0pt}{\normalbaselineskip+1pt}Not \VBF\\$\pTH>\SI{100}{\GeV}$\end{tabular}} & \multicolumn{3}{c}{$\pTH>\SI{140}{\GeV}$} \\
  &           &           & \multicolumn{3}{c}{$\dRtt<1.5$} \\[2pt]
  \cline{2-2} \cline{4-6}
  \rule{0pt}{\normalbaselineskip} & Low-\pTH   &           & \multicolumn{3}{c}{Not boosted high-\pTH} \\
 \bottomrule
 \end{tabular}
 \end{center}
\end{table}

Six control regions are defined to constrain the normalization of the dominant backgrounds in regions of phase space where their purity is high.
Their definitions are summarized in \cref{tab:cr}.
Two \Zll CRs, which are both more than 90\% pure in \Zll events, are defined by applying the same selection as for the
SF \tll \VBF and boosted inclusive regions, respectively, but with the \mll
requirement modified to $80<\mll<\SI{100}{\GeV}$.
The top-quark background is characterized by the presence of \Pqb-jets.
Four separate top CRs are defined by inverting the \Pqb-jet veto in the inclusive \VBF and boosted categories for each of the \tll and \tlhad channels.
The top CRs in the \tll channel are about 80\% pure in top-quark events.
For the top CRs in the \tlhad channel, the requirement of $\mT < \SI{70}{\GeV}$ is replaced by $\mT>\SI{40}{\GeV}$ to further enhance the purity to about 70\% in the \VBF top CR and about 60\% in the boosted top CR.
No such control regions are defined for the \thadhad channel since the top and \Zll backgrounds are negligible in this case.

One validation region is defined for each signal region (\enquote{\Ztt VRs}) to validate 
the event yields and kinematic distributions of simulated \Ztt events.
The \Ztt VRs are composed of \Zll events with kinematics similar to the \Ztt
background in the respective signal regions. 
This is achieved by starting with an event selection that is 
based on the SF \tll channel preselection with 
the following differences that account for the 
selection of light leptons instead of decay products from \Pgt{}-leptons:
The \mcoll, \met and \METHPTO requirements are dropped and the \mll requirement is inverted to
$\mll>\SI{80}{\GeV}$.
The other requirements on \Pgt{}-lepton decays are replaced with requirements on the two light leptons.
In particular, the requirements on \pTH are substituted by the \pt of the \PZ boson computed from the \pT of the light leptons (\pTZ). 
Requirements on jets are unchanged since they define the shape of most kinematic distributions for \PZ-boson production similarly in the SRs and the \Ztt VRs.
More than 99\% of the selected events are from \Zll in all \Ztt VRs.

\begin{table}
  \caption{Definitions of the six control regions (CRs) used to constrain the \Zll and top backgrounds to the event yield in data in the \tll and \tlhad channels.
    \enquote{SF} denotes a selection of same-flavor light leptons.}
  \label{tab:cr}
  \centering
  \vspace{2mm}
\begin{tabular}{ll}
  \toprule 
  Region & Selection  \\ 
  \midrule 
  \tll \VBF \Zll CR & \tll VBF incl. selection, $80<\mll<\SI{100}{\GeV}$, SF  \\ 
  \tll boosted \Zll CR & \tll boosted incl. selection, $80<\mll<\SI{100}{\GeV}$, SF  \\ 
  \tll \VBF top CR & \tll VBF incl. selection, inverted $\Pqb$-jet veto  \\ 
  \tll boosted top CR & \tll boosted incl. selection, inverted $\Pqb$-jet veto  \\ 
  \tlhad \VBF top CR & \tlhad VBF incl. selection, inverted $\Pqb$-jet veto, $\mT > \SI{40}{\GeV}$ \\ 
  \tlhad boosted top CR & \tlhad boosted incl. selection, inverted $\Pqb$-jet veto, $\mT > \SI{40}{\GeV}$ \\ 
  \bottomrule 
\end{tabular} 
\end{table}

\section{Background estimation}
\label{sec:background}

The final-state topologies of the three analysis channels have different background 
compositions, which necessitates different strategies for the background estimation. In each SR, 
the expected number of background events and the associated kinematic distributions are 
derived from a mixture of data-driven methods and simulation. 

Background contributions with \tauhadvis, with prompt light leptons and with light leptons 
from \Pgt{}-lepton decays are estimated from simulation.
If their contribution is significant, their normalization is constrained by the observed event yields in CRs.
For smaller contributions of this type, their normalization is entirely taken from the 
theoretical cross sections with the precision in QCD listed in \cref{tab:MCSamples}. 
This includes di-boson processes and a small contribution from EW production of \PW/\PZ bosons from \VBF.
Contributions from light- and heavy-flavor jets that are misidentified as
prompt, light leptons or \tauhadvis are estimated using data-driven methods. 
They are labeled as \enquote{fake-\Pl{}} and \enquote{fake-\tauhadvis{}} backgrounds, respectively, 
and collectively as \enquote{misidentified \Pgt{}}, throughout this paper. 
The contamination from \HWW decays is treated
as a background in the \tll channel, while it is negligible in other channels.

For the background sources that have
their normalization constrained using data, \cref{tab:sf} shows the
normalization factors and their uncertainties obtained from the fit (see \cref{sec:results}).
For simulated backgrounds, the factors compare the background
normalizations with 
values determined from their theoretical cross sections.  The
normalization factor for the data-driven fake-\tauhadvis background
scales the event yield of the template of events that fail the
opposite-charge requirement (see \cref{sec:bkgFakes}).
The \Ztt normalization is constrained by data in the \mMMC 
distributions of the signal regions.
Systematic uncertainties are the dominant contribution to the normalization factor uncertainties.

\begin{table}
  \caption{Normalization factors for backgrounds that have their 
    normalization constrained using data 
    in the fit, including all statistical and systematic uncertainties described in \cref{sec:systematics},
    but without uncertainties in total simulated cross sections extrapolated to the selected phase space.
    Systematic uncertainties are the dominant contribution to the normalization factor uncertainties.
    Also shown are the 
    analysis channels to which the normalization factors are applied.
    }
  \label{tab:sf}
  \centering
  \vspace{2mm}
\begin{tabular}{llll}
  \toprule 
  Background & Channel & \multicolumn{2}{c}{Normalization factors} \\ 
             &         & \multicolumn{1}{c}{\VBF} & \multicolumn{1}{c}{Boosted} \\ 
  \midrule 
  \Zll (CR) & \tll & $0.88^{+0.34}_{-0.30}$ & $1.27^{+0.30}_{-0.25}$ \\ 
  Top (CR) & \tll & $1.19\pm0.09$ & $1.07\pm0.05$ \\ 
  Top (CR) & \tlhad & $1.53^{+0.30}_{-0.27}$ & $1.13\pm0.07$ \\ 
  Fake-\tauhadvis (data-driven) & \thadhad & \multicolumn{2}{c}{$1.12\pm0.12$} \\ 
  \Ztt (fit in each SR) & \tll,\tlhad,\thadhad & $1.04^{+0.10}_{-0.09}$ & $1.11\pm0.05$ \\ 
  \bottomrule 
\end{tabular} 
\end{table}

\subsection{\Ztt background validation}
\label{sec:bkgZtautau}

The Drell--Yan process $\pp \to \PZ/\Pggx \to \Pgt\Pgt$ is a dominant irreducible background in 
all analysis categories and contributes between 50\% and 90\% of the total background depending 
on the signal region. The separation between the Drell--Yan and the \Htt signal processes
is limited by the $\mMMC$ resolution. 

\begin{figure}[htbp]
  \centering
  \subfloat[]{\includegraphics[width=0.33\textwidth]{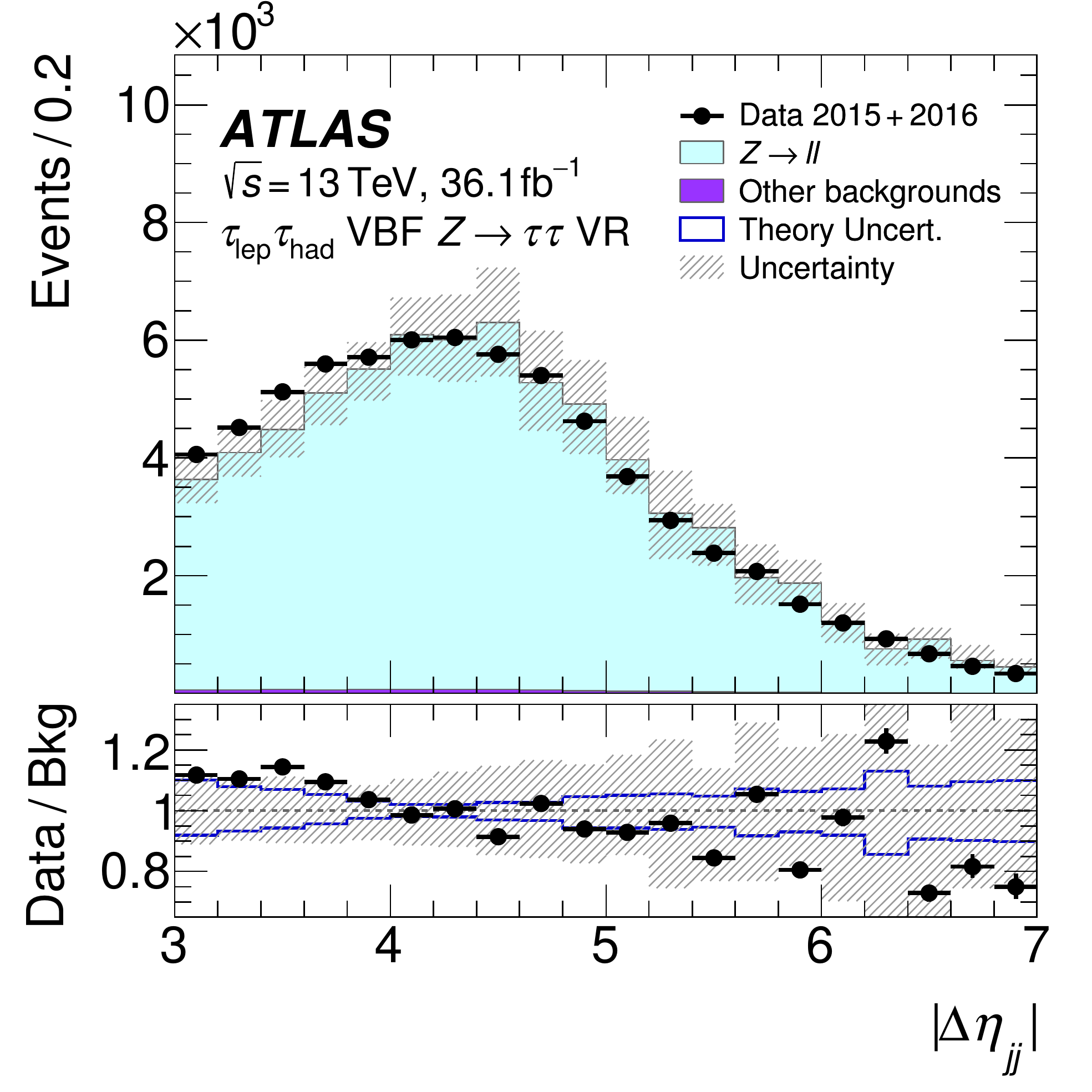} }
  \subfloat[]{\includegraphics[width=0.33\textwidth]{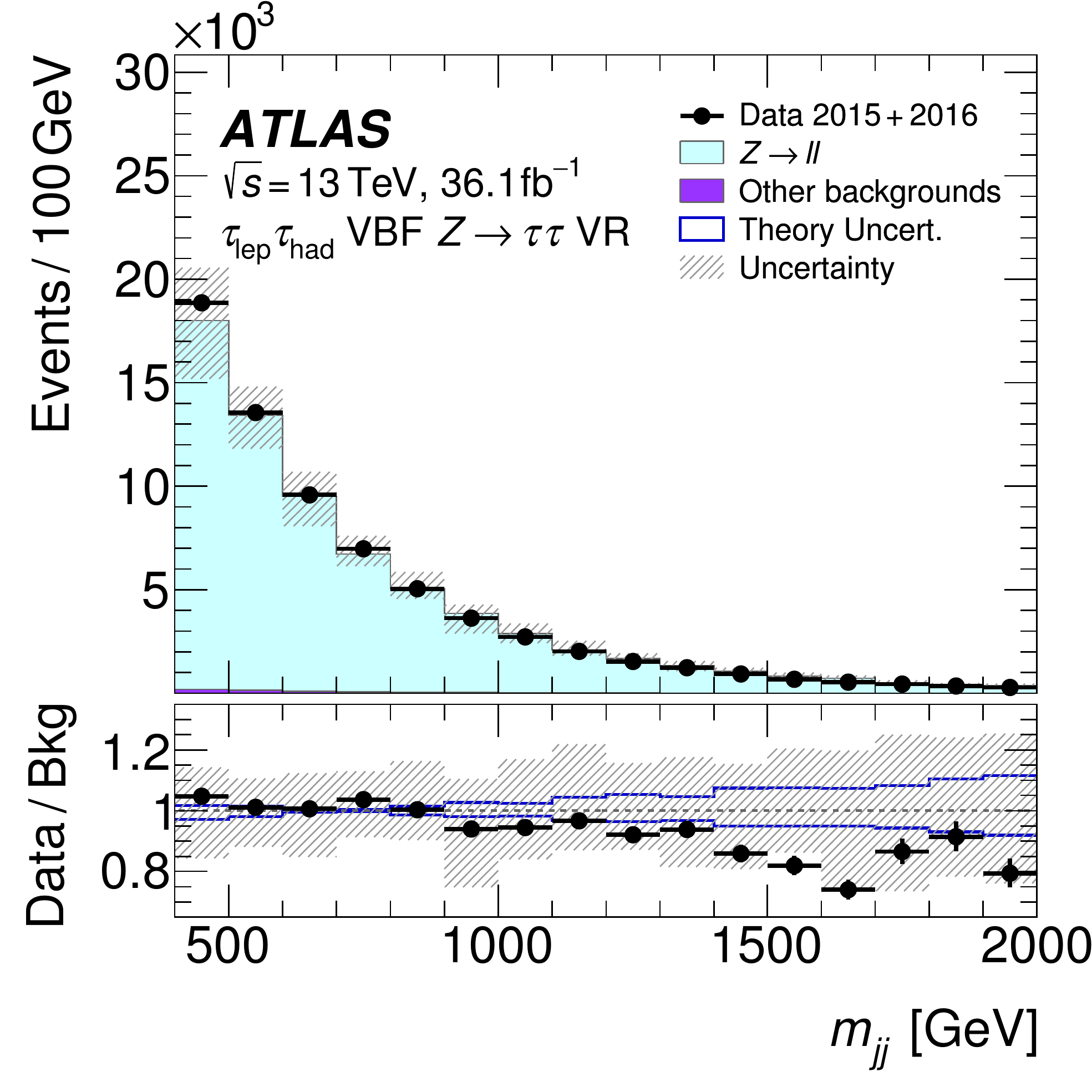} }
  \subfloat[]{\includegraphics[width=0.33\textwidth]{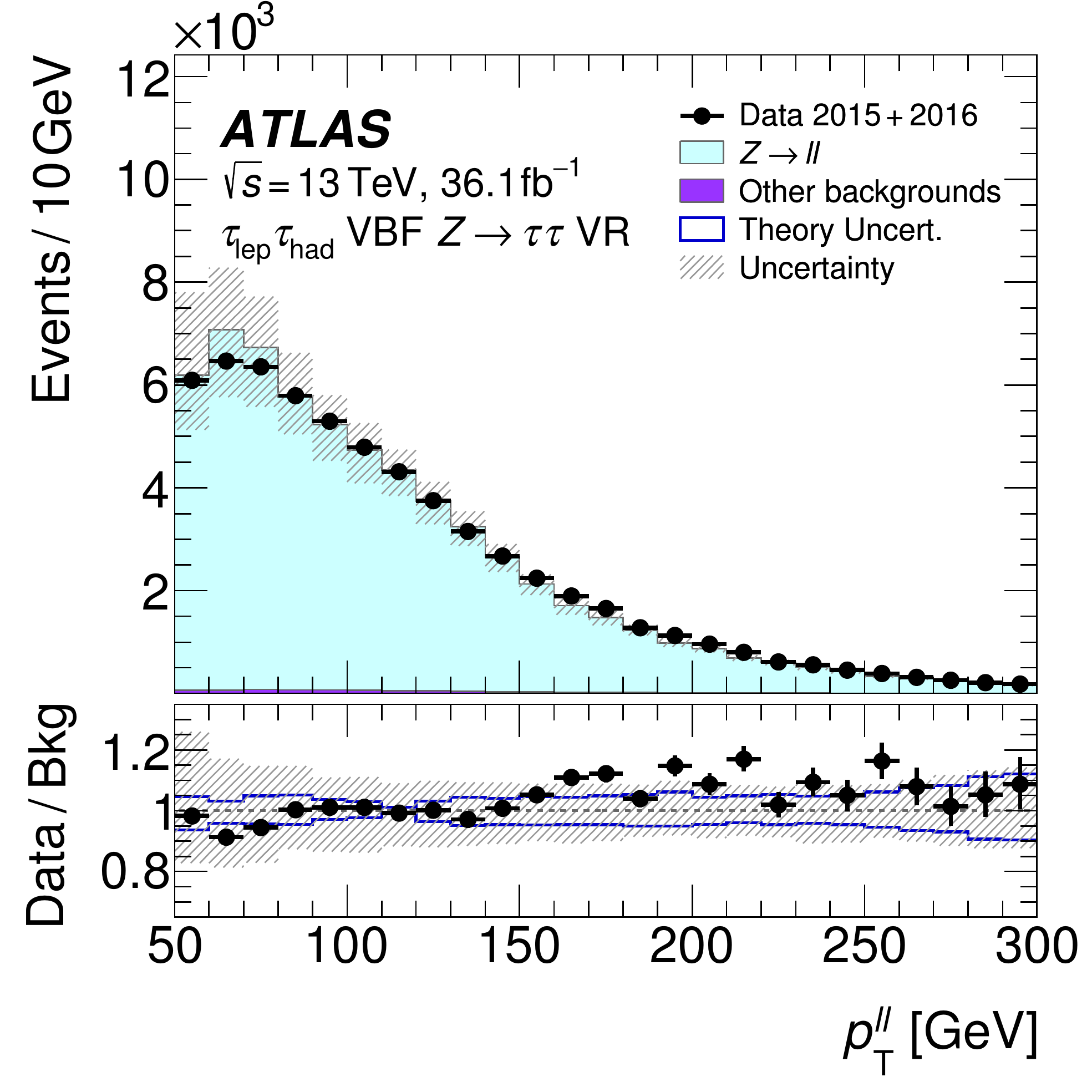} } \\
  \subfloat[]{\includegraphics[width=0.33\textwidth]{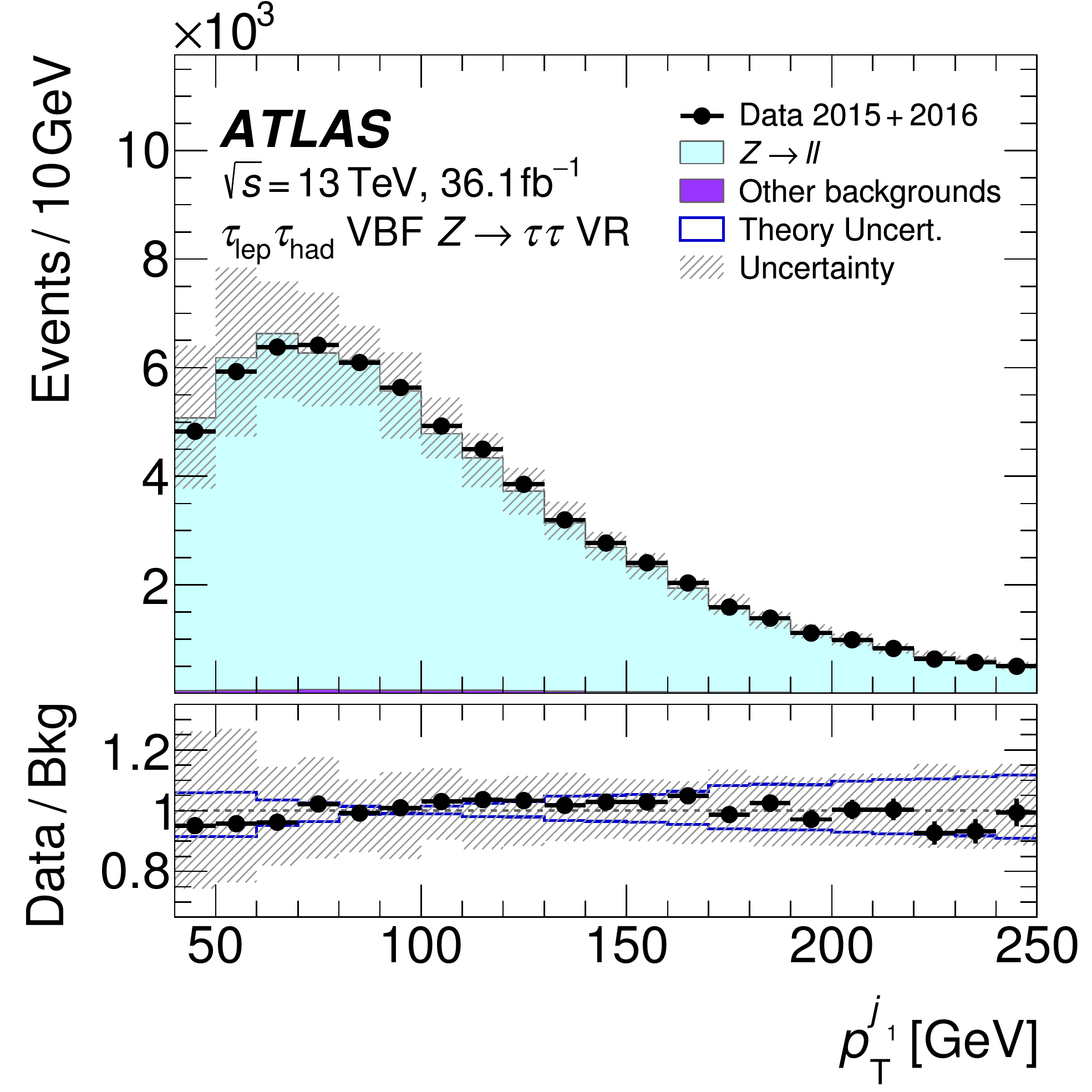} }
  \subfloat[]{\includegraphics[width=0.33\textwidth]{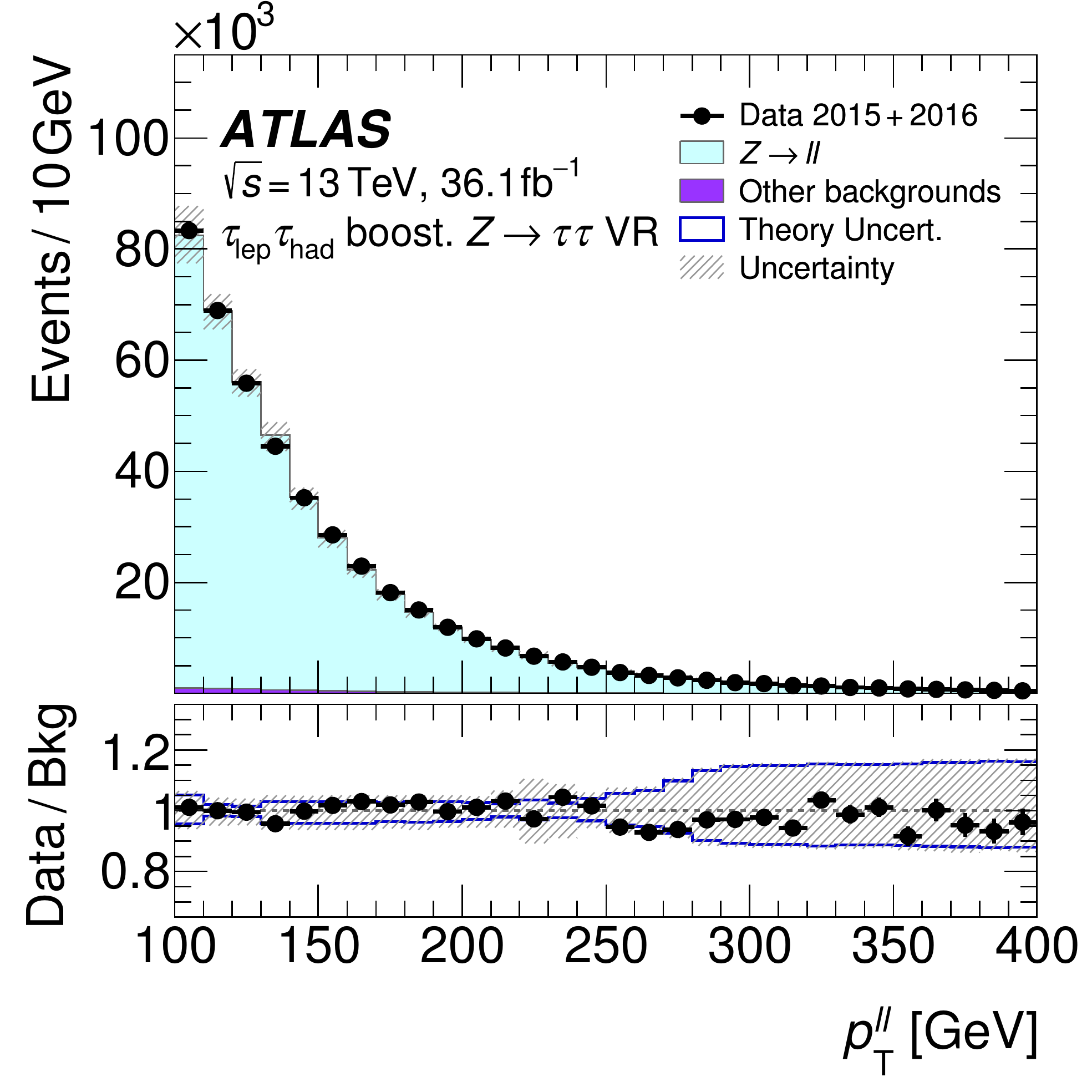} }
  \subfloat[]{\includegraphics[width=0.33\textwidth]{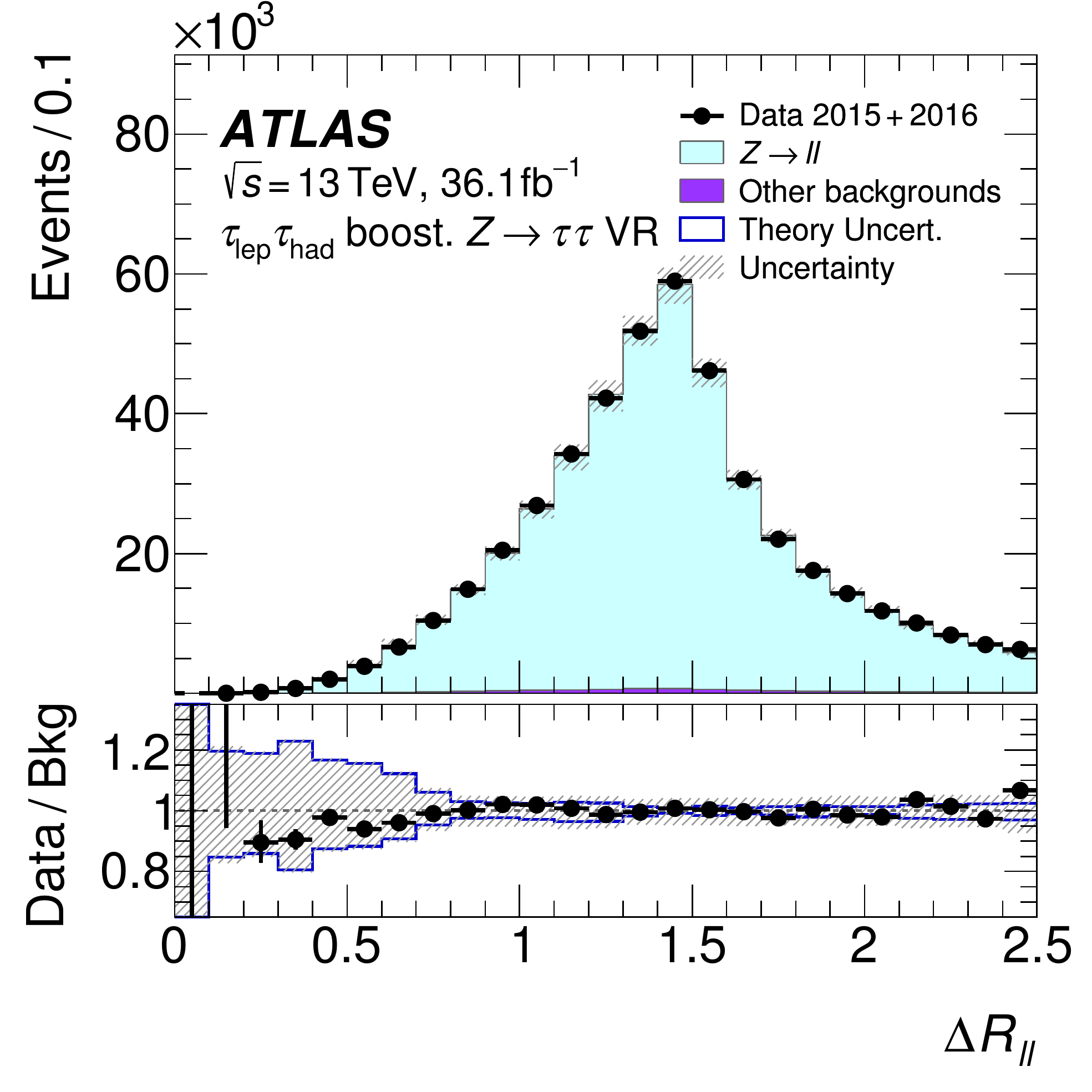} } \\
  \subfloat[]{\includegraphics[width=0.33\textwidth]{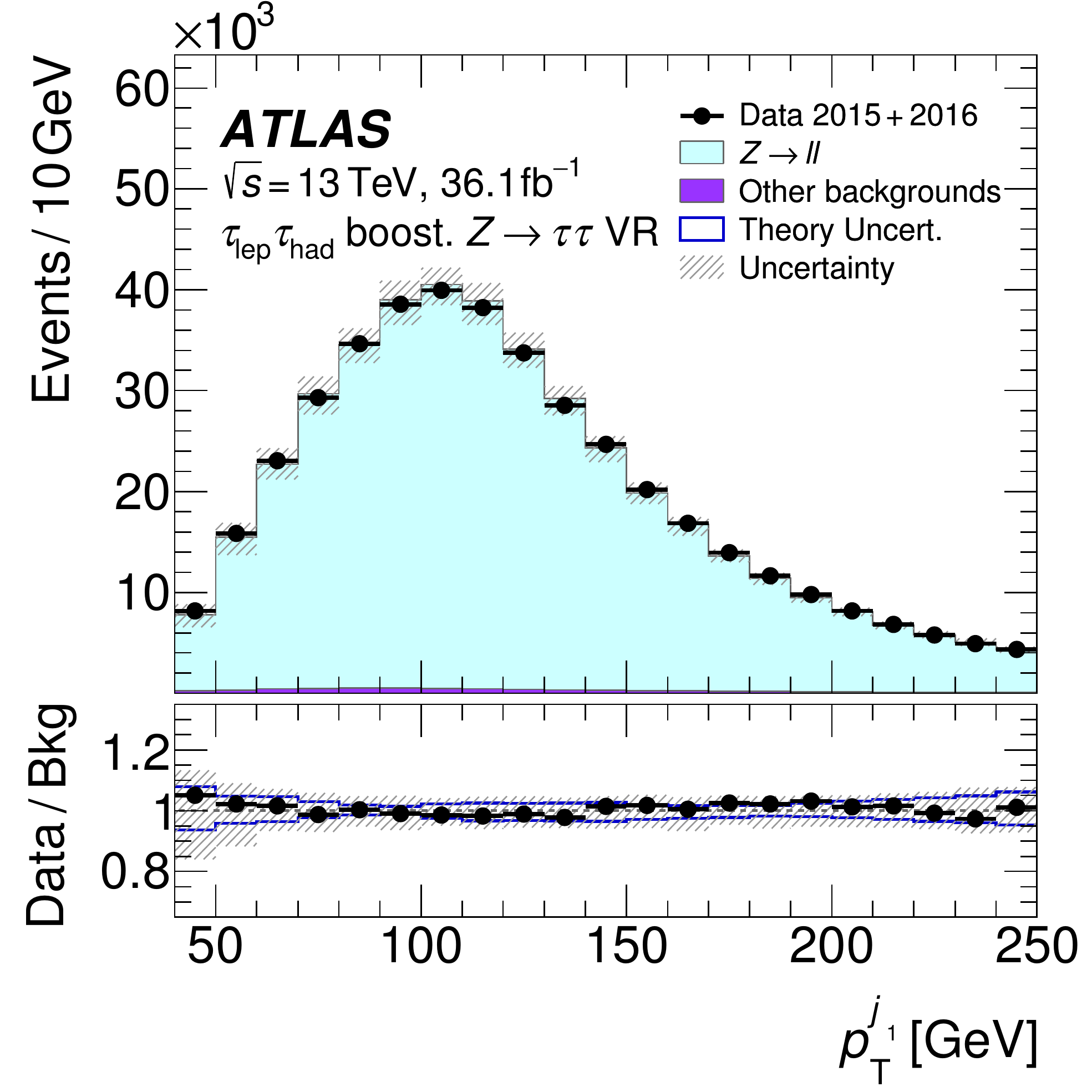} }
  \subfloat[]{\includegraphics[width=0.33\textwidth]{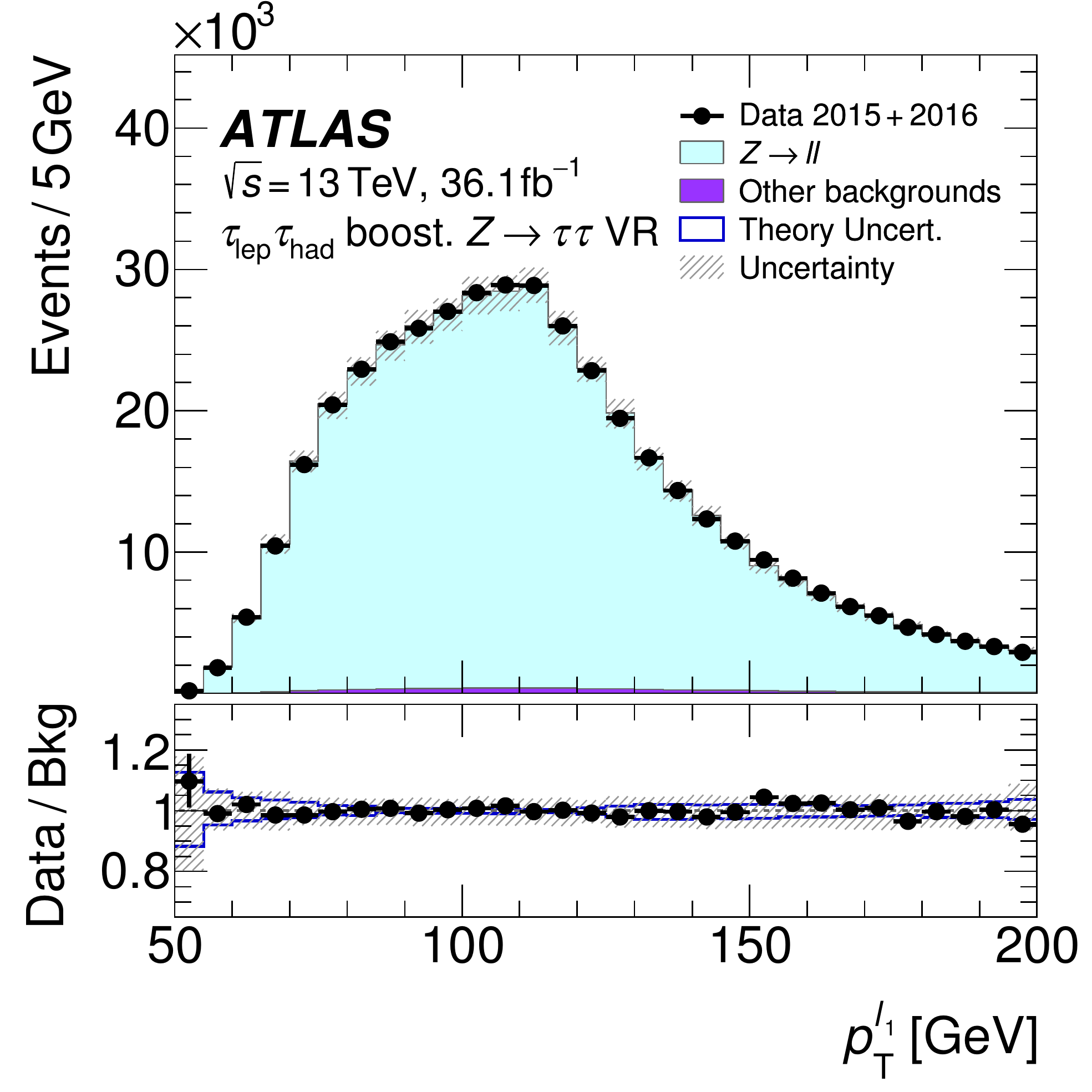} }
  \subfloat[]{\includegraphics[width=0.33\textwidth]{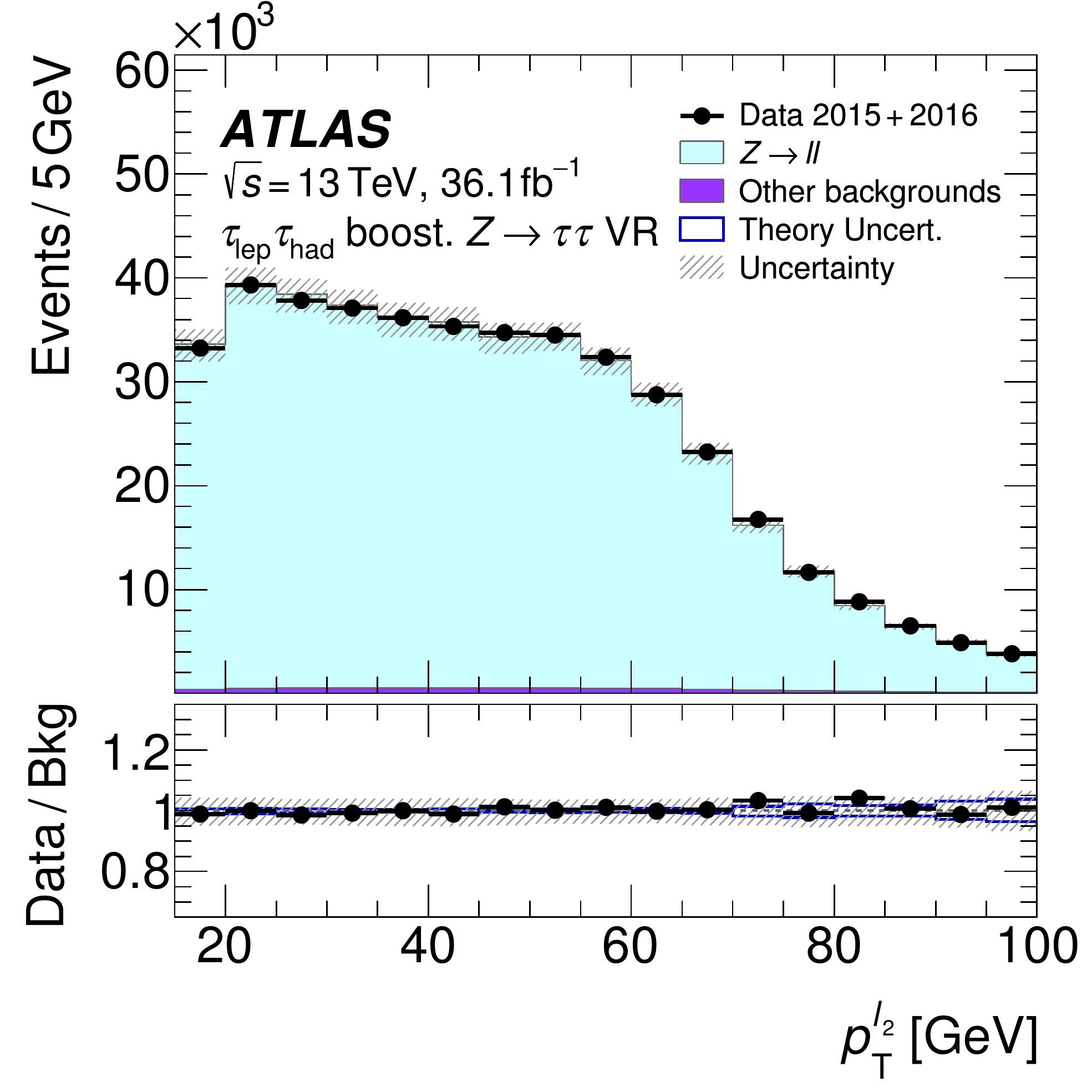} }
  \caption{Observed and expected distributions in the \Ztt validation
    regions (VRs) corresponding to (a)--(d) the \tlhad \VBF inclusive
    category and (e)--(i) the \tlhad boosted inclusive category. Shown
    are, in the respective region: (a) the pseudorapidity separation ($|\detajj|$)
    and (b) the invariant mass (\mjj) of the two highest-\pT
    jets; (c) and (e) the \pT of the di-lepton system (\pTZ); (d) and
    (g) the \pT of the highest-\pT jet (\pTjl); (f) the angular
    distance between the light leptons ($\Delta R_{\Pl\Pl}$); (h) the \pT
    of the highest-\pT light lepton (\pTll); and (i) the \pT of the
    second-highest-\pT light lepton (\pTlsl).
    The predictions in these validation regions are not computed by the fit, but 
    are simply normalized to the event yield in data. 
    The size of the combined statistical, experimental and theoretical uncertainties is indicated by the hatched bands.
    The ratios of the data to the background model are shown in the lower panels together with the theoretical uncertainties in the \sherpa simulation of \Zll, which are indicated by the blue lines.}
  \label{fig:bkg-Zvalid}
\end{figure}

The modeling of this important background is validated using \Ztt VRs that consist of \Zll events.
In \cref{fig:bkg-Zvalid}, the observed distributions of several variables
are compared with simulation normalized to the event yield in data. 
The selected observables correspond to either variables correlated with \mMMC (\pTll and \pTlsl), or to major variables used for categorization (\pTZ, $\Delta R_{\Pl\Pl}$, \detajj and \mjj), or to variables to which different requirements are applied in each decay channel (\pTjl).
Generally, the \sherpa simulation describes the shape of data distributions within the experimental and theoretical uncertainties (see \cref{sec:systematics}), 
with the exception of a slight trend in the ratio of data to simulation as a function of \detajj and \mjj shown in \cref{fig:bkg-Zvalid}.
These trends have no impact on the modeling of \mMMC.
Reweighting the simulation with the observed \mjj distribution, which is an important variable for \VBF categorization, has a negligible impact on the measurement.
In the fit, the normalization of the \Ztt background is correlated across the decay channels and
constrained by data in the \mMMC distributions of the signal regions associated with the boosted
and \VBF categories, independently.
As shown in \cref{tab:sf}, it is constrained to $\pm$5\% in the boosted category
and to $\pm$9\% in the \VBF category.
The relative acceptance of events among the signal regions within a category is validated by applying the corresponding event-selection criteria to the \Ztt VRs.
The expected relative acceptance from simulation agrees with data within uncertainties for all regions.
\Cref{fig:CBA_m-tautau_channel,fig:CBA_m-tautau} show the good modeling of the \Ztt \mMMC distribution in all signal regions.
Additional uncertainties in the relative acceptances 
and in the shape of the \mMMC distributions in the signal regions 
are evaluated from theoretical and experimental uncertainties
described in \cref{sec:systematics}.

\subsection{\Zll background}
\label{sec:bkgZll}

\begin{figure}[tbp]
  \centering
  \subfloat[]{\includegraphics[width=0.33\textwidth]{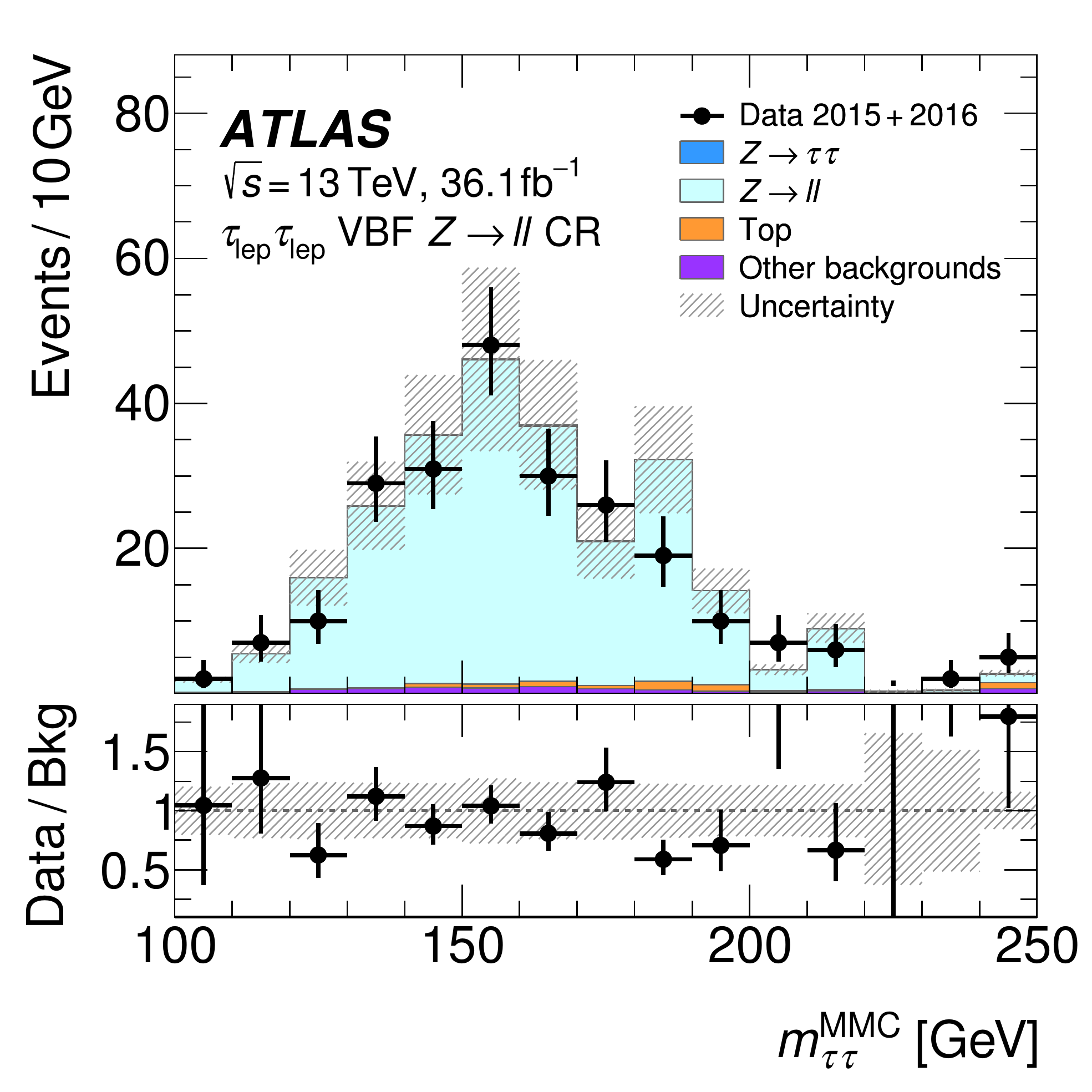} }
  \subfloat[]{\includegraphics[width=0.33\textwidth]{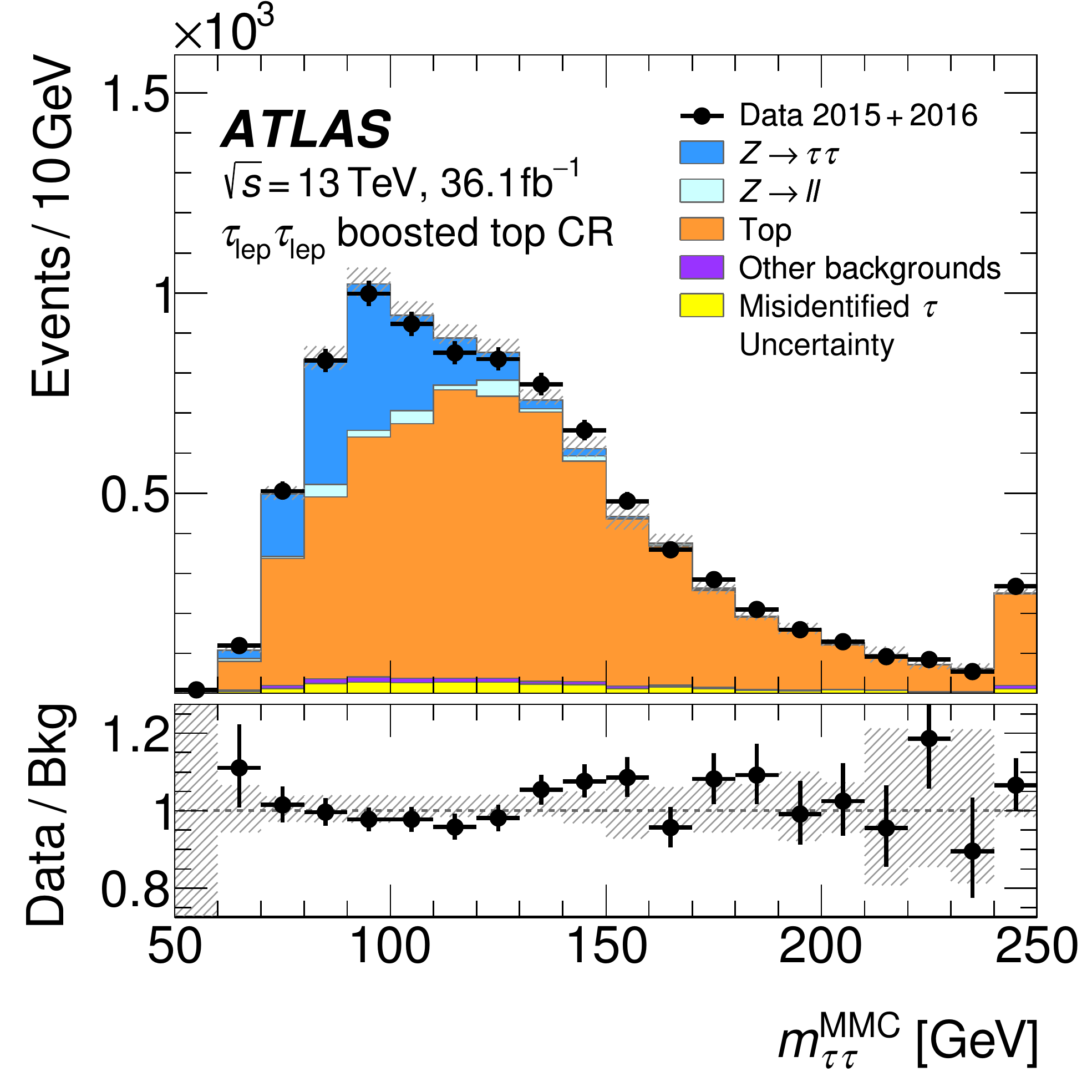} }
  \subfloat[]{\includegraphics[width=0.33\textwidth]{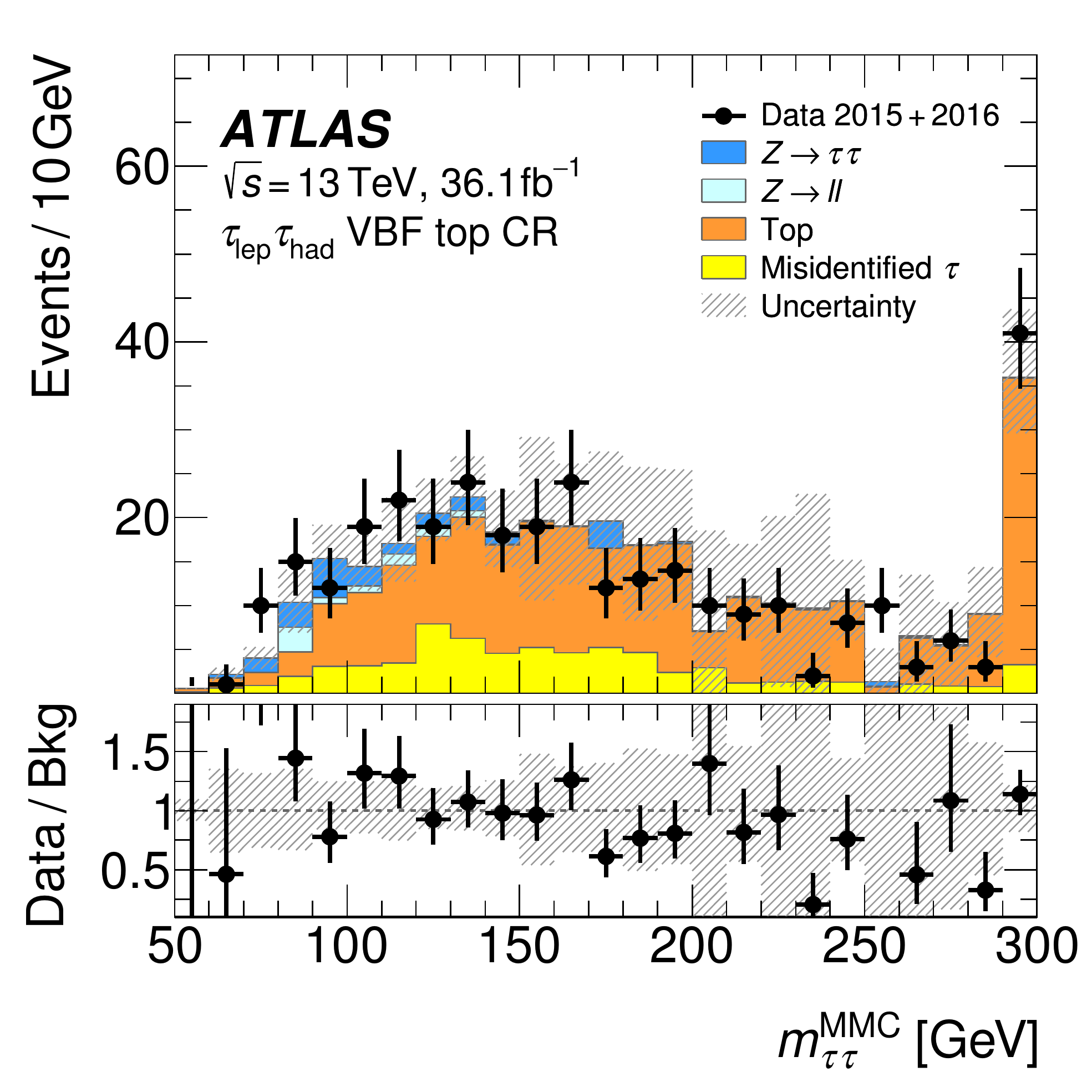} }
  \caption{For the control regions (CRs) defined in \cref{sec:selection}, comparisons between data and predictions as computed by the fit for the reconstructed di-\Pgt invariant mass ($\mMMC$). Shown are (a) the \tll \VBF \Zll control region (CR), (b) the \tll boosted top CR and (c) the \tlhad \VBF top CR.
  Entries with values that would exceed the $x$-axis range are shown in the last bin of each distribution.
  The size of the combined statistical, experimental and theoretical uncertainties in the background is indicated by the hatched bands.
  The ratios of the data to the background model are shown in the lower panels.}
  \label{fig:CRmodelling}
\end{figure}

Decays of \PZ bosons into light leptons are a significant background for 
the \tll and \tlhad channels, where mismeasured \met can bias the reconstructed \mMMC
of light-lepton pairs towards values similar to those expected for the signal.
The observed event yields in the \Zll CRs constrain the normalization of simulated \Zll 
events in the \tll channel to $\pm$40\% in the \VBF category
and to $\pm$25\% in the boosted category, as shown in \cref{tab:sf}.
The good modeling of the \mMMC distribution in the \tll \VBF \Zll CR is shown in \cref{fig:CRmodelling}(a).
In other channels, the contribution from \Zll events
is normalized
to its theoretical cross section.
In the \tlhad channel, \Zll background contributes primarily through \Zee decays where an electron is misidentified as a \tauhadvis~candidate.
Due to the dedicated electron veto algorithm applied to selected 1-prong \tauhadvis candidates (see \cref{sec:presel}), this background is small.
This and other backgrounds from light leptons misidentified as \tauhadvis
in this channel are estimated from simulation, with the 
probability for electrons misidentified as \tauhadvis~candidates scaled to match that observed in data~\cite{ATLAS-CONF-2017-029}. 

\subsection{Top-quark background}
\label{sec:bkgTop}

The production of \ttbar pairs or single top quarks is a significant background (\enquote{top background})
for the \tll and \tlhad channels, due to the production of prompt light leptons with 
associated \met in the top-quark decay chain $\Pqt \to \PW\Pqb,\, \PW \to \Pl\nu,\Pgt\nu$. 
Events where a selected \Pgt{}-lepton decay product is misidentified, are estimated using data-driven methods that are
discussed in \cref{sec:bkgFakes}. The remaining top background is estimated from simulation.
In the \tll and \tlhad channels the normalization of simulated top background is additionally constrained by the 
absolute event yields in their respective top CRs to $\pm$30\% in the \tlhad \VBF top CR 
and less than $\pm$10\% in the other top CRs, as shown in \cref{tab:sf}.
Figures~\ref{fig:CRmodelling}(b) and \ref{fig:CRmodelling}(c) show \mMMC distributions in the \tll boosted top CR and the \tlhad \VBF top CR, respectively.

\subsection{Backgrounds from misidentified \Pgt}
\label{sec:bkgFakes}

Apart from the small contribution from light leptons misidentified as
\tauhadvis described in \cref{sec:bkgZll}, hadronic jets can be misidentified
as \tauhadvis, electrons and muons.
These sources of background contribute up to half of the total background, depending 
on the signal region, and are estimated with 
data-driven techniques. Since the background sources depend on the event
topology, specific methods are applied to each individual channel.

In the \tll channel, the
main sources of the fake-\Pl background are multijets, \PW bosons in association with jets, and semileptonically decaying \ttbar events.
All these background sources are treated together.
Fake-\Pl regions are defined in data by requiring that the light lepton with the second-highest \pt does not satisfy
the \enquote{gradient} isolation criterion. 
This is referred to as \enquote{inverted} isolation. In addition, if the light lepton is an electron, its identification criteria 
are relaxed to \enquote{loose}.
Fake-\Pl templates are
created from these samples by subtracting top and \Zll
backgrounds that produce real light leptons, estimated from simulation. The
normalization of each template is then scaled by a factor that
corrects for the inverted-isolation requirement. 
These correction factors are computed for each combination of lepton flavor 
from events that pass the \tll selection but have same-charge light leptons, 
subtracting simulated top and \Zll backgrounds.
Fake-\Pl background in the top-quark CRs is estimated following the same procedure.

Systematic uncertainties in the shape and
normalization of the fake-\Pl background in the \tll channel
depend on the \pT of the second-highest-\pt lepton and are estimated as follows. 
A closure test of the background estimate is
performed using events where the leptons are required to have the same charge
and yields an uncertainty ranging between 20\% and 65\%. An uncertainty in the 
heavy-flavor content is estimated by using isolation correction factors 
that are computed from samples selected with inverted $\Pqb$-jet requirements.
This uncertainty is as large as 50\%.
Minor contributions come from the uncertainty in the fractional
composition of the fake-\Pl background in top-quark decays, multijet  
events and \PW-boson production.

\begin{figure}[tbp]
  \centering
  \subfloat[]{\includegraphics[height=0.25\textheight]{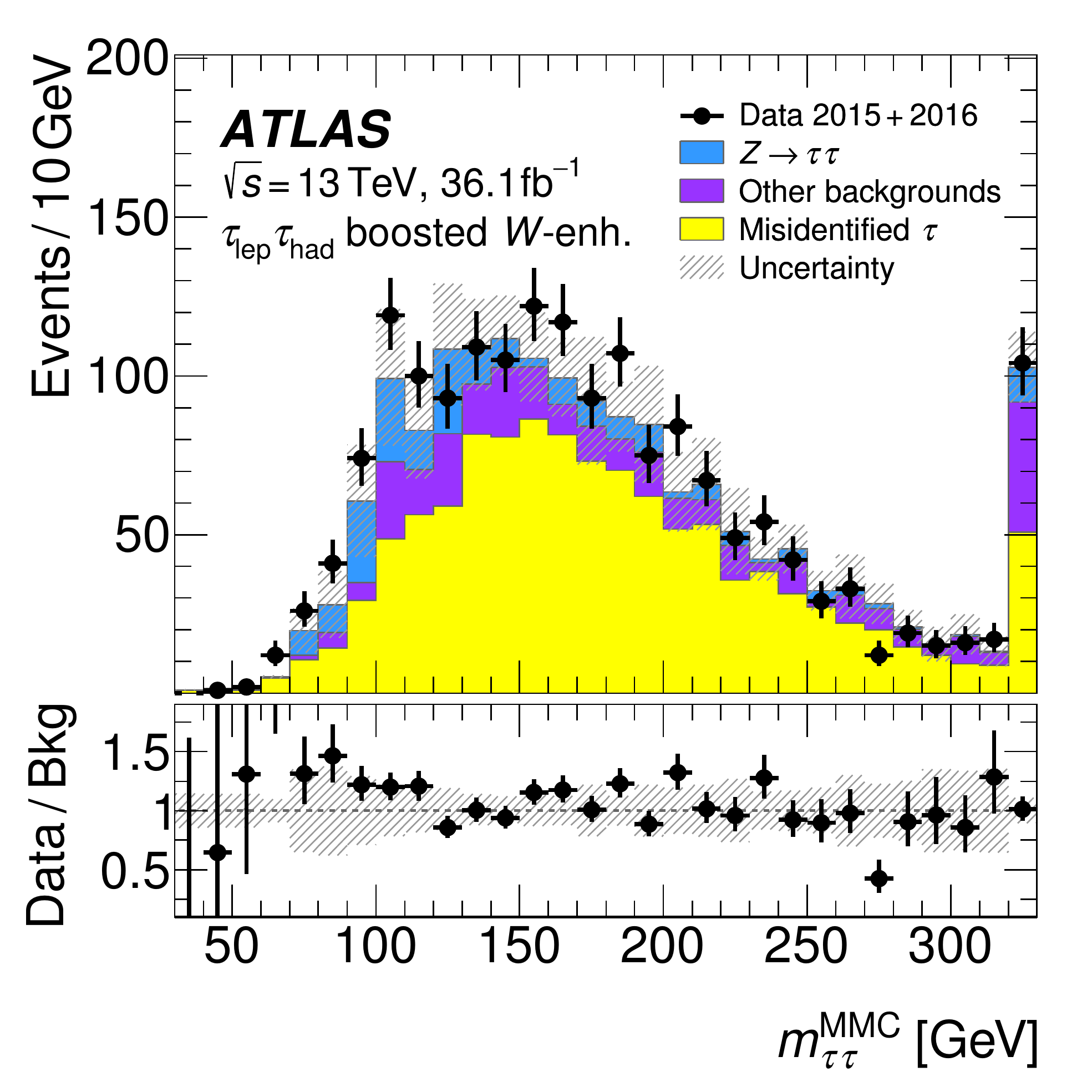} }
  \subfloat[]{\includegraphics[height=0.25\textheight]{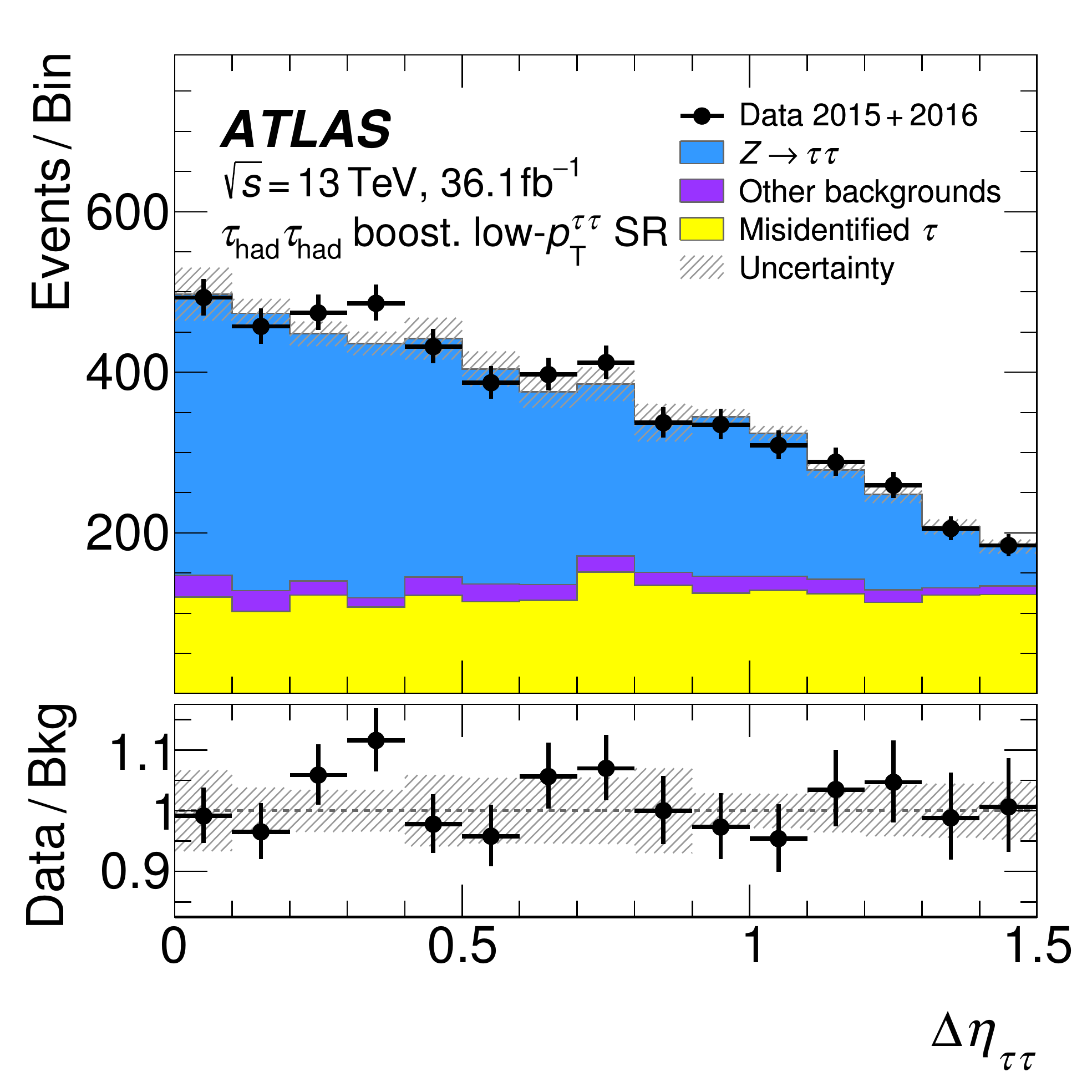} }
  \caption{Observed distributions and predictions computed by the fit for (a) \mMMC in the \PW-enhanced region of the \tlhad boosted inclusive category, 
    and (b) $\Delta\eta$ between the two \tauhadvis, for events in the boosted low-\pTH signal region (SR) of the \thadhad channel.
    Entries with values that would exceed the $x$-axis range are shown in the last bin of each distribution.
    The size of the combined statistical, experimental and theoretical uncertainties in the background is indicated by the hatched bands.
    The ratios of the data to the background model are shown in the lower panels.}
  \label{fig:bkg-fake}
\end{figure}

In the \tlhad channel, a \enquote{fake-factor} method is used to derive
estimates for fake-\tauhadvis events,
composed mainly of multijet events and
\PW-boson production in association with jets. 
A fake-factor is defined as the ratio of the number of events where the highest-\pt jet is identified
as a \enquote{medium} \tauhadvis~candidate to the number of events with a highest-\pt jet that passes 
a very loose \tauhadvis~identification but fails the \enquote{medium} one.
Fake-factors 
depend on the \pT and track multiplicity of the \tauhadvis~candidate and 
on the type of parton initiating the jet.
Therefore, they are computed depending on the \pt and the track multiplicity, 
in both quark-jet-dominated \enquote{\PW-enhanced} and gluon-jet-dominated \enquote{multijet-enhanced} regions. 
The \PW-enhanced regions are defined by inverting the $\mT<\SI{70}{\GeV}$ requirement and the multijet-enhanced regions are defined 
by inverting the light-lepton isolation, relative to the inclusive boosted and \VBF selections.
Backgrounds from \PZ-boson production with associated jets and semileptonically decaying \ttbar have fake-factors similar to those 
found in backgrounds from \PW bosons, and their contributions are negligible. 
The fake-factors are in the range 0.15--0.25 for 1-prong and 0.01--0.04 for 3-prong \tauhadvis.
To obtain the fake-\tauhadvis background estimate for the signal regions, these fake-factors are first weighted by the multijets-to-\PW fraction.
The weighted fake-factors are then applied to events in regions 
defined by the selections of the corresponding signal regions, except that the highest-\pt \tauhadvis
candidate passes a very loose \tauhadvis identification and fails the \enquote{medium} one (\enquote{anti-ID} regions).
The relative multijet contribution in each anti-ID region is estimated from the yield of events that fail the light-lepton isolation requirement, 
multiplied by a factor that corrects for this requirement.
The multijet contribution varies by more than 50\% and
depends on the lepton \pT and on the $\Delta\phi$ between \tauhadvis and \met.
The good agreement between data and background estimates is shown in \cref{fig:bkg-fake}(a)
for the main discriminant of the analysis, \mMMC, in the boosted \PW-enhanced region.

The dominant contribution to the uncertainties in the fake-\tauhadvis background 
in the \tlhad channel originates from the
statistical uncertainty in the individual fake-factors of up to
10\% in the boosted signal regions and up to 35\% in the VBF
signal regions. Minor contributions originate from the statistical
uncertainty in the anti-ID regions and uncertainties in the fractional
size of the multijet contribution to the fake-\tauhadvis background. 

In the \thadhad channel, the multijet background is modeled using a template extracted 
from data that pass the signal-region selections, but where the \tauhadvis~candidates 
are allowed to have two tracks and required to fail the opposite-charge requirement
(nOC region).
The contribution of events with true \Pgt{}-leptons from other SM processes is subtracted 
from this template using simulation. The template is then reweighted using scale factors 
dependent on the difference in $\phi$ between the \tauhadvis candidates (\dphitt). 
These scale factors are derived 
by comparing the template from an nOC selection with a region
obtained by requiring the \tauhadvis pair to have opposite charge and the second-highest-\pt \tauhadvis
to fail the \enquote{tight} but pass the \enquote{medium} identification requirements. 
As the yield of events that pass these identification requirements is small, the scale factors are derived from events 
that pass the \thadhad selection with looser $\detatt$ and $\dRtt$ requirements to gain statistical power.
The normalization of the multijet background is constrained in the fit by data in the \mMMC distribution
in the signal regions.
For this, a normalization factor is defined and it is correlated across all \thadhad signal regions.
\Cref{fig:bkg-fake}(b) shows good agreement between data and background
predictions in the distribution of $\Delta\eta$ between the 
two \tauhadvis, which has a quite different shape for the multijets than for the \Ztt process.
In this figure, events are selected that pass the \thadhad boosted low-\pTH selection.
Contributions from other backgrounds, such as \PW with associated jets, range from 2\% to 5\% in the \thadhad SRs.

The event yield of the multijet background in the \thadhad channel is
constrained by data to $\pm$15\% in the signal regions as shown in \cref{tab:sf}.
The dominant contribution to the uncertainties that affect the \mMMC shape
originates from the statistical uncertainties in the
\dphitt scale factors
and amounts to 8\%. The systematic
uncertainty in these scale factors is
estimated by comparing them with scale factors computed from the nOC
region and a CR defined by requiring opposite-charge \tauhadvis to pass
\enquote{loose} but not \enquote{medium} identification. Minor contributions arise from
the uncertainty in the extrapolation from the nOC requirement and the uncertainty
from the subtraction of simulated backgrounds. The combination of these uncertainties
leads to a total variation in the \mMMC template shape by at most 10\% between bins.

\section{Systematic uncertainties}
\label{sec:systematics}

The expected signal and background yields in the various signal and control regions
as well as the shape of the \mMMC distributions in the signal regions are affected by systematic uncertainties.
These are discussed below, grouped into three categories: theoretical uncertainties in signal, theoretical uncertainties in background, and experimental uncertainties.
The uncertainties in backgrounds from misidentified \Pgt{}-leptons, which are estimated using data-driven techniques, are discussed in \cref{sec:bkgFakes}.
The effects of all uncertainties are included in the fit model described in \cref{sec:results}.

\subsection{Theoretical uncertainties in signal}
\label{sec:systematics_theory_signal}

The procedures to estimate the uncertainty in the Higgs production cross sections
follow the recommendations by the LHC Higgs Cross Section Working
Group~\cite{Dittmaier:2011ti}.
They are briefly summarized below.
Uncertainties are evaluated separately for their impact on the total cross section,
their impact on the acceptance in different SRs, and on the shape of the \mMMC distribution in each SR.

The cross section of \ggF production in association with an exclusive number of additional jets has large uncertainties from higher-order QCD corrections~\cite{deFlorian:2016spz}.
In this analysis, the boosted and \VBF categories almost exclusively select \ggF events with one and two additional jets, respectively.
To take this effect into account, nine uncertainty sources are included.
Four sources account for uncertainties in the jet multiplicities due to missing higher-order corrections: Two sources account for yield uncertainties and two sources account for migration uncertainties of zero to one jets and one to at least two jets in the event, respectively, using the STWZ~\cite{Stewart:2013faa} and BLPTW~\cite{Stewart:2013faa,Liu:2013hba,Boughezal:2013oha} predictions as an input.
Three uncertainty sources parameterize modeling uncertainties in the Higgs-boson \pT, two of which encapsulate the migration uncertainty between the intermediate and high-\pT regions of events with at least one jet, and one which encapsulates the treatment of the top-quark mass in the loop corrections, where the difference between the LO and NLO predictions is taken as an uncertainty due to missing higher-order corrections.
Two sources account for the acceptance uncertainties of \ggF production in the \VBF phase space from selecting exactly two and at least three jets, respectively. Their size is estimated using an extension of the Stewart--Tackmann method~\cite{Stewart:2011cf,Gangal:2013nxa}.
The resulting acceptance uncertainties from these nine sources range from 1\% to 10\%, with the dominant uncertainties due to the modeling of the Higgs \pT distribution in all SRs, to the scale variations in the boosted SRs, and to the acceptance uncertainties in the \VBF signal regions.

For \VBF and \VH production cross sections, the
uncertainties due to missing higher-order QCD corrections are
estimated by varying the factorization and renormalization scales by
factors of two around the nominal scale.
The resulting uncertainties in the total cross section are below 1\% for \VBF and $\PW\PH$
production and below 5\% for $\PZ\PH$ production.
The uncertainties in the acceptance in the different SRs are about
1\% for \VBF production in all categories.
For \VH production the relative acceptance uncertainty ranges between $-10\%$ and $+20\%$
in \VBF SRs. It is below 10\% in boosted SRs.

Uncertainties related to the simulation of the underlying event, hadronization and parton shower for all signal samples are estimated by comparing the acceptance when using the default UEPS model from \pythiaEightTwo with an alternative UEPS model from \herwig.
The resulting acceptance uncertainties range from 2\% to 26\% for \ggF production and from 2\% to 18\% for \VBF production, depending on the signal region.
The PDF uncertainties are estimated using 30 eigenvector variations and two \alphas variations
that are evaluated independently relative to the default PDF set \textsc{PDF4LHC15}~\cite{Butterworth:2015oua}.
The total uncertainty due to these variations is 5\% or less depending on the SR and the Higgs production mode.
Finally, an uncertainty in the \Htt decay branching ratio
of 1\%~\cite{Dittmaier:2011ti} affects the signal rates.
All sources of theoretical uncertainties in the signal expectation are correlated across SRs.

\subsection{Theoretical uncertainties in backgrounds}
\label{sec:systematics_theory_background}

Uncertainties from missing higher-order corrections, the PDF parameterization and the UEPS modelling are also considered for the dominant \Ztt background.
The UEPS modelling uncertainties are estimated by comparing with an alternative \Ztt sample as described in \cref{sec:samples}.
Since its overall normalization is constrained separately in the \VBF and boosted SRs, variations due to these uncertainties are considered in the event migration within an analysis channel, in the \mMMC shape and in the relative change in acceptance between the three analysis channels.
These variations are treated as uncorrelated between the \VBF and boosted SRs.
In addition, the first two types of variations are treated as uncorrelated between the three analysis channels.
This treatment accounts for the differences in the corresponding event selections.
The largest uncertainties are due to the CKKW matching~\cite{ckkw} and are evaluated as a function of the number of true jets and the \PZ-boson \pT.
They vary between 1\% and 5\% depending on the SR.
The uncertainty in the measured cross section for electroweak \PZ production with two associated jets~\cite{STDM-2016-09} is found to be small compared to the other uncertainties in \PZ-boson production.

The top-quark background normalization in the \tll and \tlhad channels as well as the \Zll background normalization in the \tll channel are constrained by data in dedicated CRs.
All other simulated background contributions are normalized to their Monte~Carlo prediction.
For all simulated background contributions, other than \Ztt, no theoretical uncertainties are considered, as their impact is small compared to the uncertainties in the dominant backgrounds from \Ztt and misidentified leptons.

\subsection{Experimental uncertainties}

Experimental systematic uncertainties result from uncertainties in
efficiencies for triggering, object reconstruction and identification,
as well as from uncertainties in the energy scale and resolution of
jets, \tauhadvis, light leptons and \met.
These uncertainties affect both the event
yields and the shape of the \mMMC.
The dominant experimental
uncertainties in the final result are related to jet and \tauhadvis
reconstruction. 
The impact of the electron- and muon-related
uncertainties~\cite{PERF-2015-10,ATL-PHYS-PUB-2015-041,TRIG-2016-01}
on the measurement are small.
Uncertainties in the integrated
luminosity affect the number of predicted signal and background events, 
with the exception of processes that are normalized to data, see \cref{tab:sf}.
This uncertainty is 2.1\% for the combined 2015+2016 
dataset. It is derived using a methodology similar to that detailed in
Ref.~\cite{DAPR-2013-01}, and using the LUCID-2 detector for the baseline 
luminosity measurements~\cite{Avoni:2018iuv}, from a calibration of the luminosity 
scale using $x$-$y$ beam-separation scans.
 
The uncertainties of the
\tauhadvis~identification efficiency are in the range of 2--4.5\%
for the reconstruction efficiency~\cite{ATL-PHYS-PUB-2015-045},
3--14\% for the trigger efficiency (depending on the \tauhadvis
\pt), 5--6\% for the identification efficiency and 3--14\%
for the rate at which an electron is misidentified as 
\tauhadvis (depending on the \tauhadvis
$\eta$)~\cite{ATLAS-CONF-2017-029}. 
The uncertainties of the $\Pqb$-tagging efficiencies are measured in
dedicated calibration analyses~\cite{PERF-2012-04} and are decomposed
into uncorrelated components. 
Uncertainties in the efficiency to pass the JVT and forward JVT requirements are also considered~\cite{ATLAS-CONF-2014-018,PERF-2016-06}. 
Simulated events are corrected for
differences in these efficiencies between data and simulation and the
associated uncertainties are propagated through the analysis.

The uncertainties of the \tauhadvis energy
scale~\cite{ATLAS-CONF-2017-029} are determined
by fitting the \PZ-boson mass in \Ztt events, reconstructed using the visible \Pgt decay products. 
The precision amounts to 2--3\%, which is dominated by the
uncertainty of background modeling.
Additional uncertainties based on the modeling of the calorimeter response to single particles are added for \tauhadvis with $\pt>\SI{50}{\GeV}$~\cite{PERF-2013-06}.
The jet energy scale and its uncertainty are derived by combining
information from test-beam data, LHC collision data and
simulation~\cite{PERF-2016-04}. The uncertainties from these
measurements are factorized into eight 
principal components.
Additional uncertainties that are considered are related to jet flavor, pileup
corrections, $\eta$-dependence, and high-\pT jets, yielding a total of
20 independent sources. The uncertainties amount to 1--6\% per jet,
depending on the jet \pT. The jet energy resolution
uncertainties~\cite{ATL-PHYS-PUB-2015-015} are divided into 11
independent components and amount to 1--6\%. 

Since systematic uncertainties of the energy scales of all objects
affect the reconstructed \met, this is recalculated after each
variation is applied. The scale uncertainty of \met due to the energy
in the calorimeter cells not associated with physics objects is also
taken into account~\cite{Aaboud:2018tkc}. The uncertainty of the
resolution of \met arises from the energy resolution uncertainties of
each of the \met terms and the modeling of pileup and its effects on the soft term (see \cref{sec:reco}).

\section{Results}
\label{sec:results}

Maximum-likelihood fits are performed on data to extract parameters of
interest that probe \Htt production with increasing granularity. Firstly, a
single parameter is fitted to measure the total cross section of the
\Htt production processes.
Then, a two-parameter cross-section fit is
presented separating the \ggF and \VBF production processes. Finally,
a three-parameter fit is performed to measure \ggF production
cross sections in two exclusive regions of phase space.
For the small contribution from \HWW decays, the measurements assume
the SM predictions for production cross section and branching ratio.

A probability model is constructed that describes the \mMMC 
distributions in the 13 signal regions and the event yields in 6 control regions.
The latter are included to constrain the normalizations of the dominant backgrounds.
Each signal region is modeled by a product of Poisson distributions, where each such distribution describes the expected event count in intervals of \mMMC.
Each control region is modeled by a single Poisson distribution that 
describes the total expected event count in that region.
Signal and background predictions depend on systematic uncertainties, which are parameterized as nuisance parameters and are constrained
using Gaussian or log-normal probability distributions.
The latter are used for normalization factors (see \cref{tab:sf}) to ensure that they are
always positive.
The dependence of the predictions on nuisance parameters related to systematic uncertainties is modeled with an interpolation 
approach between yields obtained at different fixed systematic uncertainty settings.
A smoothing procedure is applied to remove occasional large local fluctuations in the \mMMC distribution templates, which encode 
systematic uncertainties of some background processes in certain regions.
For the measurements,
all theoretical uncertainties are included, except those related to 
the respective measured signal cross sections, and are correlated as described in \cref{sec:systematics_theory_signal}.
The experimental uncertainties are fully correlated across categories and
the background modeling uncertainties are generally uncorrelated, with the exception of the 
normalization factors as described in \cref{sec:background}.
Estimates of the parameters of interest and the confidence intervals are calculated with
the profile likelihood ratio~\cite{Cowan:2010js} test statistic, whereas the test statistic $\tilde{q}_0$~\cite{Cowan:2010js}
is used to compute the significances of the deviations from the background-only hypothesis.

The observed (expected) significance of the signal excess relative to the background-only hypothesis computed from the
likelihood fit is 4.4 (4.1) standard deviations, compatible with a SM Higgs boson with a mass
$m_{\PH}=\SI{125}{\GeV}$.
This result is combined with the result of the search for \Htt using data at 7 and \SI{8}{\TeV} center-of-mass energies~\cite{HIGG-2013-32}.
The combined observed (expected) significance amounts to 6.4 (5.4) standard deviations.
In this combination, all nuisance parameters are treated as uncorrelated between the two analyses.
In particular, the dominant \Ztt background is estimated differently, as mentioned in \cref{sec:intro}.

The parameter ${\POI\equiv\sigma_{\PH} \cdot \mathcal{B}(\Htt)}$ is
fitted, where $\sigma_{\PH}$ is the total cross section of the
considered Higgs-boson production processes \ggF, \VBF, \VH and \ttH, and
where $\mathcal{B}(\Htt)$ is the \Htt branching fraction.
For this measurement, the relative contributions from the various Higgs production processes are assumed as predicted by the SM
and the uncertainties related to the predicted total signal cross section are excluded.
The measured value of \POI is 
$\num[round-mode=figures,round-precision=3]{3.76558372941}\,^{+\num[round-mode=figures,round-precision=2]{0.601597224288}}_{-\num[round-mode=figures,round-precision=2]{0.591224858352}}\,\text{(stat.)}\,^{+\num[round-mode=figures,round-precision=2]{0.874736193936}}_{-\num[round-mode=figures,round-precision=2]{0.74335289208}}\,\text{(syst.)}\,\si{\pb}$,
consistent with the SM prediction, $\POIpred=\num[round-mode=figures,round-precision=3]{3.457455312} \pm \num[round-mode=figures,round-precision=2]{0.131919772196}\,\si{\pb}$~\cite{Dittmaier:2011ti}.
The signal strength \POImu is
defined as the ratio of the measured signal yield to the 
Standard Model expectation. It is computed by the fit 
described above, including uncertainties in the predicted signal 
cross section, and is evaluated to be
$1.09\,^{+0.18}_{-0.17}\,\text{(stat.)}\,^{+0.26}_{-0.22}\,\text{(syst.)}\,^{+0.16}_{-0.11}\,\text{(theory syst.)}$.

\Cref{t:yields_ll,t:yields_lh,t:yields_hh} summarize the expected signal and background yields computed by the fit in each signal region for the \POI measurement.
The signal event yields are given separately for each production
process of relevance. Within the uncertainties, good agreement is observed
between the data and the predicted sum of signal and background contributions, for 
a SM Higgs boson of mass $m_{\PH}=\SI{125}{\GeV}$ with the
measured value of \POI reported above.

\begin{table}
  \caption{Observed event yields and predictions as computed by the fit in the \tll signal regions. Uncertainties include statistical and systematic components.}
  \label{t:yields_ll}
  \centering
  \vspace{2mm}
\begin{tabular}{l%
S[table-number-alignment=right, table-figures-integer=3, table-figures-decimal=1]@{$\,\pm\,$}
S[table-number-alignment=left, table-figures-integer=2, table-figures-decimal=1]
S[table-number-alignment=right, table-figures-integer=3, table-figures-decimal=1]@{$\,\pm\,$}
S[table-number-alignment=left, table-figures-integer=2, table-figures-decimal=1]
S[table-number-alignment=right, table-figures-integer=4, table-figures-decimal=1]@{$\,\pm\,$}
S[table-number-alignment=left, table-figures-integer=2, table-figures-decimal=1]
S[table-number-alignment=right, table-figures-integer=4, table-figures-decimal=1]@{$\,\pm\,$}
S[table-number-alignment=left, table-figures-integer=2, table-figures-decimal=1]
} 
  \toprule 
	& \multicolumn{4}{c}{\tll \VBF} & \multicolumn{4}{c}{\tll boosted} \\ 
\cmidrule(lr){2-5} 
\cmidrule(lr){6-9} 
	& \multicolumn{2}{c}{Loose} & \multicolumn{2}{c}{Tight} & \multicolumn{2}{c}{Low-\pTH} & \multicolumn{2}{c}{High-\pTH} \\ 
  \midrule 
  \Ztt & 151 & 13 & 107 & 12 & 2977 & 90 & 2687 & 64 \\ 
  \Zll & 15.1 & 4.9 & 20.3 & 6.6 & 360 & 54 & 236 & 31 \\ 
  Top & 33.0 & 6.4 & 25.1 & 4.5 & 321 & 50 & 189 & 29 \\ 
  \VV & 11.8 & 2.2 & 10.7 & 1.5 & 194.1 & 8.5 & 195.3 & 8.8 \\ 
  Misidentified \Pgt & 18.3 & 9.6 & 9.6 & 4.8 & 209 & 92 & 80 & 35 \\ 
  \ggF, \HWW & 1.2 & 0.2 & 1.4 & 0.3 & 11.8 & 2.6 & 16.4 & 1.7 \\ 
  \VBF, \HWW & 1.7 & 0.2 & 4.1 & 0.5 & 2.9 & 0.3 & 2.9 & 0.3 \\ 
  \midrule 
  \ggF, \Htt & 2.6 & 0.9 & 1.8 & 0.9 & 34.4 & 9.2 & 33.8 & 9.5 \\ 
  \VBF, \Htt & 5.3 & 1.5 & 11.3 & 3.0 & 7.7 & 2.1 & 8.2 & 2.3 \\ 
  \WH, \Htt & \multicolumn{2}{l}{$<0.1$} & \multicolumn{2}{l}{$<0.1$} & 2.5 & 0.7 & 3.1 & 0.9 \\ 
  \ZH, \Htt & \multicolumn{2}{l}{$<0.1$} & \multicolumn{2}{l}{$<0.1$} & 1.3 & 0.4 & 1.6 & 0.4 \\ 
  \ttH, \Htt & \multicolumn{2}{l}{$<0.1$} & 0.1 & 0.1 & 1.5 & 0.5 & 1.2 & 0.4 \\ 
  \midrule 
  Total background & 232 & 13 & 178 & 12 & 4075 & 61 & 3408 & 54 \\ 
  Total signal & 8.0 & 2.2 & 13.2 & 3.5 & 47 & 12 & 48 & 12 \\ 
  \midrule 
  Data & \multicolumn{2}{l}{237} & \multicolumn{2}{l}{188} & \multicolumn{2}{l}{4124} & \multicolumn{2}{l}{3444} \\ 
  \bottomrule 
\end{tabular} 
\end{table}

\begin{table}
  \caption{Observed event yields and predictions as computed by the fit in the \tlhad signal regions. Uncertainties include statistical and systematic components.}
  \label{t:yields_lh}
  \centering
  \vspace{2mm}
\begin{tabular}{l%
S[table-number-alignment=right, table-figures-integer=3, table-figures-decimal=1]@{$\,\pm\,$}
S[table-number-alignment=left, table-figures-integer=2, table-figures-decimal=1]
S[table-number-alignment=right, table-figures-integer=3, table-figures-decimal=1]@{$\,\pm\,$}
S[table-number-alignment=left, table-figures-integer=2, table-figures-decimal=1]
S[table-number-alignment=right, table-figures-integer=4, table-figures-decimal=1]@{$\,\pm\,$}
S[table-number-alignment=left, table-figures-integer=2, table-figures-decimal=1]
S[table-number-alignment=right, table-figures-integer=4, table-figures-decimal=1]@{$\,\pm\,$}
S[table-number-alignment=left, table-figures-integer=2, table-figures-decimal=1]
} 
  \toprule 
	& \multicolumn{4}{c}{\tlhad \VBF} & \multicolumn{4}{c}{\tlhad boosted} \\ 
\cmidrule(lr){2-5} 
\cmidrule(lr){6-9} 
	& \multicolumn{2}{c}{Loose} & \multicolumn{2}{c}{Tight} & \multicolumn{2}{c}{Low-\pTH} & \multicolumn{2}{c}{High-\pTH} \\ 
  \midrule 
  \Ztt & 178 & 18 & 323 & 21 & 4187 & 92 & 5347 & 82 \\ 
  \Zll & 10.0 & 3.0 & 12.7 & 3.1 & 130 & 37 & 115 & 16 \\ 
  Top & 5.8 & 1.6 & 17.9 & 4.6 & 121 & 20 & 57 & 10 \\ 
  Misidentified \Pgt & 103 & 16 & 101 & 15 & 1895 & 80 & 605 & 29 \\ 
  Other backgrounds & 4.0 & 1.6 & 9.3 & 1.9 & 115.0 & 7.8 & 129.0 & 8.8 \\ 
  \midrule 
  \ggF, \Htt & 3.8 & 1.1 & 7.1 & 1.9 & 62 & 16 & 66 & 22 \\ 
  \VBF, \Htt & 7.6 & 2.2 & 24.7 & 6.8 & 11.9 & 3.4 & 14.0 & 4.0 \\ 
  \WH, \Htt & \multicolumn{2}{l}{$<0.1$} & 0.1 & 0.0 & 3.9 & 1.1 & 5.4 & 1.4 \\ 
  \ZH, \Htt & \multicolumn{2}{l}{$<0.1$} & \multicolumn{2}{l}{$<0.1$} & 1.8 & 0.5 & 2.8 & 0.7 \\ 
  \ttH, \Htt & \multicolumn{2}{l}{$<0.1$} & \multicolumn{2}{l}{$<0.1$} & 0.1 & 0.0 & 0.2 & 0.1 \\ 
  \midrule 
  Total background & 301 & 17 & 463 & 21 & 6448 & 81 & 6253 & 80 \\ 
  Total signal & 11.5 & 3.2 & 32.0 & 8.2 & 80 & 20 & 89 & 26 \\ 
  \midrule 
  Data & \multicolumn{2}{l}{318} & \multicolumn{2}{l}{496} & \multicolumn{2}{l}{6556} & \multicolumn{2}{l}{6347} \\ 
  \bottomrule 
\end{tabular} 
\end{table}

\begin{table}
  \caption{Observed event yields and predictions as computed by the fit in the \thadhad signal regions. Uncertainties include statistical and systematic components.}
  \label{t:yields_hh}
  \vspace{2mm}
  \centering
\begin{tabular}{l%
S[table-number-alignment=right, table-figures-integer=3, table-figures-decimal=1]@{$\,\pm\,$}
S[table-number-alignment=left, table-figures-integer=1, table-figures-decimal=1]
S[table-number-alignment=right, table-figures-integer=3, table-figures-decimal=1]@{$\,\pm\,$}
S[table-number-alignment=left, table-figures-integer=2, table-figures-decimal=1]
S[table-number-alignment=right, table-figures-integer=3, table-figures-decimal=1]@{$\,\pm\,$}
S[table-number-alignment=left, table-figures-integer=2, table-figures-decimal=1]
S[table-number-alignment=right, table-figures-integer=4, table-figures-decimal=1]@{$\,\pm\,$}
S[table-number-alignment=left, table-figures-integer=3, table-figures-decimal=1]
S[table-number-alignment=right, table-figures-integer=4, table-figures-decimal=1]@{$\,\pm\,$}
S[table-number-alignment=left, table-figures-integer=2, table-figures-decimal=1]
} 
  \toprule 
	& \multicolumn{6}{c}{\thadhad \VBF} & \multicolumn{4}{c}{\thadhad boosted} \\ 
\cmidrule(lr){2-7} 
\cmidrule(lr){8-11} 
	& \multicolumn{2}{c}{Loose} & \multicolumn{2}{c}{Tight} & \multicolumn{2}{c}{High-\pTH} & \multicolumn{2}{c}{Low-\pTH} & \multicolumn{2}{c}{High-\pTH} \\ 
  \midrule 
  \Ztt & 67.3 & 9.2 & 100 & 12 & 141 & 12 & 3250 & 130 & 3582 & 82 \\ 
  Misidentified \Pgt & 45.0 & 5.4 & 96.4 & 9.2 & 20.0 & 2.9 & 1870 & 140 & 364 & 53 \\ 
  Other backgrounds & 4.4 & 1.4 & 11.6 & 1.7 & 4.4 & 0.7 & 281 & 21 & 109.9 & 9.2 \\ 
  \midrule 
  \ggF, \Htt & 1.1 & 0.4 & 2.0 & 0.7 & 3.5 & 1.0 & 41 & 11 & 48 & 14 \\ 
  \VBF, \Htt & 1.4 & 0.5 & 6.4 & 1.8 & 11.2 & 3.0 & 9.0 & 3.4 & 10.7 & 2.9 \\ 
  \WH, \Htt & \multicolumn{2}{l}{$<0.1$} & \multicolumn{2}{l}{$<0.1$} & \multicolumn{2}{l}{$<0.1$} & 3.3 & 0.9 & 4.4 & 1.2 \\ 
  \ZH, \Htt & \multicolumn{2}{l}{$<0.1$} & \multicolumn{2}{l}{$<0.1$} & \multicolumn{2}{l}{$<0.1$} & 2.4 & 0.7 & 2.9 & 0.8 \\ 
  \ttH, \Htt & \multicolumn{2}{l}{$<0.1$} & \multicolumn{2}{l}{$<0.1$} & \multicolumn{2}{l}{$<0.1$} & 1.6 & 0.5 & 1.9 & 0.5 \\ 
  \midrule 
  Total background & 116.7 & 9.4 & 208 & 12 & 165 & 12 & 5401 & 78 & 4057 & 64 \\ 
  Total signal & 2.6 & 0.8 & 8.6 & 2.4 & 14.9 & 3.8 & 57 & 15 & 68 & 18 \\ 
  \midrule 
  Data & \multicolumn{2}{l}{121} & \multicolumn{2}{l}{220} & \multicolumn{2}{l}{179} & \multicolumn{2}{l}{5455} & \multicolumn{2}{l}{4103} \\ 
  \bottomrule 
\end{tabular} 
\end{table}

\begin{table}
  \caption{Summary of different sources of uncertainty in decreasing
    order of their impact on \POI.
    Their observed
    and expected fractional (\%) impacts, both computed by the fit, are given, relative to the \POI value.
    Experimental uncertainties in reconstructed objects combine efficiency and energy/momentum scale and resolution uncertainties.
    Background statistics includes the bin-by-bin statistical uncertainties in the simulated backgrounds as well as statistical uncertainties in misidentified \Pgt backgrounds, which are estimated using data.
    Background normalization describes the combined impact of all background normalization uncertainties.
    }
  \label{tab:systs}
  \centering
  \vspace{2mm}
  \begin{tabular}{l%
    S[table-number-alignment=right, table-figures-integer=2, table-figures-decimal=1, table-space-text-pre=$+$, retain-explicit-plus]@{$\,/\,$}
    S[table-number-alignment=left, table-figures-integer=2, table-figures-decimal=1, table-space-text-pre=$-$]
    S[table-number-alignment=right, table-figures-integer=2, table-figures-decimal=1, table-space-text-pre=$+$, retain-explicit-plus]@{$\,/\,$}
    S[table-number-alignment=left, table-figures-integer=2, table-figures-decimal=1, table-space-text-pre=$-$]}
      \toprule
      Source of uncertainty & \multicolumn{4}{c}{Impact $\Delta\sigma/\POI$ [\%]} \\
                            & \multicolumn{2}{c}{Observed}   & \multicolumn{2}{c}{Expected} \\
      \midrule
      Theoretical uncert. in signal      & +13.4 & -8.7 & +12.0 & -7.8 \\
      Background statistics              & +10.8 & -9.9 & +10.1 & -9.7 \\
      Jets and \met                      & +11.2 & -9.1 & +10.4 & -8.4 \\
      Background normalization           & +6.3 & -4.4  & +6.3 & -4.4 \\
      Misidentified \Pgt                 & +4.5 & -4.2  & +3.4 & -3.2 \\
      Theoretical uncert. in background  & +4.6 & -3.6  & +5.0 & -4.0 \\
      Hadronic \Pgt decays               & +4.4 & -2.9  & +5.5 & -4.0 \\
      Flavor tagging                     & +3.4 & -3.4  & +3.0 & -2.3 \\
      Luminosity                         & +3.3 & -2.4  & +3.1 & -2.2 \\
      Electrons and muons                & +1.2 & -0.9  & +1.1 & -0.8 \\
      \midrule
      Total systematic uncert.           & +23 & -20    & +22 & -19 \\
      Data statistics & \multicolumn{2}{l}{$\qquad\pm 16$} & \multicolumn{2}{l}{$\qquad\pm 15$}\\
      Total                              & +28 & -25    & +27 & -24 \\
      \bottomrule
  \end{tabular}
\end{table}

\begin{figure}[tb!]
  \begin{center}
    \includegraphics[width=0.6\textwidth]{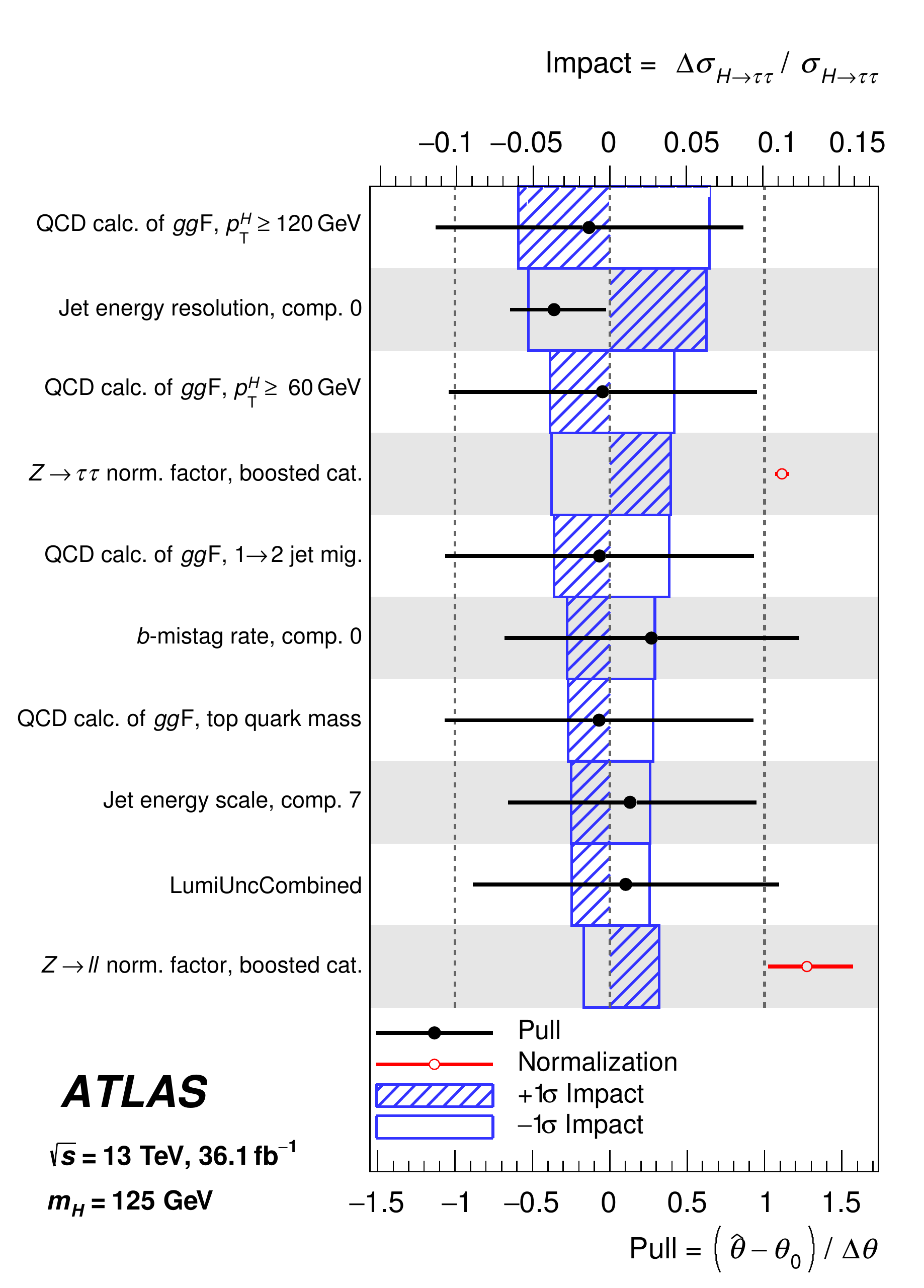}
  \end{center}
  \caption{Fractional impact of systematic uncertainties in \POI as computed by
    the fit. The systematic uncertainties are listed in decreasing
    order of their impact on \POI on the $y$-axis. The hatched
    blue and open blue boxes show the variations of \POI
    referring to the
    top $x$-axis (impact), as described in the text.
    The filled circles, referring to the bottom $x$-axis,
    show the pulls of the fitted nuisance parameters, i.e.\ the
    deviations of the fitted parameters $\hat\theta$ from their
    nominal values $\theta_0$, normalized to their nominal
    uncertainties $\Delta\theta$. The black lines show the
    uncertainties of the nuisance parameters resulting from the fit. 
    Several sources of uncertainties such as the jet energy scale and resolution as well as the \Pqb-mistag rate are described by their principal components in the fit.
    The open circles, also referring to the bottom $x$-axis, show the
    values of the fitted \Ztt and \Zll normalization factors in the
    boosted category as listed in \cref{tab:sf}.
    Their uncertainties do not include uncertainties in total simulated cross sections extrapolated to the selected phase space.
    }
  \label{fig:nuisance-parameters}
\end{figure}

\Cref{tab:systs} shows a summary of the dominant uncertainties in
\POI, grouped by their respective sources. \Cref{fig:nuisance-parameters} shows the systematic
uncertainties with the largest impact, together with a comparison with their
nominal values used as input to the fit.
In both the table and the figure the shown uncertainties are 
ranked by their fractional impact on the measurement of \POI. 
To compute the impact for each 
nuisance parameter, a separate fit is performed again with the parameter fixed to its
fitted value, and the resulting uncertainty in \POI
is subtracted from the uncertainty obtained in the original fit via variance subtraction.
The dominant uncertainties 
are related to the limited number of events in the simulated samples,
the missing higher-order QCD corrections to the signal
process cross sections, the jet energy resolution, the \tauhadvis
identification and the normalizations of the \Ztt and \Zll backgrounds. 
\Cref{fig:nuisance-parameters} also shows that in most cases the fitted parameters are in agreement 
with the nominal values,
except for the uncertainties related to jet
energy resolution and scale.
In the case of real di-\Pgt events, the distribution of \mMMC is sensitive to the jet-related uncertainties because
selected di-\Pgt events in the \VBF and boosted categories are characterized by one or more high-\pt jets that
recoil against the two \Pgt{}-leptons.
The main contributions to \met are thus the neutrinos in the \Pgt{}-lepton decays and the impact of the jet energy resolution when projected onto the \met direction.
Applying both the jet energy resolution and scale uncertainties causes a shift in the mean jet \pT, which therefore translates directly into a shift of the reconstructed \met.
This, in turn, translates into a shift of the reconstructed \mMMC that is constrained by data in the region of the \Ztt mass peak.

\begin{figure}[htbp]
  \centering
  \subfloat[]{\includegraphics[width=0.49\textwidth]{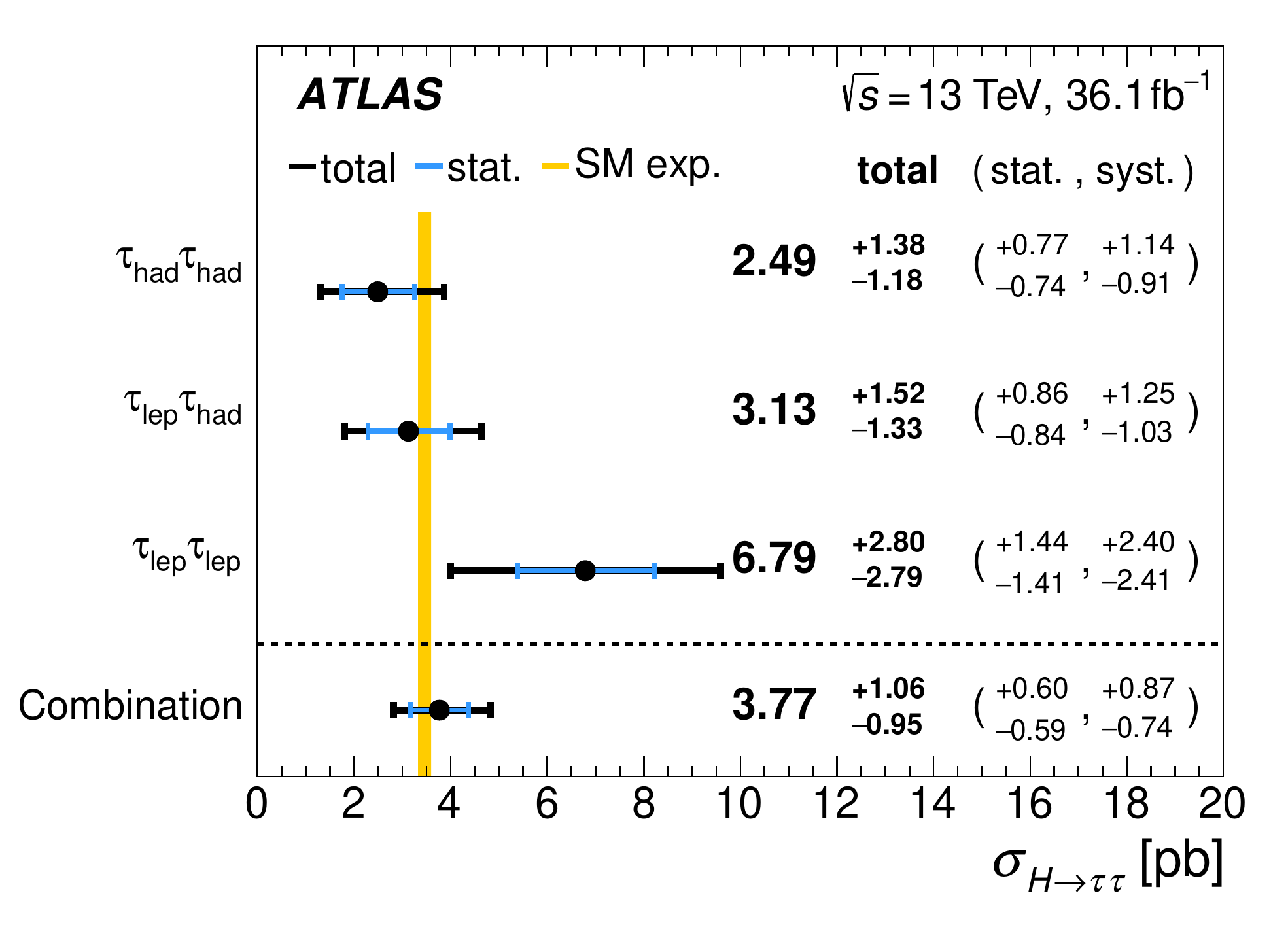} } 
  \subfloat[]{\includegraphics[width=0.49\textwidth]{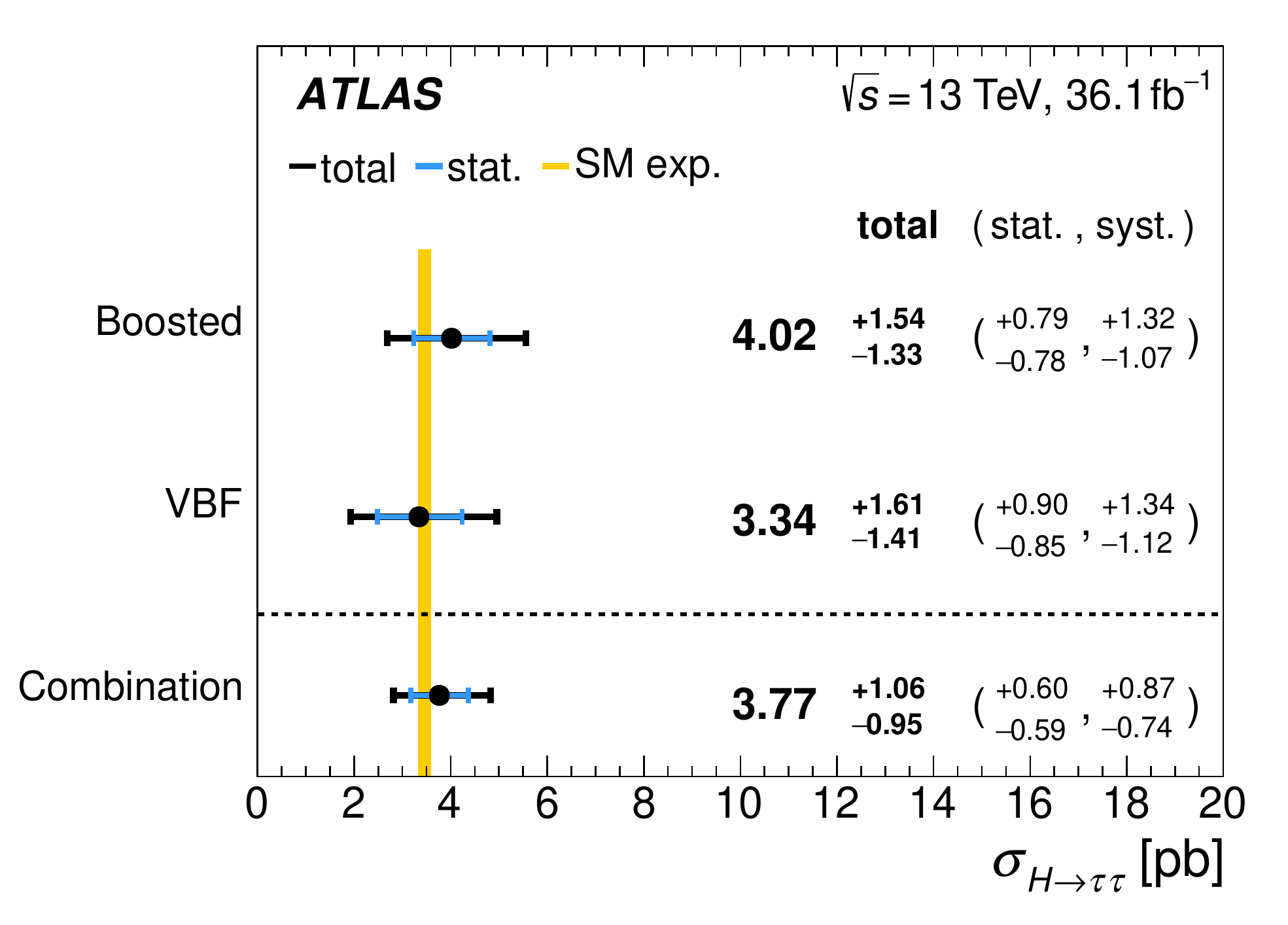} }
  \caption{The measured values for
    \POI when only the data of (a) individual channels
    or (b) individual categories are used.
    Also shown is the result from the
    combined fit. The total $\pm{}1\sigma$ uncertainty in the measurement is indicated by
    the black error bars, with the individual contribution from the
    statistical uncertainty in blue. The theory uncertainty 
    in the predicted signal cross section is shown by the yellow band.}
  \label{fig:x-section_summary}
\end{figure}

\begin{figure}[htbp]
   \centering
   \includegraphics[width=0.32\textwidth]{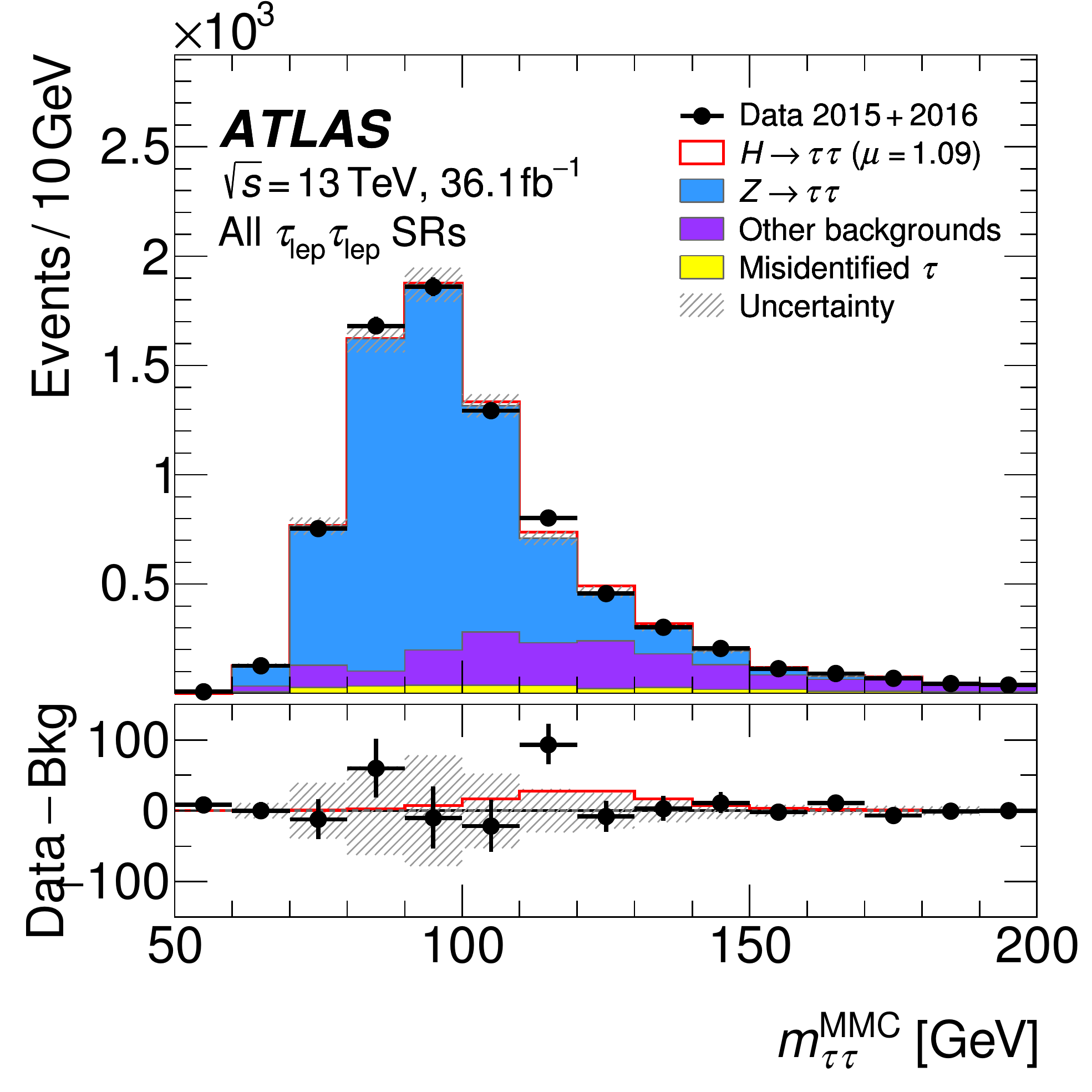}
   \includegraphics[width=0.32\textwidth]{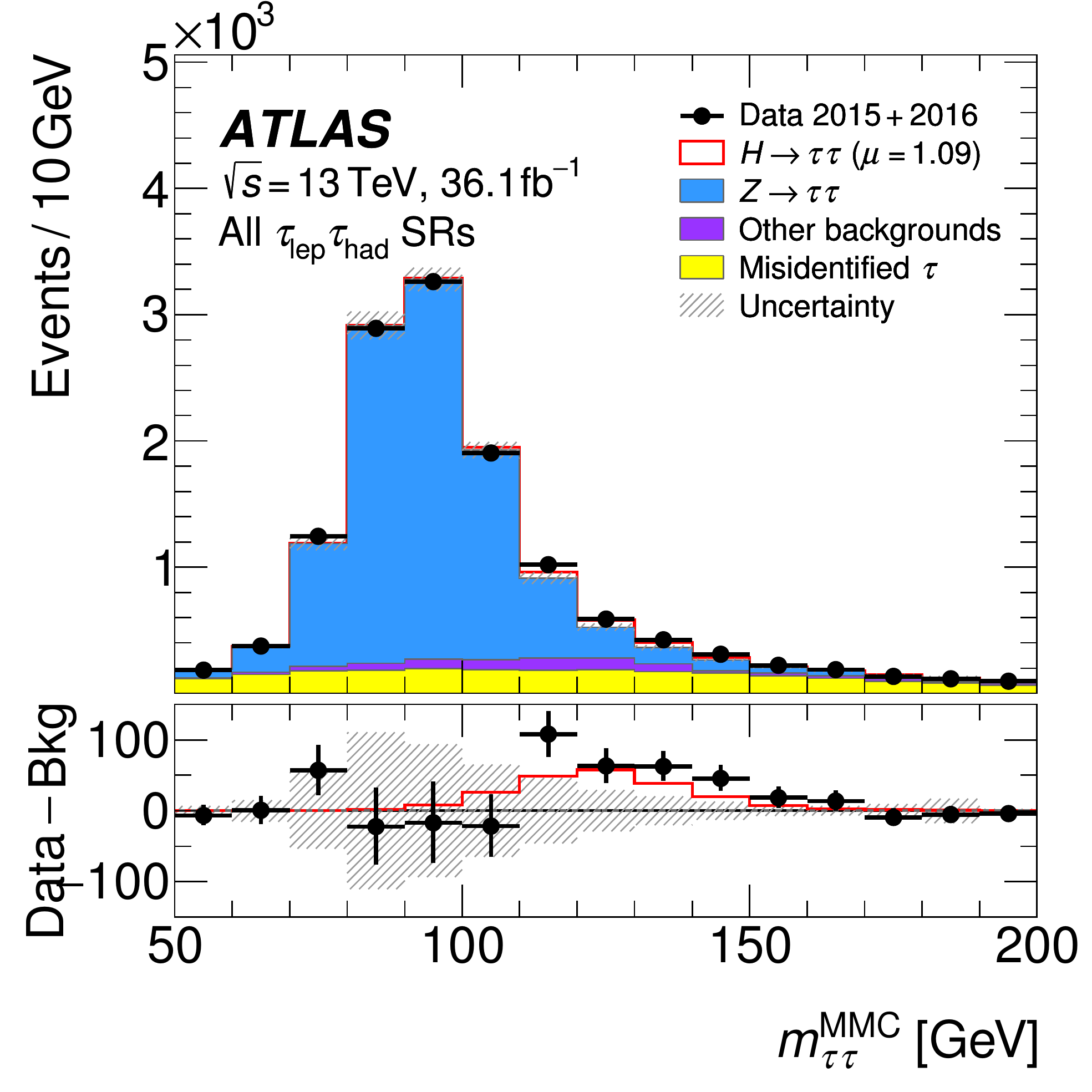}
   \includegraphics[width=0.32\textwidth]{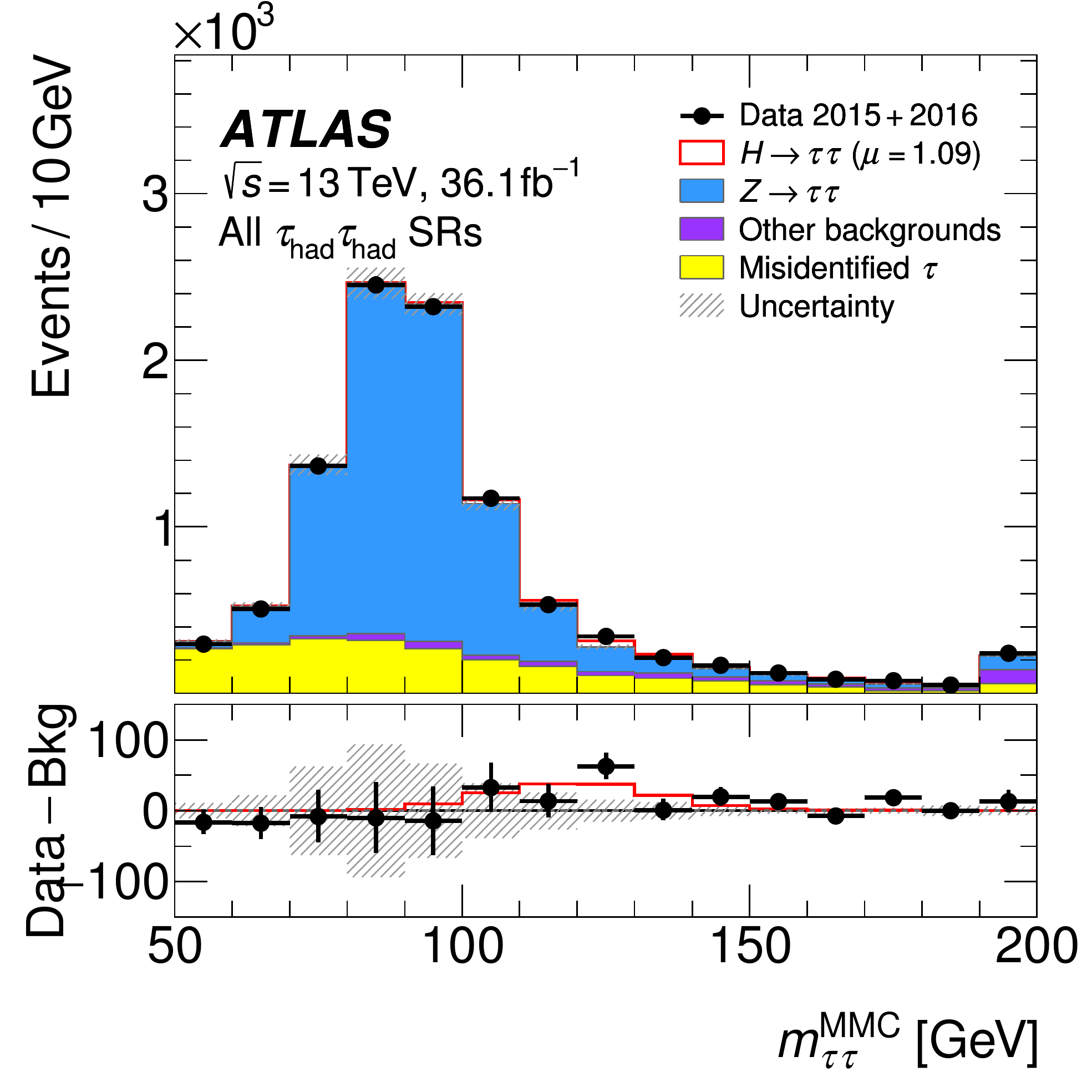}
   \caption{Distributions of the reconstructed di-\Pgt invariant mass ($\mMMC$) for the sum of
     (left) all \tll, (center) all \tlhad and (right) all \thadhad signal regions (SRs).
     The bottom panels show the differences between observed data events and expected background
     events (black points).
     The observed Higgs-boson signal ($\mu = 1.09$) is shown with the solid red line.
     Entries with values that would exceed the $x$-axis range are shown in the last bin of each distribution.
     The signal and background predictions are determined in the likelihood fit.
     The size of the combined statistical, experimental and theoretical uncertainties in the background is indicated by the hatched bands.}
   \label{fig:CBA_m-tautau_channel}
 \end{figure}

\begin{figure}[htbp]
   \centering
   \subfloat[]{\includegraphics[width=0.49\textwidth]{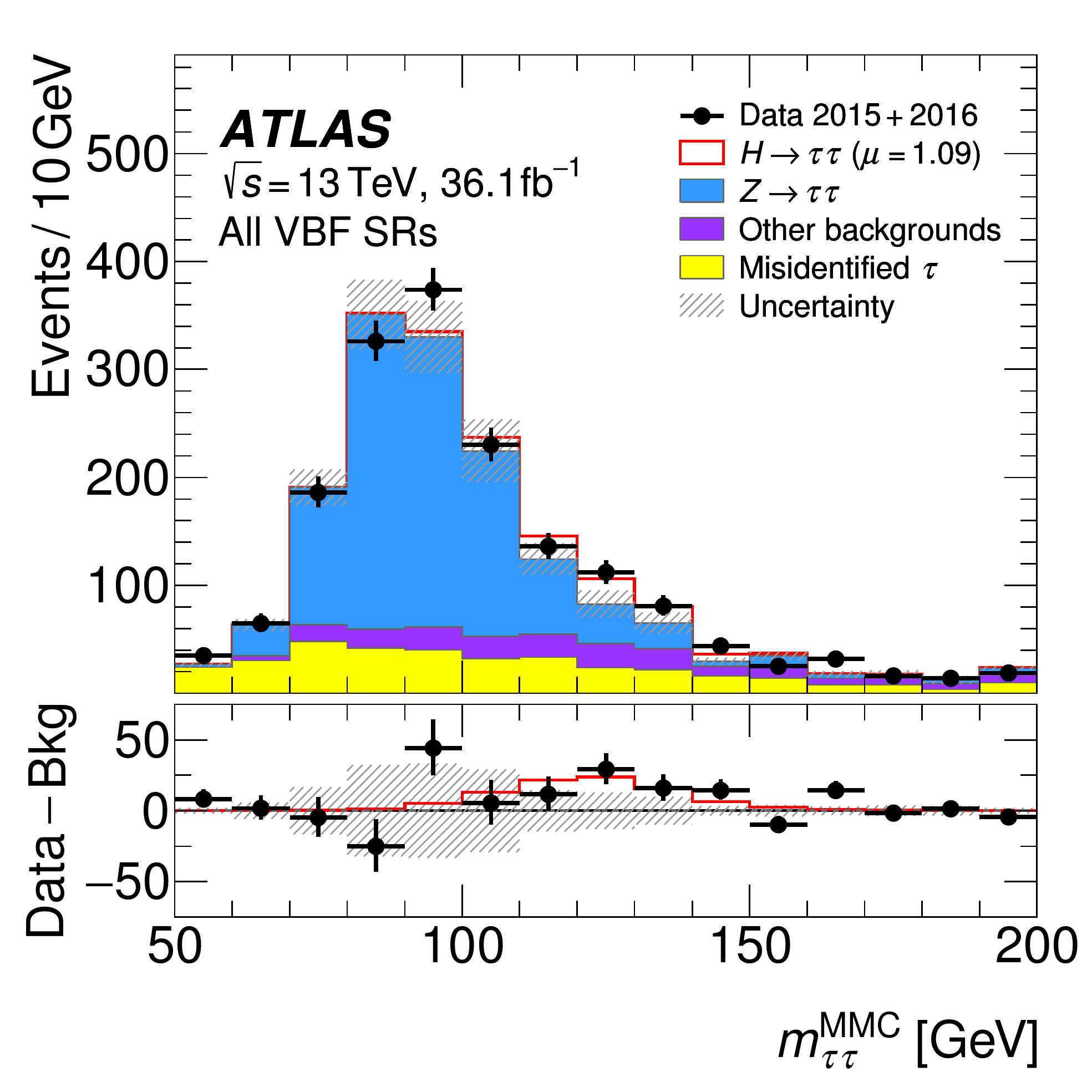}}
   \subfloat[]{\includegraphics[width=0.49\textwidth]{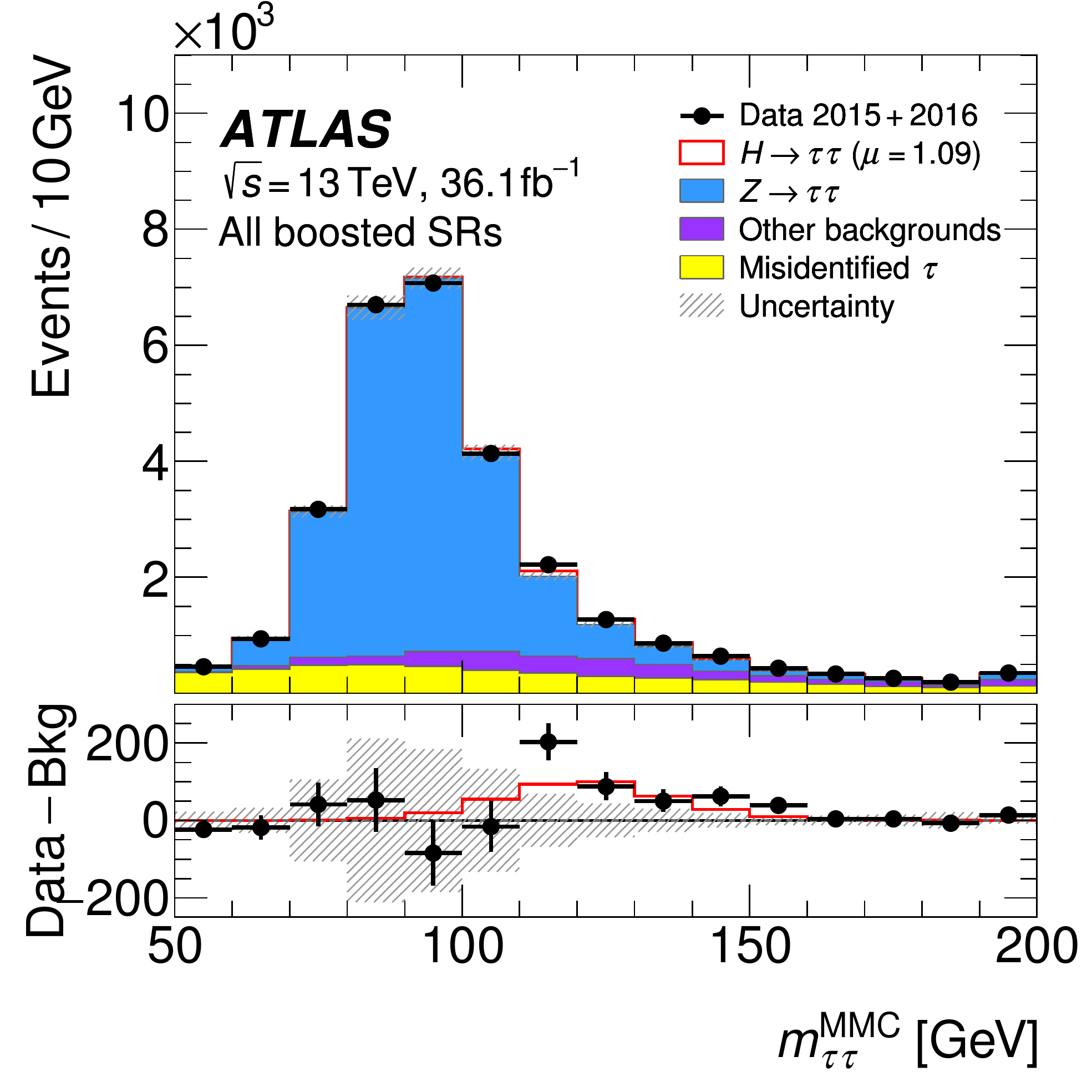}}
   \caption{Distribution of the reconstructed di-\Pgt invariant mass ($\mMMC$) for the sum of
     (a) all \VBF and (b) all boosted signal regions (SRs).
     The bottom panels show the differences between observed data events and expected background
     events (black points).
     The observed Higgs-boson signal ($\mu = 1.09$) is shown with the solid red line.
     Entries with values that would exceed the $x$-axis range are shown in the last bin of each distribution.
     The signal and background predictions are determined in the likelihood fit.
     The size of the combined statistical, experimental and theoretical uncertainties in the background is indicated by the hatched bands.}
   \label{fig:CBA_m-tautau}
 \end{figure}

\begin{figure}[htbp]
   \centering
   \includegraphics[width=0.49\textwidth]{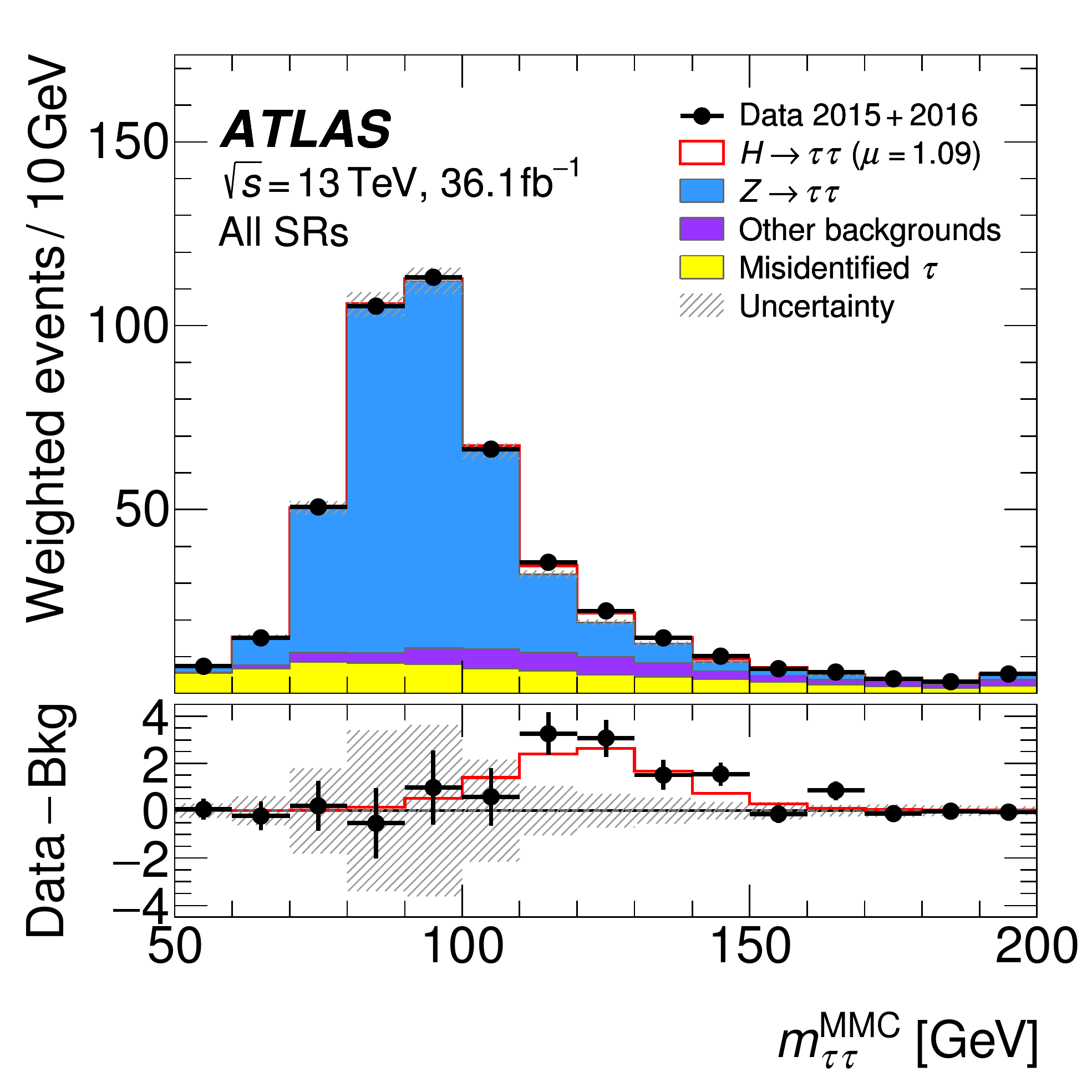}
   \caption{Distribution of the reconstructed di-\Pgt invariant mass ($\mMMC$) for the sum of
     all signal regions (SRs).
     The contributions of the different SRs are weighted by a factor of $\ln(1+S/B)$, where $S$ and $B$ are the expected numbers of signal and background events in that region, respectively.
     The bottom panel shows the differences between observed data events and expected background
     events after applying the same weights (black points).
     The observed Higgs-boson signal ($\mu = 1.09$) is shown with the solid red line.
     Entries with values that would exceed the $x$-axis range are shown in the last bin of each distribution.
     The signal and background predictions are determined in the likelihood fit. 
     The size of the combined statistical, experimental and theoretical uncertainties in the background is indicated by the hatched bands.}
   \label{fig:CBA_m-tautau_weighted}
 \end{figure}

Results of the fit when only the data of an individual channel or of an individual category are used,
are shown in \cref{fig:x-section_summary}. Also shown is the result from the 
fit and the uncertainty in \POIpred.
All results are consistent with the SM expectations.
The simple combination of the individual fit results does not agree exactly with
the result of the combined fit because the values of the nuisance 
parameters are different.
The \mMMC distributions in all signal regions with background predictions adjusted by the likelihood fit are shown in \cref{fig:MMC-signal_vbf,fig:MMC-signal_boosted} in the Appendix.
The \mMMC distributions for the predicted signal plus background are compared with the data in \cref{fig:CBA_m-tautau_channel}, separately for the combined signal regions of \thadhad, \tlhad and \tll analysis channels, and in \cref{fig:CBA_m-tautau}, separately for the combined \VBF and the combined boosted signal regions.
A weighted combination of the \mMMC distributions in all signal regions is shown in \cref{fig:CBA_m-tautau_weighted}.
The events are weighted by a factor of $\ln(1+S/B)$ which enhances the events compatible with the signal hypothesis.
Here, $S/B$ is the expected signal-to-background ratio in the corresponding signal region.

\begin{figure}[htbp]
  \begin{center}
    \includegraphics[width=11cm]{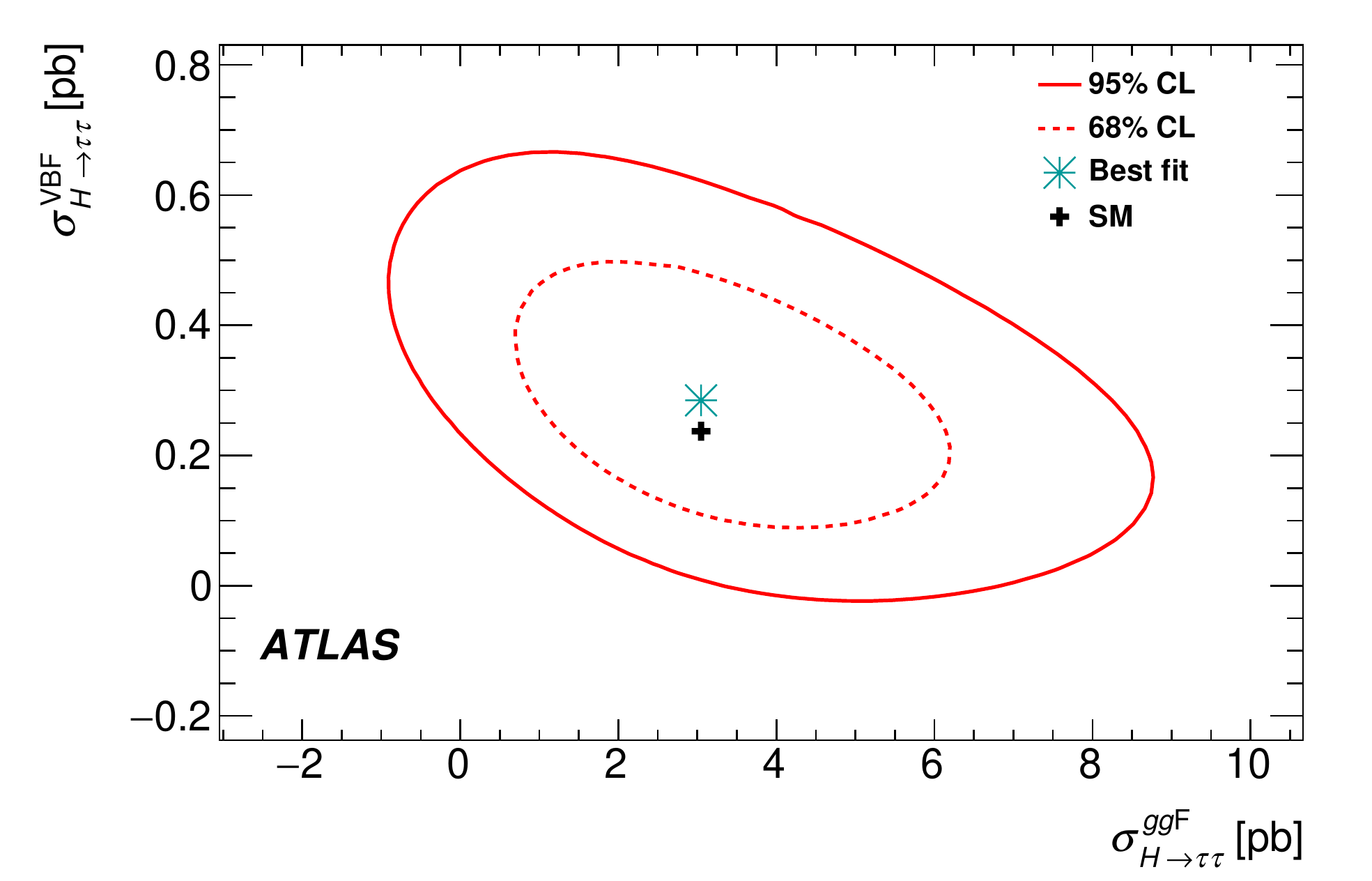}
  \end{center}
  \caption{Likelihood contours
    for the combination of all channels in the
    (\xsecggF, \xsecVBF) plane. The 68\% and 95\% CL contours are shown as dashed
    and solid lines, respectively, for $m_{\PH}=\SI{125}{\GeV}$.  The SM
    expectation is indicated by a plus symbol and the best fit to
    the data is shown as a star.}
  \label{fig:2D-contour}
\end{figure}

\Cref{fig:x-section_summary} illustrates that the \VBF and boosted categories provide good sensitivity, respectively, to \VBF and \ggF Higgs-boson production.
A two-parameter fit is therefore performed to determine the cross sections of these production processes by exploiting the sensitivity offered by the use of the event categories in the analysis of the three channels.
Two cross-section parameters \xsecVBF and \xsecggF are introduced
and the data are fitted to these parameters, separating 
the vector-boson-mediated
\VBF process from the fermion-mediated \ggF process, while the contributions from other Higgs production
processes are set to their predicted SM values. 
The two-dimensional $68\%$ and $95\%$ confidence level
(CL) contours in the plane of \xsecVBF and \xsecggF
are shown in \cref{fig:2D-contour}. The best-fit values are
$\xsecVBF = \num[round-mode=figures,round-precision=2]{0.2800453818} \pm \num[round-mode=figures,round-precision=1]{0.09}\,\text{(stat.)}\,^{+\num[round-mode=figures,round-precision=2]{0.106}}_{-\num[round-mode=figures,round-precision=1]{0.093}}\,\text{(syst.)}\,\si{\pb}$
and
$\xsecggF = \num[round-mode=figures,round-precision=2]{3.096016028} \pm \num[round-mode=figures,round-precision=2]{1.0}\,\text{(stat.)}\,^{+\num[round-mode=figures,round-precision=2]{1.63}}_{-\num[round-mode=figures,round-precision=2]{1.258}}\,\text{(syst.)}\,\si{\pb}$,
in agreement with the predictions from the Standard Model of
$\xsecpredVBF=\num[round-mode=figures,round-precision=3]{0.23721}\pm 0.006\,\si{\pb}$
and
$\xsecpredggF=\num[round-mode=figures,round-precision=3]{3.0469}\pm\num[round-mode=figures,round-precision=2]{0.128994749633}\,\si{\pb}$~\cite{Dittmaier:2011ti}. The two
results are strongly anti-correlated (correlation coefficient of
$-52$\%), as can be seen in \cref{fig:2D-contour}.

The \ggF signal provides enough events to measure \ggF cross
sections in mutually exclusive regions of the \ggF phase space. Two
\ggF regions are defined by particle-level events with at least one
jet where a jet is required to have $\pT>\SI{30}{\GeV}$: events with a
Higgs-boson \pT of $60<\pTHtruth<\SI{120}{\GeV}$ and events with
$\pTHtruth>\SI{120}{\GeV}$.
A cross-section parameter for each of the two \ggF regions is
introduced, along with a parameter for \VBF production in an inclusive
region, and a combined three-parameter fit is performed using the
event categories in the analysis of the three channels. The
particle-level definitions of all three phase-space regions closely follow
the framework of simplified template cross
sections~\cite{deFlorian:2016spz} where the Higgs-boson rapidity $y_H$
is required to satisfy $|y_H|<2.5$.
The \ggF and \VBF production cross sections outside the respective particle-level region requirements are set to the measured values reported above.
Cross sections of other Higgs-boson production processes are set to their SM values.
\Cref{tab:stxs} shows the resulting cross sections along with the
SM predictions in the respective particle-level region.
The measurements in all regions have a precision similar to that of
the inclusive ggF and VBF measurements reported above.

\begin{table}
  \caption{Measurement of the \VBF and \ggF production cross sections 
    in three mutually exclusive regions of phase space 
    of particle-level events.
    The number of jets $N_\text{jets}$ in \ggF events comprises all jets with $\pT>\SI{30}{\GeV}$.
    The cross section of \ggF events that
    fail the particle-level requirements of the two \ggF regions is set to the measured \xsecggF value. 
    Results are shown along with  
    the SM predictions in the respective particle-level regions. 
    The definitions of the regions closely follow the framework of simplified
    template cross sections~\cite{deFlorian:2016spz}.}
  \label{tab:stxs}
  \centering
  \vspace{2mm}
  \begin{tabular}{lc%
S[table-number-alignment=right, table-figures-integer=1, table-figures-decimal=2]@{$\,\pm\,$}
S[table-number-alignment=right, table-figures-integer=1, table-figures-decimal=2]@{$\,\text{(stat.)} \pm \,$}
S[table-number-alignment=right, table-figures-integer=1, table-figures-decimal=2]@{$\,\text{(syst.)}\ \ \ $}
S[table-number-alignment=right, table-figures-integer=1, table-figures-decimal=2]@{$\,\pm\,$}
S[table-number-alignment=right, table-figures-integer=1, table-figures-decimal=2]}
    \toprule
    Process & Particle-level selection & \multicolumn{3}{c}{$\sigma$ [pb]} & \multicolumn{2}{c}{$\sigma^\text{SM}$ [pb]} \\
    \midrule
    \ggF & $N_\text{jets} \geq 1$, $60<\pTHtruth<\SI{120}{\GeV}$, $|y_H|<2.5$ & 1.79 & 0.53 & 0.74 & 0.40 & 0.05 \\
    \ggF & $N_\text{jets} \geq 1$, $\pTHtruth>\SI{120}{\GeV}$, $|y_H|<2.5$ & 0.12 & 0.05 & 0.05 & 0.14 & 0.03 \\
    \VBF & $|y_H|<2.5$ & 0.25 & 0.08 & 0.08 & 0.22 & 0.01 \\
    \bottomrule
  \end{tabular}
\end{table}

\FloatBarrier

\section{Conclusions}
\label{sec:conclusions}

A measurement of total production cross sections of the Higgs boson
in proton--proton collisions is presented in the
\Htautau decay channel. The analysis was performed using
\SI{36.1}{\per\fb} of data recorded by the ATLAS experiment at the LHC
at a center-of-mass energy of $\sqrt{s}=\SI{13}{\TeV}$.
All combinations of leptonic and hadronic \Pgt
decays were considered. An excess of events over the expected
background from other Standard Model processes was found with an
observed (expected) significance of 4.4 (4.1) standard deviations.
Combined with results using data taken
at $\sqrt s$ of~7 and \SI{8}{\TeV}, the observed (expected) significance amounts to 6.4 (5.4)
standard deviations and constitutes an observation of \Htt decays by the ATLAS experiment.
Using the data taken at $\sqrt{s}=\SI{13}{\TeV}$, the $\pp \to \Htt$ total cross section 
is measured to be
$\num[round-mode=figures,round-precision=3]{3.76558372941}\,^{+\num[round-mode=figures,round-precision=2]{0.601597224288}}_{-\num[round-mode=figures,round-precision=2]{0.591224858352}}\,\text{(stat.)}\,^{+\num[round-mode=figures,round-precision=2]{0.874736193936}}_{-\num[round-mode=figures,round-precision=2]{0.74335289208}}\,\text{(syst.)}\,\si{\pb}$,
for a Higgs boson of mass \SI{125}{\GeV}.
A two-dimensional fit was performed to
separate the vector-boson-mediated \VBF process from the
fermion-mediated \ggF process.
The cross sections of the Higgs boson decaying into two \Pgt leptons are measured to be 
$\xsecVBF = \num[round-mode=figures,round-precision=2]{0.2800453818} \pm \num[round-mode=figures,round-precision=1]{0.09}\,\text{(stat.)}\,^{+\num[round-mode=figures,round-precision=2]{0.106}}_{-\num[round-mode=figures,round-precision=1]{0.093}}\,\text{(syst.)}\,\si{\pb}$ and 
$\xsecggF = \num[round-mode=figures,round-precision=2]{3.096016028} \pm \num[round-mode=figures,round-precision=2]{1.0}\,\text{(stat.)}\,^{+\num[round-mode=figures,round-precision=2]{1.63}}_{-\num[round-mode=figures,round-precision=2]{1.258}}\,\text{(syst.)}\,\si{\pb}$, respectively, for the two production processes.
Similarly, a three-dimensional fit was performed in the framework of simplified template cross sections.
Results are reported for the \VBF cross section in an inclusive phase space and
\ggF cross sections in two exclusive regions of phase space 
defined by particle-level requirements on the Higgs-boson \pT.
All measurements are consistent with SM predictions.

\clearpage

\appendix
\part*{Appendix: Distributions of \mMMC in signal regions}
\addcontentsline{toc}{part}{Appendix}

\begin{figure}[htbp]
  \centering
    \hspace*{0.64\linewidth}
    \includegraphics[width=0.32\linewidth]{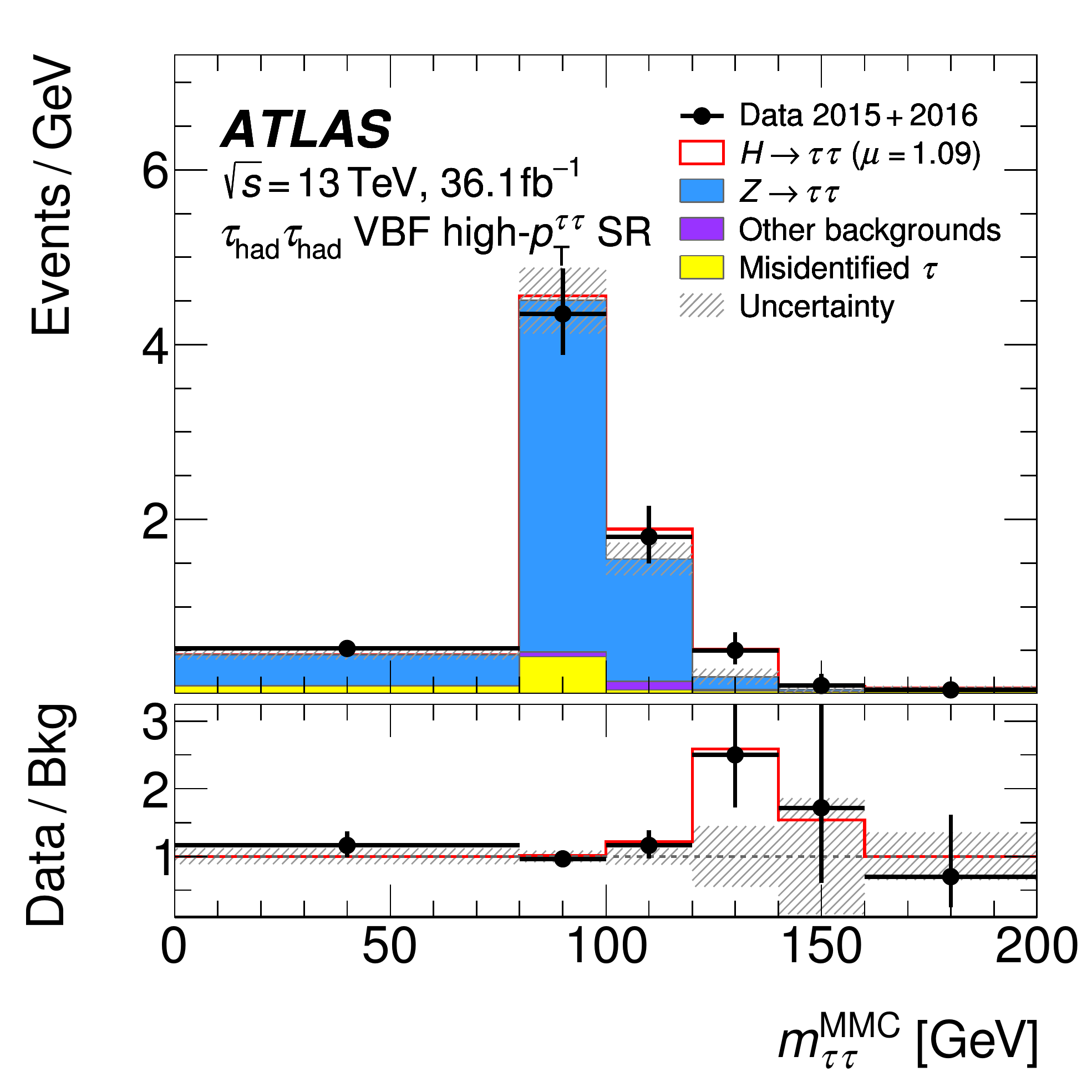}\\
    \includegraphics[width=0.32\linewidth]{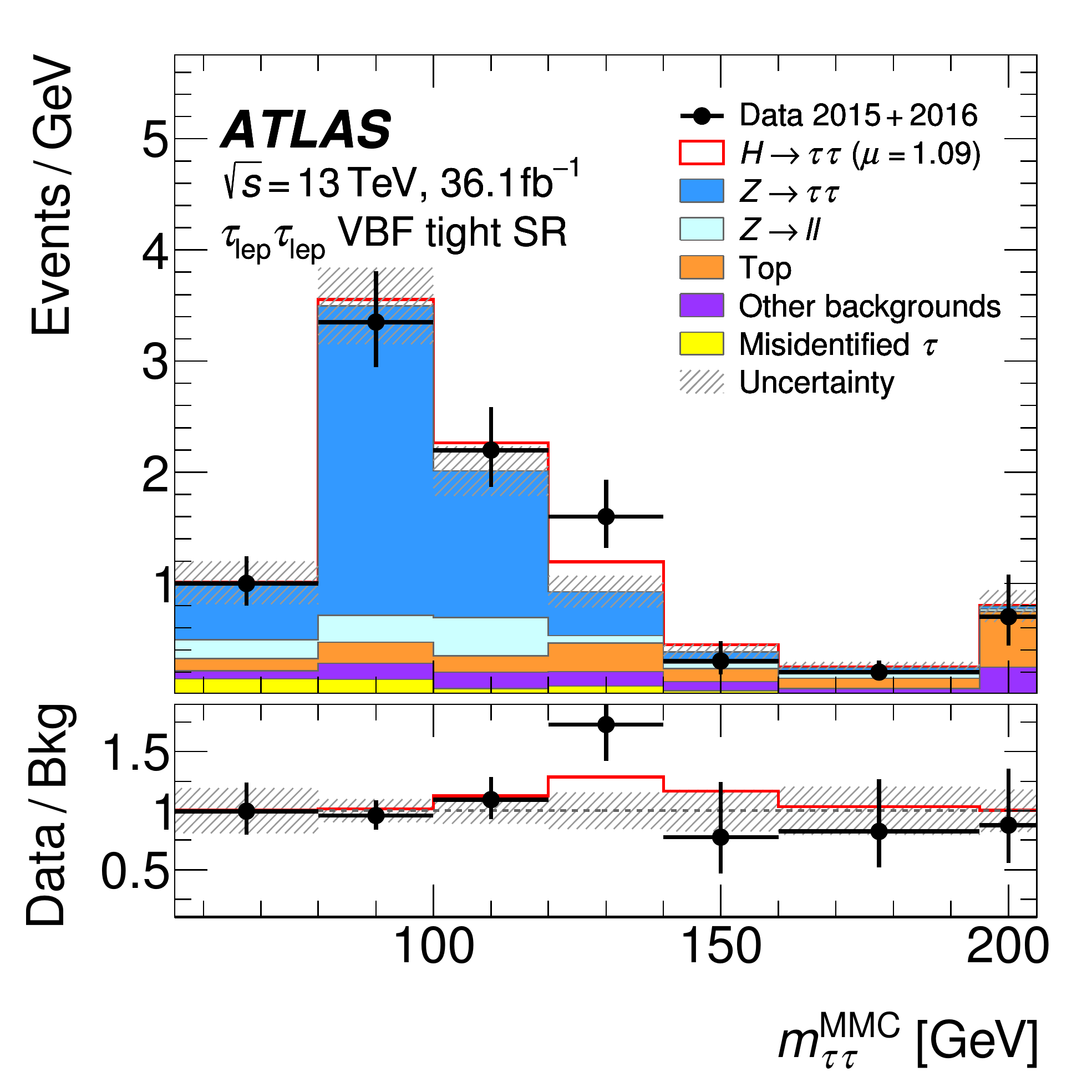}
    \includegraphics[width=0.32\linewidth]{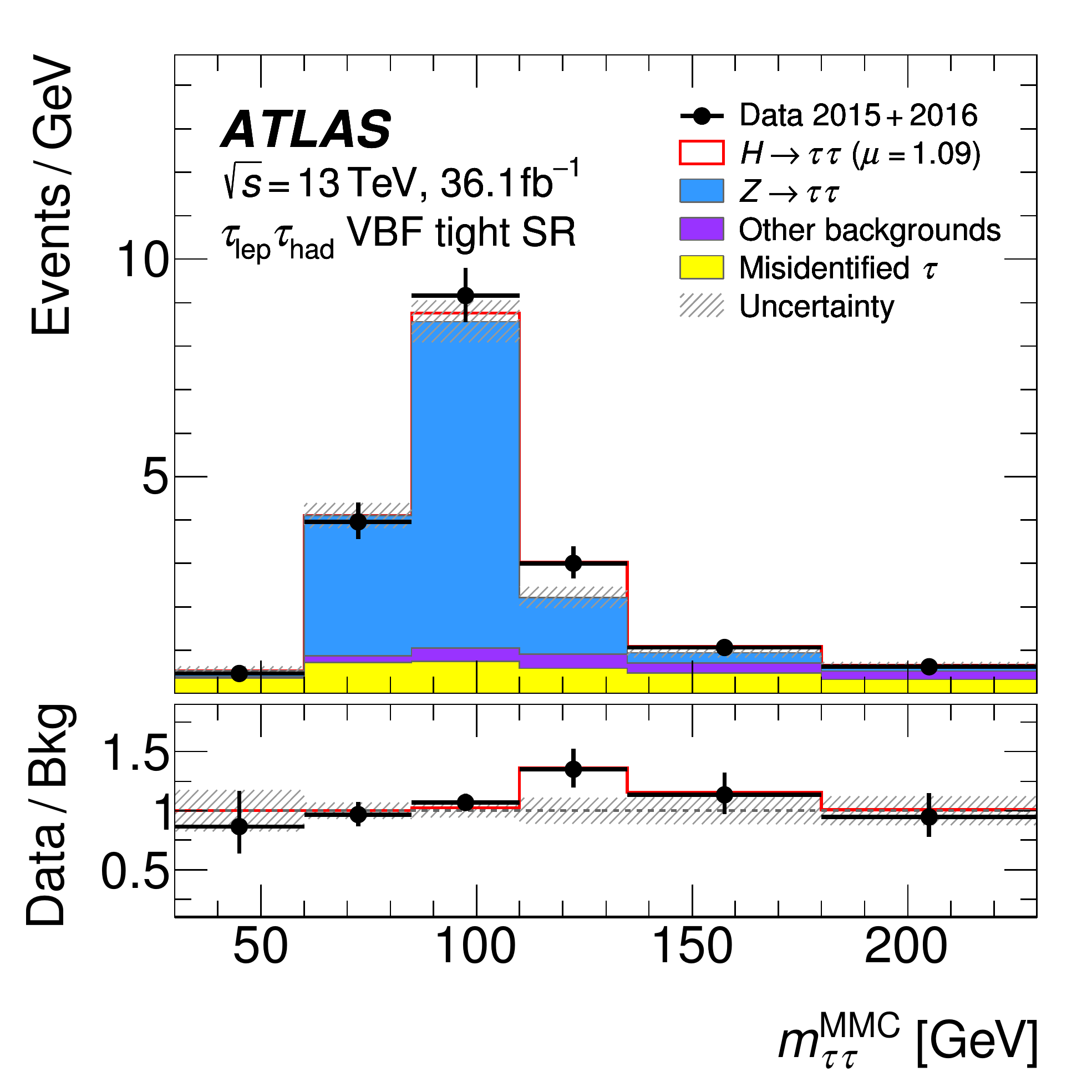}
    \includegraphics[width=0.32\linewidth]{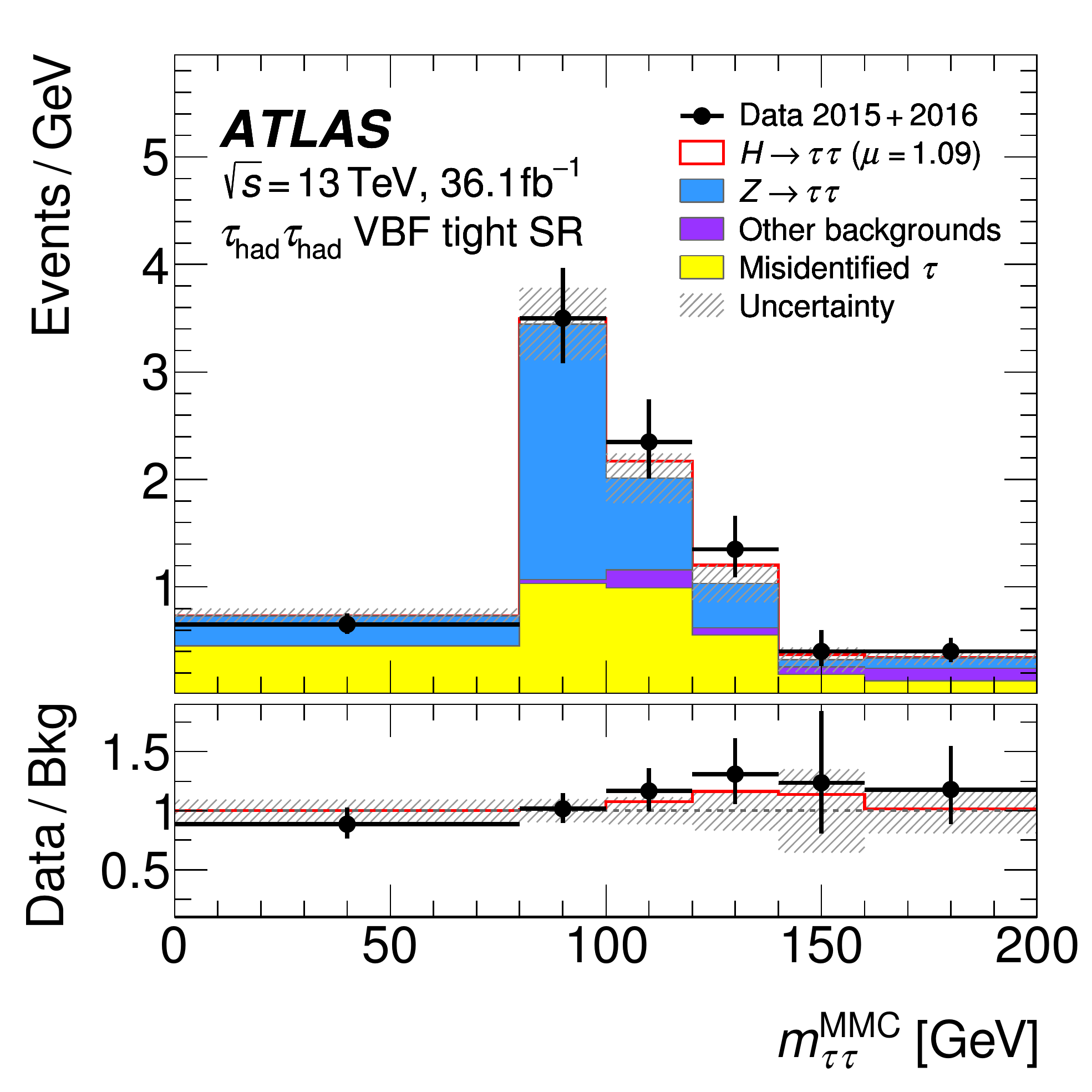}\\
    \includegraphics[width=0.32\linewidth]{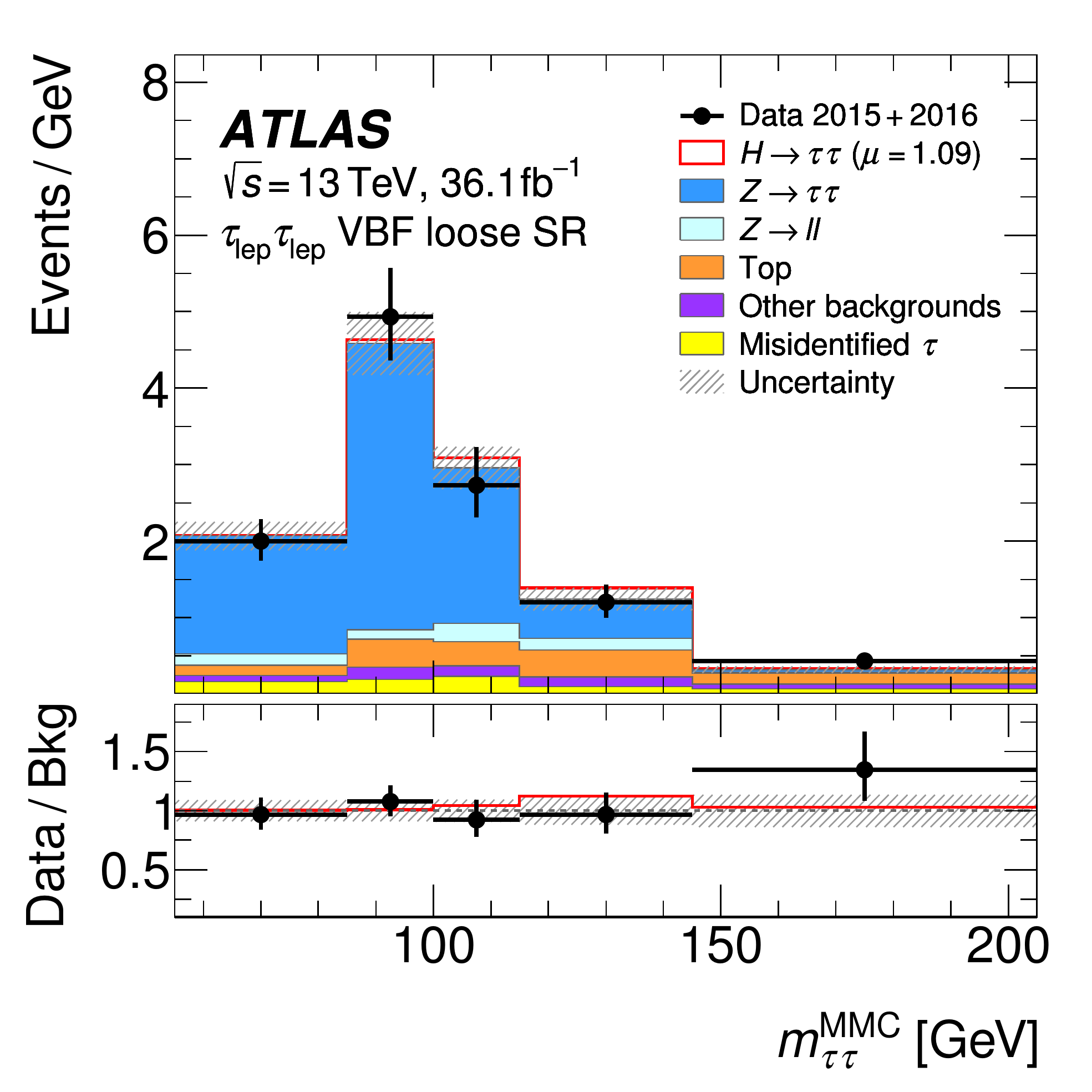}
    \includegraphics[width=0.32\linewidth]{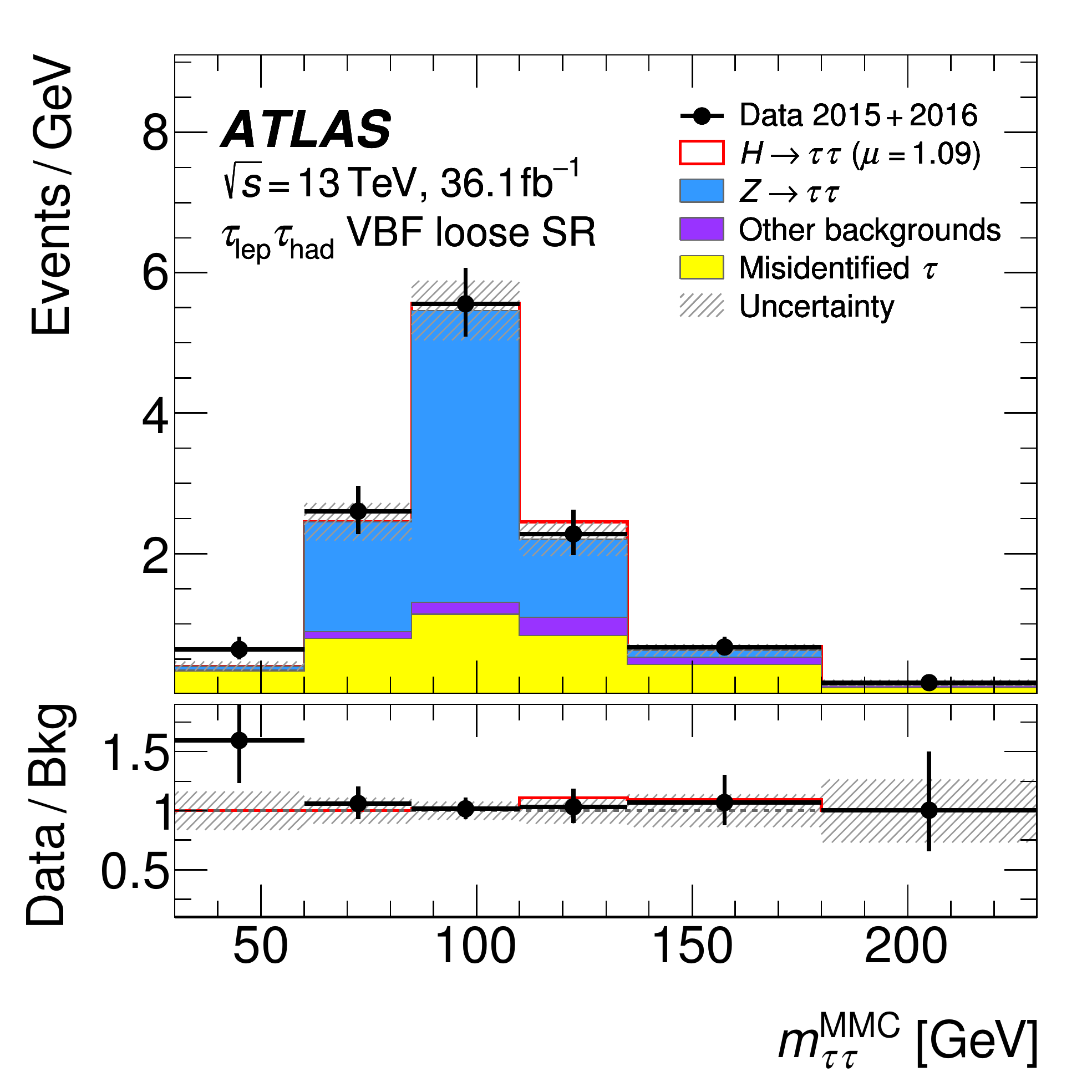}
    \includegraphics[width=0.32\linewidth]{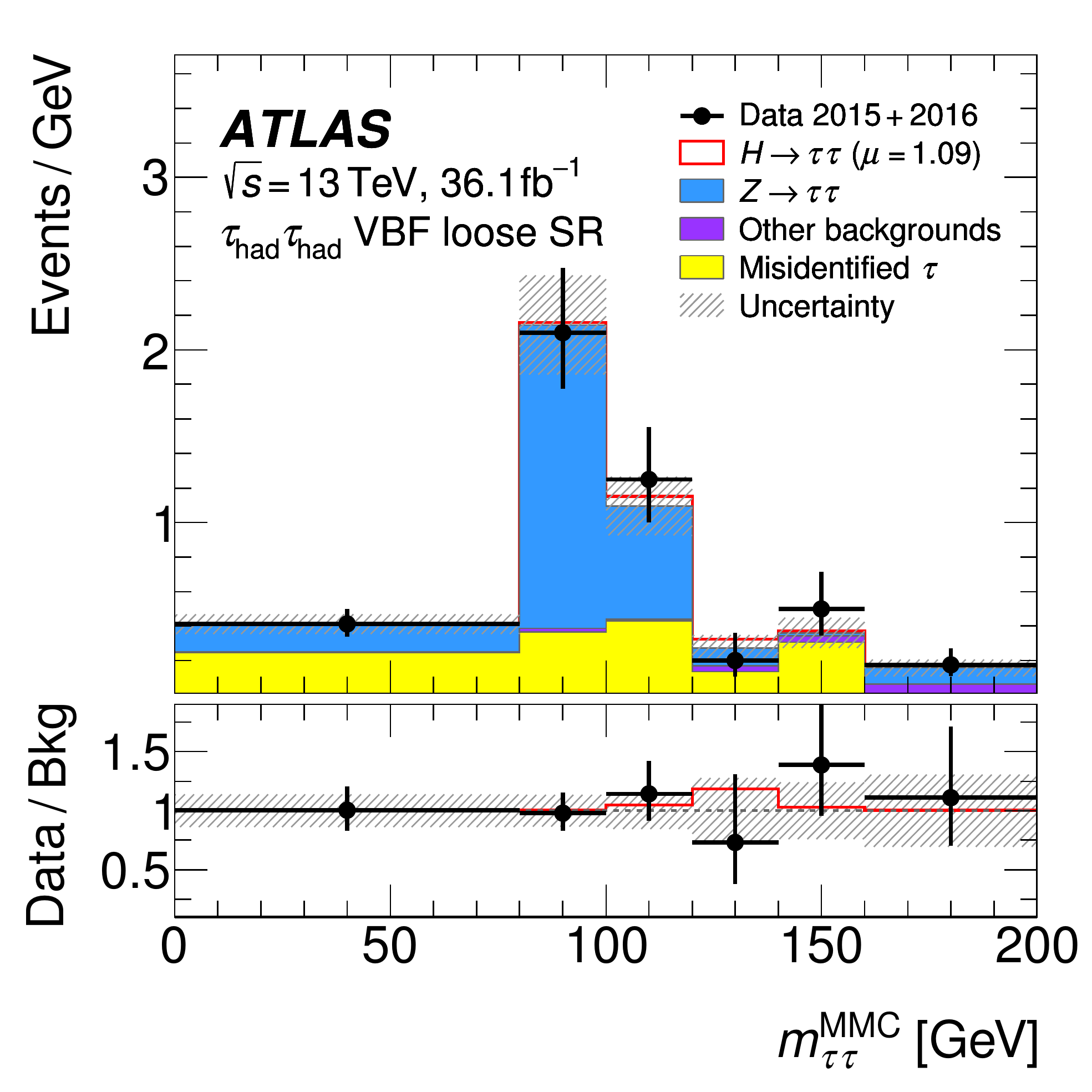}
  \caption{Observed and expected \mMMC distributions as used in the fit in all signal regions (SRs) in the \VBF
     category for the \tll (left), \tlhad (middle) and \thadhad (right) analysis channels.
     The bottom panels show the ratio of observed data events to expected background
     events (black points).
     The observed Higgs-boson signal ($\mu = 1.09$) is shown with the solid red line.
     Entries with values that would exceed the $x$-axis range are shown in the last bin of each distribution.
     The signal and background predictions are determined in the likelihood fit.
     The size of the combined statistical, experimental and theoretical uncertainties in the background is indicated by the hatched bands.}
  \label{fig:MMC-signal_vbf}
\end{figure}

\begin{figure}[htbp]
  \centering
    \includegraphics[width=0.32\linewidth]{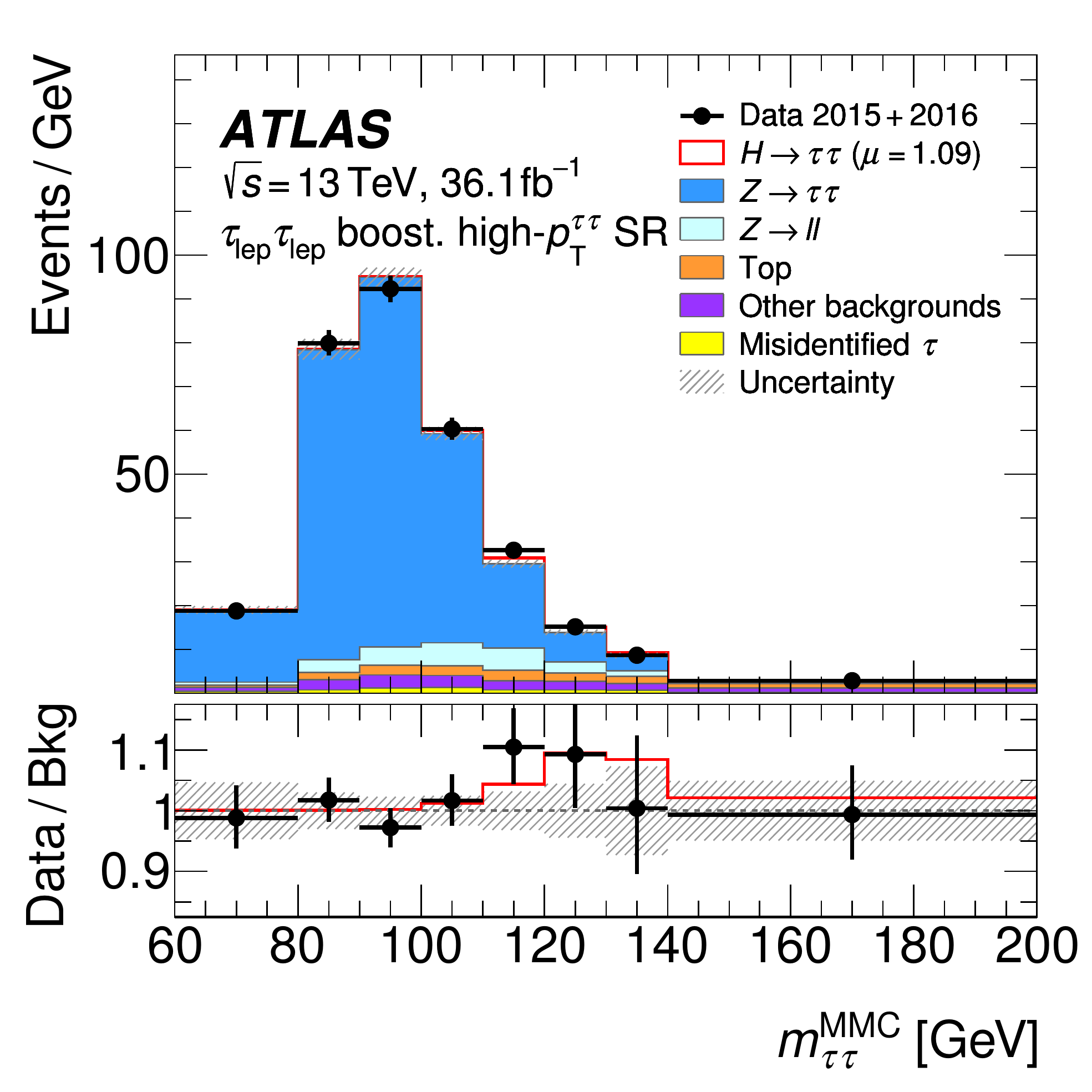}
    \includegraphics[width=0.32\linewidth]{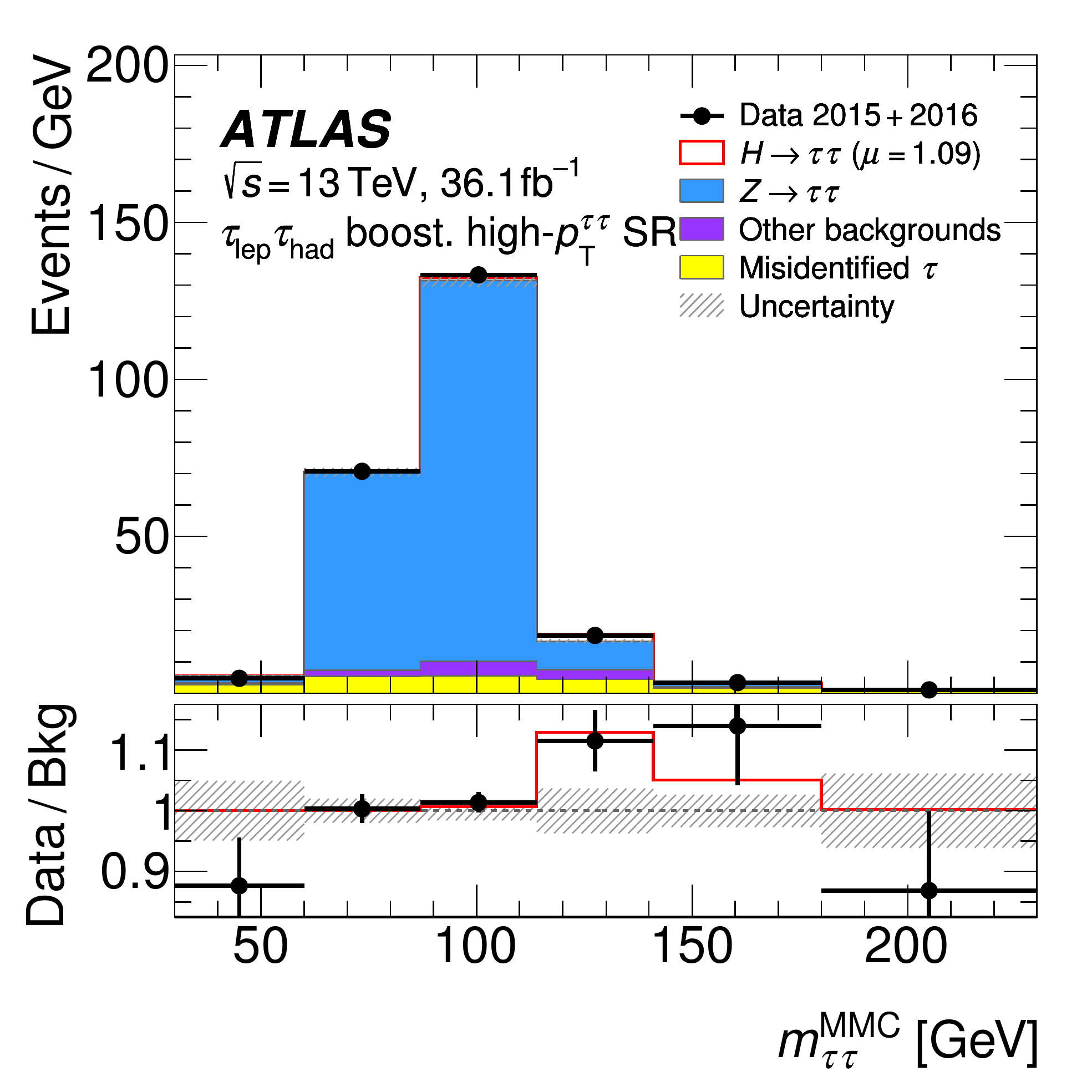}
    \includegraphics[width=0.32\linewidth]{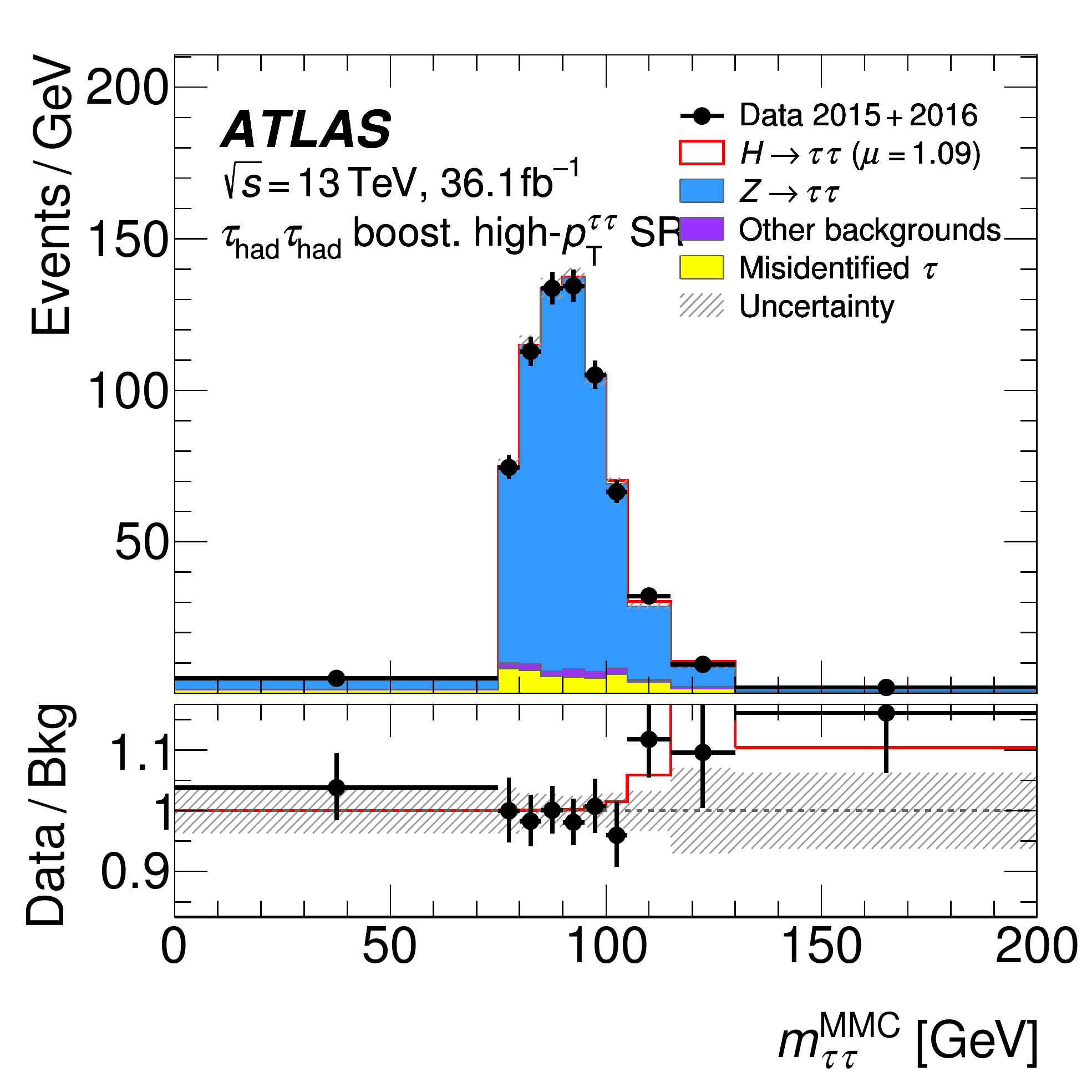}\\
    \includegraphics[width=0.32\linewidth]{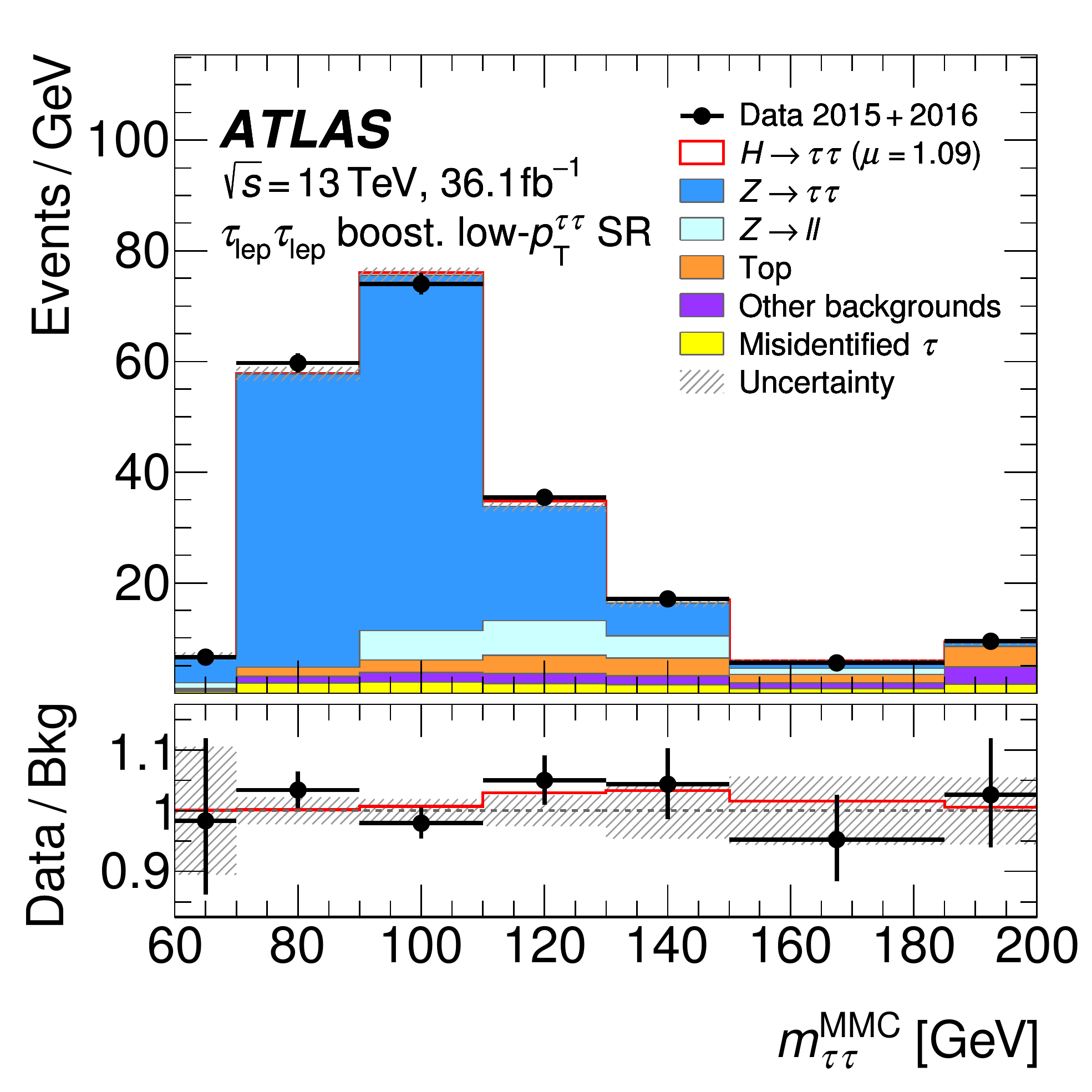}
    \includegraphics[width=0.32\linewidth]{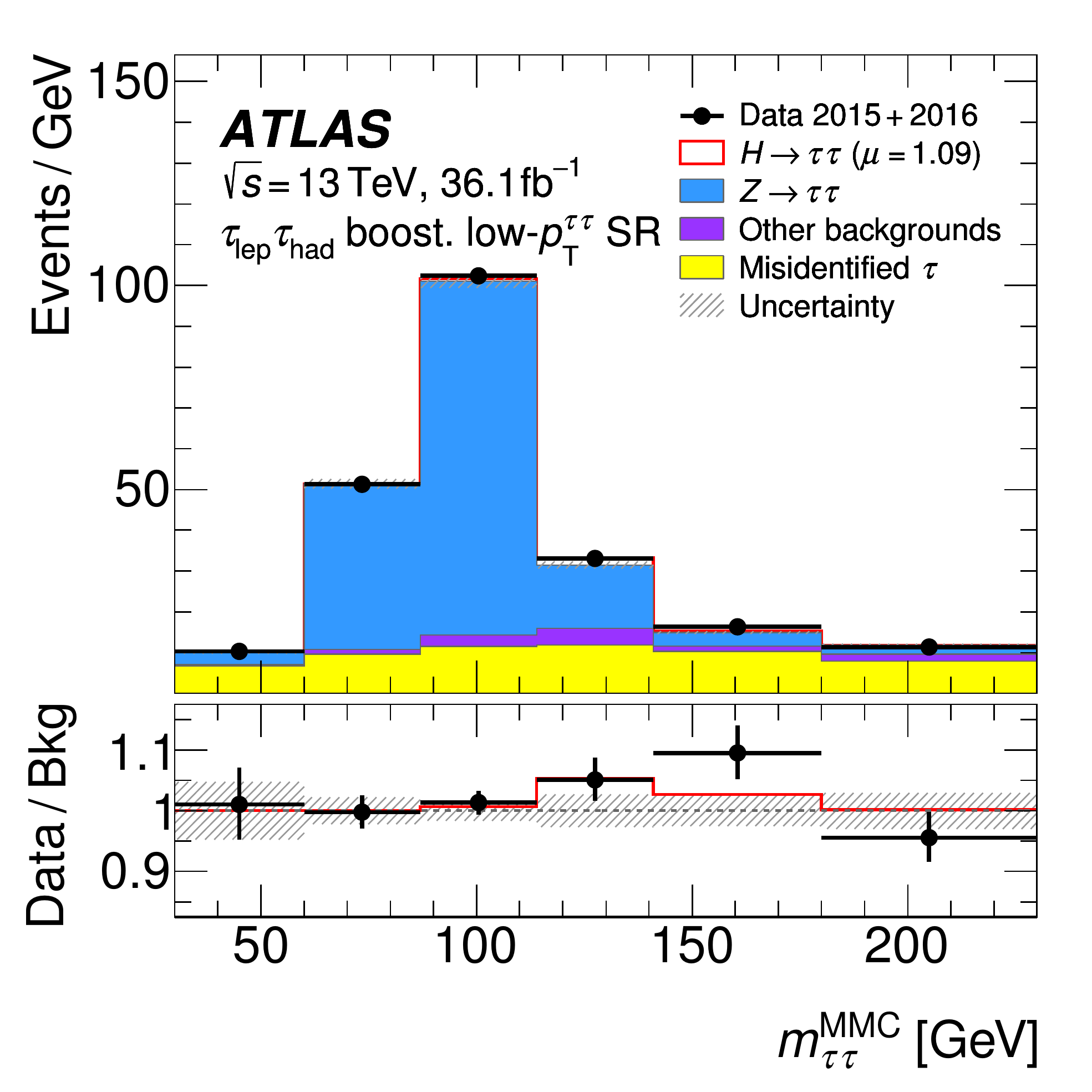}
    \includegraphics[width=0.32\linewidth]{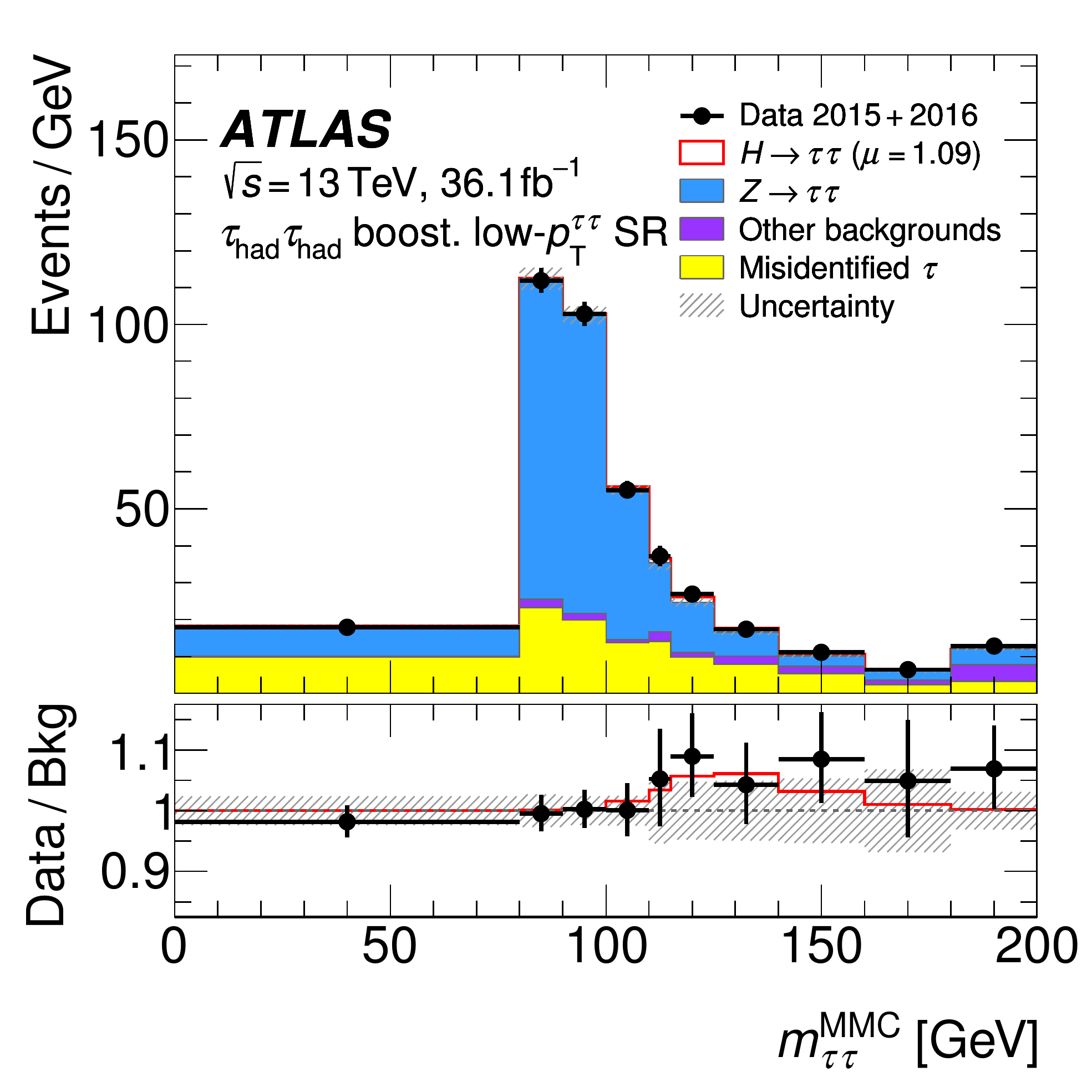}
  \caption{Observed and expected \mMMC distributions as used in the fit in all signal regions (SRs) in the boosted
     category for the \tll (left), \tlhad (middle) and \thadhad (right) analysis channels.
     The bottom panels show the ratio of observed data events to expected background
     events (black points).
     The observed Higgs-boson signal ($\mu = 1.09$) is shown with the solid red line.
     Entries with values that would exceed the $x$-axis range are shown in the last bin of each distribution.
     The signal and background predictions are determined in the likelihood fit.
     The size of the combined statistical, experimental and theoretical uncertainties in the background is indicated by the hatched bands.}
  \label{fig:MMC-signal_boosted}
\end{figure}

\Cref{fig:MMC-signal_vbf,fig:MMC-signal_boosted} show the \mMMC distributions in all signal regions with background predictions adjusted by the likelihood fit.

\FloatBarrier
\clearpage

\section*{Acknowledgments}


We thank CERN for the very successful operation of the LHC, as well as the
support staff from our institutions without whom ATLAS could not be
operated efficiently.

We acknowledge the support of ANPCyT, Argentina; YerPhI, Armenia; ARC, Australia; BMWFW and FWF, Austria; ANAS, Azerbaijan; SSTC, Belarus; CNPq and FAPESP, Brazil; NSERC, NRC and CFI, Canada; CERN; CONICYT, Chile; CAS, MOST and NSFC, China; COLCIENCIAS, Colombia; MSMT CR, MPO CR and VSC CR, Czech Republic; DNRF and DNSRC, Denmark; IN2P3-CNRS, CEA-DRF/IRFU, France; SRNSFG, Georgia; BMBF, HGF, and MPG, Germany; GSRT, Greece; RGC, Hong Kong SAR, China; ISF and Benoziyo Center, Israel; INFN, Italy; MEXT and JSPS, Japan; CNRST, Morocco; NWO, Netherlands; RCN, Norway; MNiSW and NCN, Poland; FCT, Portugal; MNE/IFA, Romania; MES of Russia and NRC KI, Russian Federation; JINR; MESTD, Serbia; MSSR, Slovakia; ARRS and MIZ\v{S}, Slovenia; DST/NRF, South Africa; MINECO, Spain; SRC and Wallenberg Foundation, Sweden; SERI, SNSF and Cantons of Bern and Geneva, Switzerland; MOST, Taiwan; TAEK, Turkey; STFC, United Kingdom; DOE and NSF, United States of America. In addition, individual groups and members have received support from BCKDF, CANARIE, CRC and Compute Canada, Canada; COST, ERC, ERDF, Horizon 2020, and Marie Sk{\l}odowska-Curie Actions, European Union; Investissements d' Avenir Labex and Idex, ANR, France; DFG and AvH Foundation, Germany; Herakleitos, Thales and Aristeia programmes co-financed by EU-ESF and the Greek NSRF, Greece; BSF-NSF and GIF, Israel; CERCA Programme Generalitat de Catalunya, Spain; The Royal Society and Leverhulme Trust, United Kingdom. 

The crucial computing support from all WLCG partners is acknowledged gratefully, in particular from CERN, the ATLAS Tier-1 facilities at TRIUMF (Canada), NDGF (Denmark, Norway, Sweden), CC-IN2P3 (France), KIT/GridKA (Germany), INFN-CNAF (Italy), NL-T1 (Netherlands), PIC (Spain), ASGC (Taiwan), RAL (UK) and BNL (USA), the Tier-2 facilities worldwide and large non-WLCG resource providers. Major contributors of computing resources are listed in Ref.~\cite{ATL-GEN-PUB-2016-002}.

\printbibliography

\clearpage
 
\begin{flushleft}
{\Large The ATLAS Collaboration}

\bigskip

M.~Aaboud$^\textrm{\scriptsize 35d}$,    
G.~Aad$^\textrm{\scriptsize 100}$,    
B.~Abbott$^\textrm{\scriptsize 127}$,    
O.~Abdinov$^\textrm{\scriptsize 13,*}$,    
B.~Abeloos$^\textrm{\scriptsize 131}$,    
D.K.~Abhayasinghe$^\textrm{\scriptsize 92}$,    
S.H.~Abidi$^\textrm{\scriptsize 166}$,    
O.S.~AbouZeid$^\textrm{\scriptsize 40}$,    
N.L.~Abraham$^\textrm{\scriptsize 155}$,    
H.~Abramowicz$^\textrm{\scriptsize 160}$
% ...
\bigskip
\\

% ...
$^{13}$Institute of Physics, Azerbaijan Academy of Sciences, Baku; Azerbaijan.\\
% ...
$^{35}$$^{(a)}$Facult\'e des Sciences Ain Chock, R\'eseau Universitaire de Physique des Hautes Energies - Universit\'e Hassan II, Casablanca;$^{(b)}$Facult\'{e} des Sciences, Universit\'{e} Ibn-Tofail, K\'{e}nitra;$^{(c)}$Facult\'e des Sciences Semlalia, Universit\'e Cadi Ayyad, LPHEA-Marrakech;$^{(d)}$Facult\'e des Sciences, Universit\'e Mohamed Premier and LPTPM, Oujda;$^{(e)}$Facult\'e des sciences, Universit\'e Mohammed V, Rabat; Morocco.\\
% ...
$^{40}$Niels Bohr Institute, University of Copenhagen, Copenhagen; Denmark.\\
% ...
$^{92}$Department of Physics, Royal Holloway University of London, Egham; United Kingdom.\\
% ...
$^{100}$CPPM, Aix-Marseille Universit\'e, CNRS/IN2P3, Marseille; France.\\
% ...
$^{127}$Homer L. Dodge Department of Physics and Astronomy, University of Oklahoma, Norman OK; United States of America.\\
% ...
$^{131}$LAL, Universit\'e Paris-Sud, CNRS/IN2P3, Universit\'e Paris-Saclay, Orsay; France.\\
% ...
$^{155}$Department of Physics and Astronomy, University of Sussex, Brighton; United Kingdom.\\
% ...
$^{160}$Raymond and Beverly Sackler School of Physics and Astronomy, Tel Aviv University, Tel Aviv; Israel.\\
% ...
$^{166}$Department of Physics, University of Toronto, Toronto ON; Canada.\\
% ...
$^{a}$ Also at Borough of Manhattan Community College, City University of New York, New York NY; United States of America.\\
$^{b}$ Also at Centre for High Performance Computing, CSIR Campus, Rosebank, Cape Town; South Africa.\\
$^{c}$ Also at CERN, Geneva; Switzerland.\\
$^{d}$ Also at CPPM, Aix-Marseille Universit\'e, CNRS/IN2P3, Marseille; France.\\
$^{e}$ Also at D\'epartement de Physique Nucl\'eaire et Corpusculaire, Universit\'e de Gen\`eve, Gen\`eve; Switzerland.\\
% ...
$^{*}$ Deceased

\end{flushleft}


\end{document}